\documentclass[a4paper,14pt]{article}
\usepackage[english]{babel}
\pdfoutput=1 

\usepackage{jheppub} 

\usepackage[T1,T2A]{fontenc}
\usepackage[utf8]{inputenc}

\usepackage{comment}

\usepackage{cases}

\newcommand{\bi}{\begin{itemize}}
    \newcommand{\ei}{\end{itemize}}
\newcommand{\bea}{\begin{eqnarray}}
    \newcommand{\eea}{\end{eqnarray}}
\newcommand{\bt}{\begin{tabular}}
    \newcommand{\et}{\end{tabular}}
\newcommand{\bc}{\begin{center}}
    \newcommand{\ec}{\end{center}}

\newcommand{\be}{\begin{equation}}
    \newcommand{\ee}{\end{equation}}
\newcommand{\ba}{\begin{array}}
    \newcommand{\ea}{\end{array}}

\newcommand{\lb}[1]{\label{#1}}

\def\bbox{{\,\lower0.9pt\vbox{\hrule \hbox{\vrule height 0.2 cm
                \hskip 0.2 cm \vrule height 0.2 cm}\hrule}\,}}
\newcommand{\dsl}{\pa \kern-0.5em /}

\makeatletter \@addtoreset{equation}{section} \makeatother

\def\slashchar#1{\setbox0=\hbox{$#1$}           
    \dimen0=\wd0                                 
    \setbox1=\hbox{/} \dimen1=\wd1               
    \ifdim\dimen0>\dimen1                        
    \rlap{\hbox to \dimen0{\hfil/\hfil}}      
    #1                                        
    \else                                        
    \rlap{\hbox to \dimen1{\hfil$#1$\hfil}}   
    /                                         
    \fi}

\pdfoutput=1

\title{\boldmath  Linearized $\mathcal{N}=2$ conformal supergravity in the harmonic approach}

\author[a]{Evgeny~Ivanov,}
\author[a,b]{Nikita~Zaigraev}

\affiliation[a]{Bogoliubov Laboratory of Theoretical Physics, JINR,\\141980 Dubna, Moscow region, Russia}
\affiliation[b]{Moscow Institute of Physics and Technology,\\ 141700 Dolgoprudny, Moscow region, Russia}

\emailAdd{eivanov@theor.jinr.ru}
\emailAdd{nikita.zaigraev@phystech.edu}

\abstract{ Using the harmonic superspace approach, we construct the superconformal harmonic action for $\mathcal{N}=2$ Weyl supermultiplet.
The fundamental objects of the theory are unconstrained analytic potentials
$h^{++\alpha\dot{\alpha}}, h^{++\alpha+}, h^{++\dot{\alpha}+}, h^{(+4)}$, which distinguishes our construction among the previously known ones.
An important role is played by the ``half-analyticity''
conditions introduced in \href{https://arxiv.org/abs/2407.08524}{arxiv:2407.08524 [hep-th]}.
The structure of the harmonic linearized $\mathcal{N}=2$ Weyl action to large extent repeats the structure of the $\mathcal{N}=2$ Maxwell action,
which suggests a conjecture on the possible structure
of the complete nonlinear $\mathcal{N}=2$ Weyl theory action in the harmonic superspace. We provide a detailed study of the rigid superconformal properties
of the proposed action  and prove its invariance in both harmonic and chiral superspaces. First steps are also undertaken towards constructing the nonlinear $\mathcal{N}=2$ Weyl action, based
on an analogy with $\mathcal{N}=2$ Maxwell action just mentioned and a generalization of the concept of half-analyticity to curved harmonic superspace.

\vspace{1.5cm}

\begin{flushright}
    \textit{To the memory of Luca Mezincescu and Kelly Stelle}
\end{flushright}
 }

\arxivnumber{2511.16325 }

\makeatletter
\gdef\@fpheader{}
\makeatother

\begin{document}

\maketitle
\flushbottom


\section{Introduction}

Harmonic superspace (HSS) is the most efficient method to deal with theories exhibiting the manifest extended $\mathcal{N}=2$ supersymmetry \cite{Galperin:1984av, 18}. In particular, $\mathcal{N}=2$ superconformal group admits a
geometric realization on the harmonic superspace coordinates \cite{18, Galperin:1985zv}, which was successfully used in a number of works (e.g., \cite{Eden:2000qp, Ivanov:2005qf}). In a recent paper
\cite{Buchbinder:2024pjm}, based on our previous results on the harmonic formulation of $\mathcal{N}=2$ higher-spin theories \cite{Buchbinder:2021ite, Buchbinder:2022kzl, Buchbinder:2022vra} (see \cite{Buchbinder:2025ceg}
for a short review), we proposed the harmonic description of $\mathcal{N}=2$ higher-spin superconformal multiplets\footnote{The superfield formulation of $\mathcal{N}=2$ superconformal higher-spin multiplets in the conventional superspace was first
elaborated in \cite{Kuzenko:2021pqm} and for $\mathcal{N}=2$  conformal gravitino in \cite{Hutchings:2023iza}. The corresponding superfield prepotentials  can be derived from  the harmonic analytic prepotentials by imposing a specific harmonic-independent gauge. For this
reason, it seems preferable to operate at once in terms of harmonic superspace. }. These multiplets are the natural higher-spin generalization of the $\mathcal{N}=2$ conformal supergravity
 Weyl multiplet \cite{Galperin:1987ek}  and can be used to construct a number of $\mathcal{N}=2$
 higher-spin theories\footnote{The gauge $\mathcal{N}=2$ multiplets with the highest half-integer
 spin have different structure and origin in harmonic superspace, as was shown for the $\mathcal{N}=2$ superconformal gravitino multiplet \cite{Ivanov:2024bsb}.}. In particular,
 generalizing the method of superconformal compensators \cite{Galperin:1987ek, Galperin:1987em, 18, Ivanov:2022vwc, Sokatchev:1988aa}, one can construct various non-conformal theories on flat or AdS backgrounds.
Moreover, taking an infinite tower of such multiplets one can construct a fully consistent interacting theory of the hypermultiplet and, using it, formally define a consistent $\mathcal{N}=2$ higher-spin conformal supergravity
as the logarithmically divergent part of the hypermultiplet effective action \cite{Buchbinder:2024pjm, Buchbinder:2025ceg, Kuzenko:2024vms}.

One of the important open problems posed in \cite{Buchbinder:2024pjm} is the construction of \textit{dynamical harmonic superspace actions} for $4D, \mathcal{N}=2$ higher-spin superconformal multiplets. It seems rather surprising
that the harmonic actions of this kind  were never presented before in the literature, even for $\mathcal{N}=2$ conformal supergravity multiplet at the linearized level, despite the extensive study of invariants of the
extended conformal supergravities (see, e.g., \cite{Butter:2013lta, Butter:2015nza, Butter:2016qkx}).

In this paper we present the linearized harmonic action for the $\mathcal{N}=2$ conformal supergravity multiplet, proceeding solely from the description of the latter
by harmonic analytic prepotentials, and study the rigid superconformal
invariance of the action constructed\footnote{Note that the superfield structure of such an action (and its higher-spin generalization) was already sketched in the review \cite{Buchbinder:2025ceg}.}.
The structure of the action almost literally
repeats that of the linearized Weyl-squared invariant in the ``minimal'' $\mathcal{N}=2$ Einstein supergravity, constructed in \cite{Ivanov:2024gjo} (see also \cite{Ivanov:2025mld}).
However, verification of its
superconformal invariance requires a separate study and reveals a number of the previously unknown peculiarities of the harmonic superspace realization of ${\cal N}=2$ superconformal group.
The basic geometric principle underlying the linearized harmonic superspace $\mathcal{N}=2$ Weyl action (and, hopefully, its full nonlinear version) is the harmonic ``half-analyticity''.

\medskip

It is worth noting that the superfield action for $\mathcal{N}=2$ conformal supergravity in the conventional superspace approaches has been known for quite a long time,
both at the linear and at the nonlinear levels (see, e.g., \cite{Howe:1981qj, Gates:1981qq, Howe:1981gz, Fradkin:1985am, Muller:1989uhj, Kuzenko:2008ep}).
 A generalization of the linearized action to arbitrary spins
was constructed in refs. \cite{Kuzenko:2021pqm, Hutchings:2023iza} in the framework of
the $\mathcal{N}=2$ conformal superspace approach \cite{Butter:2011sr} which provides the manifest superconformal invariance.
Our goal here is to construct such actions in the harmonic superspace,
which implies a natural geometric way of defining unconstrained gauge prepotentials, as well as of their superconformal and gauge transformations,
the notions which are just postulated in other approaches. Moreover, understanding of the harmonic structure of such theories, even at the linearized level, is extremely important for
constructing their most general actions, setting up quantum effective actions, etc. All the previously known results are recovered after fixing a special
harmonic-independent gauge (see, e.g., discussion in \cite{Buchbinder:2024pjm, Ivanov:2024bsb, Kuzenko:2024vms} and in Appendix \ref{Appendix}).

\medskip

The paper is organized as follows. In the rest of this section we briefly recall some necessary details of the harmonic superspace approach.
In section \ref{eq: sec 2} we present the harmonic superspace realization of $\mathcal{N}=2$ superconformal group, including the superconformal transformations of derivatives
and superspace integration measures.  Section \ref{eq: sec 3} deals with the well-known example of free $\mathcal{N}=2$ Maxwell theory and its superconformal invariance,
in both the harmonic superspace and the chiral superspace. This example illustrates some peculiarities of our consideration, which
are then generalized to $\mathcal{N}=2$ conformal supergravity. In section \ref{eq: sec 4} we present the detailed discussion of the linearized $\mathcal{N}=2$ super Weyl action
in harmonic superspace and prove its rigid superconformal invariance.  Section \ref{sec 5} is devoted to an attempt to generalize the linearized $\mathcal{N}=2$ super Weyl action to
the full nonlinear level. It is based on the generalization of the half-analyticity principle to curved harmonic superspace in the so called ``hybrid'' basis \cite{Galperin:1987em}.
In the concluding section \ref{sec 6} we highlight our  main results and
mention their possible generalizations to $\mathcal{N}=2$ higher-spins.
Appendix \ref{App: sc conf} is devoted to the discussion of superconformal transformations in different superspace bases: analytic, central and chiral.
In Appendix  \ref{Appendix} we specify another, the Mezincescu-type prepotential of $\mathcal{N}=2$ conformal supergravity, give the explicit expression for super Weyl
tensor in a harmonic-independent gauge and derive the corresponding Bianchi identity. In the Appendix  \ref{eq: App C}, we present various useful relations for the superparameters of superconformal transformations.

\subsection*{ABC of $4D,\mathcal{N}=2$ harmonic superspace}

Throughout the paper we will use the analytical parametrization of harmonic superspace,
\begin{equation}
    Z_A = \left\{\zeta^M, \theta^{-\alpha}, \bar{\theta}^{-\dot{\alpha}},  u^\pm_i  \right\},
\end{equation}
where $\zeta^M$ denotes the analytic   coordinates
\begin{equation}
    \zeta^M = \left\{ x^{\alpha\dot{\alpha}}, \theta^{+\alpha}, \bar{\theta}^{+\dot{\alpha}} \right\}.
\end{equation}
The harmonic coordinates $u^\pm_i$ parametrize the internal 2-sphere $SU(2)/U(1)$ and satisfy the condition $u^{+i} u^-_i =1$.

$\mathcal{N}=2$ supersymmetry is realized on the above coordinates as
\begin{equation}\label{eq: rig susy}
    \delta_\epsilon x^{\alpha\dot{\alpha}} = -4i \left(\epsilon^{-\alpha} \bar{\theta}^{+\dot{\alpha}} + \theta^{+\alpha} \bar{\epsilon}^{-\dot{\alpha}} \right),
    \quad
    \delta_\epsilon \theta^{\pm\alpha} = \epsilon^{\pm\alpha},
    \quad
    \delta_\epsilon \bar{\theta}^{\pm\dot{\alpha}}
    = \bar{\epsilon}^{\pm\dot{\alpha}},
    \quad
    \delta_\epsilon u^\pm_i = 0,
\end{equation}
where $\epsilon^{\pm \hat{\alpha}} := \epsilon^{\hat{\alpha}i} u^\pm_i$.
The analytic superspace
$$
\mathbb{HA}^{4+2|4} = \{\zeta^M,u^\pm\}
$$
is closed under these transformations.

In the analytic basis, the harmonic derivatives read :
\begin{equation}\label{eq:D0}
    \begin{split}
        &\mathcal{D}^{++} = \partial^{++} - 4i \theta^{+\rho} \bar{\theta}^{+\dot{\rho}} \partial_{\rho\dot{\rho}} + \theta^{+\hat{\rho}} \partial^+_{\hat{\rho}},
        \\
        &\mathcal{D}^{--} = \partial^{--} - 4i \theta^{-\rho} \bar{\theta}^{-\dot{\rho}} \partial_{\rho\dot{\rho}} + \theta^{-\hat{\rho}} \partial^-_{\hat{\rho}},
        \\
        &\mathcal{D}^0 = \partial^0 + \theta^{+\hat{\rho}} \partial^-_{\hat{\rho}}
        -
        \theta^{-\hat{\rho}} \partial^+_{\hat{\rho}},
    \end{split}
\end{equation}
where we used the notations
\begin{equation}
    \partial^{++} = u^{+i} \frac{\partial}{\partial u^-_i},
    \qquad
    \partial^{--} = u^{-i} \frac{\partial}{\partial u^+_i},
    \qquad
    \partial^0 = u^{+i} \frac{\partial}{\partial u^+_i} - u^{-i} \frac{\partial}{\partial u^-_i}
\end{equation}
and $\hat{\alpha} := (\alpha, \dot\alpha)$. The harmonic derivatives satisfy the $\mathfrak{su}(2)$ algebra relations $[\mathcal{D}^{++}, \mathcal{D}^{--}] = \mathcal{D}^0$, $[\mathcal{D}^0, \mathcal{D}^{\pm\pm}] = \pm 2 \mathcal{D}^{\pm\pm} $.

The covariant spinor derivatives in the analytic basis are expressed as:
\begin{equation}\label{eq: cov der}
    \begin{split}
        &\mathcal{D}^+_{\hat{\alpha}} = \partial^+_{\hat{\alpha}} = (\partial^+_{{\alpha}},\;\bar{\partial}^+_{\dot{\alpha}}),
        \\
        &\mathcal{D}^-_\alpha = - \partial^-_\alpha + 4i \bar{\theta}^{-\dot{\alpha}}\partial_{\alpha\dot{\alpha}},
        \\
        &\bar{\mathcal{D}}^-_{\dot{\alpha}} = - \bar{\partial}^-_{\dot{\alpha}} - 4i \theta^{-\alpha}\partial_{\alpha\dot{\alpha}}\,,
    \end{split}
\end{equation}
and satisfy the algebra:
\begin{equation}\label{eq: rel cov derivatives}
    \begin{split} &\mathcal{D}^\pm_{\hat{\alpha}} = [\mathcal{D}^{\pm\pm}, \mathcal{D}^{\mp}_{\hat{\alpha}}],
        \\
        & [\mathcal{D}^{\pm\pm}, \mathcal{D}^\pm_{\hat{\alpha}}] = 0,
        \\& \{\mathcal{D}^+_{\alpha}, \bar{\mathcal{D}}^-_{\dot{\alpha}}\} = - \{\mathcal{D}^-_{\alpha}, \bar{\mathcal{D}}^+_{\dot{\alpha}}\}   = -4i \partial_{\alpha\dot{\alpha}}.
    \end{split}
\end{equation}
The analytic harmonic superfields are subjected to the Grassmann analyticity constraints
\bea
\mathcal{D}^+_{\hat{\alpha}}\Phi(Z) =0\,. \label{AnalConstr}
\eea
In the analytic basis, due to the ``shortness'' of $\mathcal{D}^+_{\hat{\alpha}}$, eq. \eqref{AnalConstr} implies
$$
\Phi(Z) \, \Rightarrow \, \Phi(\zeta, u)\,.
$$
The covariant derivatives are invariant under rigid $\mathcal{N}=2$ supersymmetry, but exhibit non-trivial transformation properties with respect to $\mathcal{N}=2$
superconformal symmetry (see below).


\section{$\mathcal{N}=2$ superconformal symmetry}\label{eq: sec 2}

$\mathcal{N}=2$  superconformal symmetry is realized by special coordinate transformations which are a particular case of those leaving the analytic subspace
$\mathbb{HA}^{4+2|4} = \{\zeta^M,u^\pm\}$ invariant:
\begin{equation}\label{eq: an diff}
    \delta \zeta^M =  \lambda^M(\zeta, u),
    \quad
    \delta \theta^{-\hat{\alpha}} = \lambda^{- \hat{\alpha}} (\zeta, \theta^- ,u),
    \quad
    \delta u^+_i = \lambda^{++} (\zeta, u) u^-_i,
    \quad
    \delta u^-_i = 0.
\end{equation}

It is convenient to represent these transformations through the differential operator
\begin{equation}\label{eq: hat Lambda}
    \hat{\Lambda} = \lambda \cdot \partial = \lambda^{M} \partial_{M} + \lambda^{-\hat{\alpha}} \partial^+_{\hat{\alpha}}
    +
    \lambda^{++}\partial^{--},
    \qquad
    \delta Z^A := [\hat{\Lambda}, Z^A]\,,
\end{equation}
which preserves the analyticity condition for an arbitrary analytic superfield $\Phi(\zeta, u)$,
\begin{equation}\label{eq: Lambda op an}
    [\mathcal{D}^+_{\hat{\alpha}}, \hat{\Lambda}] \Phi(\zeta, u) = 0.
\end{equation}
The rigid $\mathcal{N}=2$ superconformal transformations correspond to the particular case when this operator  satisfies the equation:
\begin{equation}\label{eq: N=2 sc}
    \delta_{sc}\mathcal{D}^{++} := [ \hat{\Lambda}_{sc}, \mathcal{D}^{++}] = -  \lambda^{++}_{sc} \mathcal{D}^0\,.
\end{equation}
It determines the super-Killing vectors of the flat $\mathcal{N}=2$ conformal supergravity background and amounts
to the system of superfield equations for the $\lambda$-parameters:
\begin{equation}\label{eq: sc equations}
    \begin{cases}
        \mathcal{D}^{++} \lambda_{sc}^{\alpha\dot{\alpha}}
        +
        4i \lambda_{sc}^{+\alpha} \bar{\theta}^{+\dot{\alpha}}
        +
        4i \theta^{+\alpha} \bar{\lambda}_{sc}^{+\dot{\alpha}} = 0,
        \\
        \mathcal{D}^{++} \lambda_{sc}^{+\hat{\alpha}} - \theta^{+\hat{\alpha}} \lambda_{sc}^{++} = 0,
        \\
        \mathcal{D}^{++} \lambda_{sc}^{-\hat{\alpha}} - \lambda_{sc}^{+\hat{\alpha}}
        + \theta^{-\hat{\alpha}} \lambda_{sc}^{++}  = 0,
        \\
        \mathcal{D}^{++} \lambda_{sc}^{++} = 0.
    \end{cases}
\end{equation}
The most general solution of eqs. \eqref{eq: sc equations} reads \cite{Buchbinder:2024pjm}:
\begin{equation}\label{eq:superconformal symmetry}
    \begin{split}
            \lambda_{sc}^{\alpha\dot{\alpha}}
            =\;&
            x^{\dot{\alpha}\beta} k_{\beta\dot{\beta}} x^{\dot{\beta}\alpha}
            +
            {\rm d} x^{\alpha\dot{\alpha}}
            -
            4i \theta^{+\alpha} \bar{\theta}^{+\dot{\alpha}} \lambda^{(ij)}u^-_i u^-_j
            -
            4i \left(x^{\alpha\dot{\beta}}\bar{\eta}_{\dot{\beta}}^i \bar{\theta}^{+\dot{\alpha}}
            +
            \theta^{+\alpha} \eta^i_{\beta} x^{\beta\dot{\alpha}}
            \right) u^-_i,
            \\
            \lambda^{+\alpha}_{sc}
            =\;&
            \frac{1}{2}
            ({\rm d} + i {\rm r})\theta^{+\alpha}
            +
            x^{\alpha\dot{\beta}} k_{\beta\dot{\beta}} \theta^{+\beta}
            +
            x^{\alpha\dot{\beta}}  \bar{\eta}^i_{\dot{\beta}} u^+_i
            +
            \theta^{+\alpha}
            \left( \lambda^{(ij)}u^+_i u^-_j
            +
            4i \theta^{+\beta} \eta^i_{\beta} u^-_i
            \right),
            \\
            \bar{\lambda}^{+\dot{\alpha}}_{sc}
            =\;&
            \frac{1}{2}
            ({\rm d} - i {\rm r})\bar{\theta}^{+\dot{\alpha}}
            +
            x^{\dot{\alpha}\beta} k_{\beta\dot{\beta}} \bar{\theta}^{+\dot{\beta}}
            +
            x^{\beta\dot{\alpha}}  \eta^i_{\beta} u^+_i
            +
            \bar{\theta}^{+\dot{\alpha}}
            \left( \lambda^{(ij)}u^+_i u^-_j
            -
            4i \bar{\theta}^{+\dot{\beta}} \bar{\eta}^i_{\dot{\beta}} u^-_i
            \right),
            \\
            \lambda^{-\alpha}_{sc} =\;&
            \frac{1}{2} ({\rm d}+i{\rm r})\theta^{-\alpha}
            +
            x^{\alpha\dot{\beta}}
            k_{\beta\dot{\beta}}   \theta^{-\beta}
            -
            2i (\theta^-)^2 \bar{\theta}^+_{\dot{\beta}} k^{\dot{\beta}\alpha}
            + \left( x^{\alpha\dot{\beta}} + 4i \theta^{-\alpha}\bar{\theta}^{+\dot{\beta}} \right) \bar{\eta}^i_{\dot{\beta}} u^-_i
            \\& +
            4i \eta^i_{\beta} \theta^{-\beta} \left( \theta^{-\alpha} u^+_i - \theta^{+\alpha} u^-_i \right)
            +
            \lambda^{ij} u^-_i \left( u^-_j \theta^{+\alpha} - u^+_j \theta^{-\alpha} \right),
            \\
            \bar{\lambda}^{-\dot{\alpha}}_{sc}
            =\;&
            \frac{1}{2}({\rm d}-i{\rm r}) \bar{\theta}^{-\dot{\alpha}}
            +
            x^{\dot{\alpha}\beta} k_{\beta\dot{\beta}} \bar{\theta}^{-\dot{\beta}}
            -
            2i (\bar{\theta}^-)^2 \theta^+_{\beta} k^{\beta\dot{\alpha}}
            +
            \left( x^{\beta\dot{\alpha}} + 4i \theta^{+\beta} \bar{\theta}^{-\dot{\alpha}} \right) \eta^i_{\beta} u^-_i
            \\&
            -
            4i \bar{\eta}^i_{\dot{\beta}} \bar{\theta}^{-\dot{\beta}} \left( \bar{\theta}^{-\dot{\alpha}} u^+_i - \bar{\theta}^{+\dot{\alpha}} u^-_i \right)
            +
            \lambda^{ij} u^-_i \left( u^-_j \bar{\theta}^{+\dot{\alpha}} - u^+_j \bar{\theta}^{-\dot{\alpha}} \right),
            \\
            \lambda^{++}_{sc} =\;&
            \lambda^{ij} u^+_i u^+_j
            +
            4i \theta^{+\beta} \bar{\theta}^{+\dot{\beta}} k_{\beta\dot{\beta}}
            +
            4i \left( \theta^{+\beta} \eta^i_{\beta} + \bar{\eta}^i_{\dot{\beta}} \bar{\theta}^{+\dot{\beta}}  \right) u^+_i.
    \end{split}
\end{equation}
The parameters $k_{\alpha\dot{\alpha}}$,  ${\rm d}$, ${\rm r}$, $\lambda^{(ij)}$ and $\eta^i_\alpha, \bar{\eta}^i_{\dot{\alpha}}$ correspond to the special conformal transformations, dilatations, $U(1)_R$ symmetry, $SU(2)_R$
symmetry and the conformal supersymmetry, respectively. The transformations  \eqref{eq: rig susy} and \eqref{eq:superconformal symmetry}, together with 4-translations, supersymmetry and Lorentz~rotations,
\begin{equation}
    \begin{split}
        &\lambda^{\alpha\dot{\alpha}}
        =
        a^{\alpha\dot{\alpha}}
        +
        l^{(\alpha}_{\;\beta)}x^{\beta\dot{\alpha}}
        +
        \bar{l}^{(\dot{\alpha}}_{\;\dot{\beta})} x^{\alpha\dot{\beta}}
        -4i \left(\epsilon^{-\alpha} \bar{\theta}^{+\dot{\alpha}} + \theta^{+\alpha} \bar{\epsilon}^{-\dot{\alpha}} \right),
        \\
        &\lambda^{\pm\alpha} = \epsilon^{\pm \alpha} + l^{(\alpha}_{\;\beta)} \theta^{\pm\beta},
        \quad
        \bar{\lambda}^{\pm\dot{\alpha}} = \bar{\epsilon}^{\pm \dot{\alpha}} + \bar{l}^{(\dot{\alpha}}_{\;\dot{\beta})} \bar{\theta}^{\pm\dot{\beta}}\,,
    \end{split}
\end{equation}~form $4D, \mathcal{N}=2$ superconformal algebra $\mathfrak{su}(2,2|2)$.

The full set of the parameters of $\mathcal{N}=2$ superconformal transformations satisfies a number of useful identities. From the explicit form of $\lambda^{-\hat{\alpha}}$, we obtain
\begin{subequations}\label{eq: sc der prop}
    \begin{equation}\label{eq: 28a}
        \mathcal{D}^{--}    \lambda^{-\hat{\alpha}}_{sc}    = 0.
    \end{equation}
    Taking this property into account and acting on the third equation of the system \eqref{eq: sc equations}, we obtain
    \begin{equation}\label{eq: lambda - repr}
        \lambda_{sc}^{-\hat{\alpha}} = \mathcal{D}^{--} \left( \lambda_{sc}^{+\hat{\alpha}}
        - \theta^{-\hat{\alpha}} \lambda_{sc}^{++}\right).
    \end{equation}
    This property has a simple explanation in the covariant basis; see the appendix \ref{eq: App C}.
    The parameters $\lambda^{-\hat{\alpha}}$ also satisfy the chirality conditions
    \begin{equation}\label{eq: sc b}
        \bar{\mathcal{D}}^\pm_{\dot{\alpha}} \lambda_{sc}^{-\alpha} = 0,
        \qquad
        \mathcal{D}^\pm_\alpha \bar{\lambda}_{sc}^{-\dot{\alpha}} = 0\,,
    \end{equation}
    as well as the relations:
    \begin{equation}\label{eq: sc c}
        \mathcal{D}^{-(\alpha}\lambda_{sc}^{-\beta)} = 0,
        \qquad
        \bar{\mathcal{D}}^{-(\dot{\alpha}}\bar{\lambda}_{sc}^{-\dot{\beta})} = 0.
    \end{equation}
    Together with the equation \eqref{eq: 28a} and third equation in \eqref{eq: sc equations},  it ensures the consistency of the homogeneous transformation law $\delta \Phi_{\alpha} = - \left( \mathcal{D}^+_{(\alpha} \lambda^{-\beta)}_{sc} \right)  \Phi_{\beta} $ for the harmonically independent superfield~$\mathcal{D}^{\pm\pm} \Phi_{\alpha} = 0 $.
    Applying $\bar{\mathcal{D}}^+_{\dot{\beta}}$ and $\mathcal{D}^+_\beta$ to the equations   \eqref{eq: sc c} yields
    \begin{equation}\label{eq: sc d}
        \partial^{(\alpha\dot{\beta}} \lambda^{\pm\beta)}_{sc} = 0,
        \qquad
        \partial^{(\dot{\alpha}\beta} \bar{\lambda}^{\pm\dot{\beta})}_{sc} = 0.
    \end{equation}
    Relations \eqref{eq: sc b} and \eqref{eq: sc d} also imply the corollaries:
    \begin{equation}
        \bar{\mathcal{D}}^{\pm}_{\dot{\beta}} \mathcal{D}^+_{(\alpha} \lambda^{-\beta)}_{sc} = 0,
            \qquad
        \mathcal{D}^{\pm}_\beta  \bar{\mathcal{D}}^+_{(\dot{\alpha}} \bar{\lambda}^{-\dot{\beta})}_{sc} = 0.
    \end{equation}
    These relations show that the homogeneous transformation law $\delta \Phi_{\alpha} = - \left( \mathcal{D}^+_{(\alpha} \lambda^{-\beta)}_{sc} \right)  \Phi_{\beta}$ is compatible with $\mathcal{N}=2$ chirality.
    Action of $(\mathcal{D}^+)^2$ on \eqref{eq: lambda - repr}  yields
    \begin{equation}\label{eq: sc e}
        \;\;\;   (\mathcal{D}^+)^2 \lambda_{sc}^{-\alpha} = 2 \mathcal{D}^{-\alpha} \lambda^{++}_{sc},
        \qquad
        (\bar{\mathcal{D}}^+)^2 \bar{\lambda}_{sc}^{-\dot{\alpha}}
        =
        - 2 \bar{\mathcal{D}}^{-\dot{\alpha}} \lambda_{sc}^{++}.
    \end{equation}
    One can also derive an equality relating the $\lambda^-$ parameters of different chirality:
    \begin{equation}\label{eq: sc f}
        \begin{split}
            &\mathcal{D}^{--}\mathcal{D}^-_\alpha \lambda^{++}_{sc}
            =
            +(\mathcal{D}^+ \mathcal{D}^-) \lambda^-_{sc\,\alpha}
            =
            -2i \partial_{\alpha\dot{\beta}} \bar{\lambda}^{-\dot{\beta}}_{sc}
            =
            -4i \left( k_{\alpha\dot{\beta}} \bar{\theta}^{-\dot{\beta}} + \eta^i_\alpha u^-_i \right),
            \\
            & \mathcal{D}^{--}\bar{\mathcal{D}}^-_{\dot{\alpha}} \lambda^{++}_{sc}
            =
            -(\bar{\mathcal{D}}^+ \bar{\mathcal{D}}^-) \bar{\lambda}^-_{sc\,\dot{\alpha}}
            =
            +2i \partial_{\beta\dot{\alpha}} \lambda^{-\beta}_{sc}
            =
            +4i\, \Big(k_{\beta\dot{\alpha}} \theta^{-\beta} + \bar{\eta}^i_{\dot{\alpha}} u^-_i  \Big).
        \end{split}
    \end{equation}
From these relations, using the property $\partial_{\alpha\dot{\alpha}} \lambda^{++}_{sc} = 0$, we obtain
    \begin{equation}\label{eq: sc g}
        \partial_{\alpha\dot{\beta}} \bar{\mathcal{D}}^+_{\dot{\alpha}} \bar{\lambda}^{-\dot{\beta}}_{sc}
        =
        \partial_{\beta\dot{\alpha}} \mathcal{D}^+_{\alpha} \lambda^{-\beta}_{sc} = \frac{i}{2} \mathcal{D}^-_\alpha \bar{\mathcal{D}}^-_{\dot{\alpha}} \lambda^{++}_{sc}
        =
        2 k_{\alpha\dot{\alpha}}.
    \end{equation}
\end{subequations}

In the next subsections we will need the superconformal transformation laws of the harmonic, spinor and vector derivatives.

\subsection{Superconformal transformations of harmonic derivatives}

Under superconformal transformations the harmonic derivatives \eqref{eq:D0} transform as
\begin{equation}
    \begin{split}
        \delta_{sc}\mathcal{D}^{++} &= [\hat{\Lambda}_{sc}, \mathcal{D}^{++}] =  - \lambda_{sc}^{++} \mathcal{D}^0,
        \\
        \delta_{sc} \mathcal{D}^{--} &= [\hat{\Lambda}_{sc}, \mathcal{D}^{--}]  =  - (\mathcal{D}^{--}\lambda_{sc}^{++}) \mathcal{D}^{--} ,
        \\
        \delta_{sc} \mathcal{D}^0 &= [\hat{\Lambda}_{sc}, \mathcal{D}^{0}]  =  0.
    \end{split}
\end{equation}
They preserve the algebra $[\mathcal{D}^{++}, \mathcal{D}^{--}] = \mathcal{D}^0, \; [\mathcal{D}^0, \mathcal{D}^{\pm\pm}] = \pm 2 \mathcal{D}^{\pm\pm} $.

\subsection{Superconformal transformations of spinor and vector derivatives}

$\mathcal{N}=2$ superconformal transformations of covariant derivatives can be deduced as follows. First, the shortened  spinor derivatives have the transformation law
\begin{equation}\label{eq: D+ sc}
    \begin{split}
   & \delta_{sc} \mathcal{D}^+_{\alpha}
    :=
    [\hat{\Lambda}_{sc},  \mathcal{D}^+_{\alpha} ]
    =
    -
    \left( \mathcal{D}^+_{\alpha} \lambda_{sc}^{-\beta} \right) \mathcal{D}^+_{\beta}\,,
   \\
    &\delta_{sc} \bar{\mathcal{D}}^+_{\dot{\alpha}}
   : =
        [\hat{\Lambda}_{sc},  \bar{\mathcal{D}}^+_{\dot{\alpha}} ]
    = - \left( \bar{\mathcal{D}}^+_{\dot{\alpha}} \bar{\lambda}^{-\dot{\beta}}_{sc}  \right) \bar{\mathcal{D}}^+_{\dot{\beta}}.
    \end{split}
\end{equation}
Since this derivatives transforms homogeneously, the analyticity condition \eqref{AnalConstr} is superconformally covariant.
Using this property and the first relation in \eqref{eq: rel cov derivatives},  we obtain:
\begin{equation}\label{eq: D- sc}
    \begin{split}
        \delta_{sc} \mathcal{D}^-_\alpha =&\; [\delta_{sc} \mathcal{D}^{--}, \mathcal{D}^+_{\alpha}] +  [\mathcal{D}^{--}, \delta_{sc} \mathcal{D}^+_{\alpha}]
             \\
              = & -
        (\mathcal{D}^{--}\lambda_{sc}^{++}) \mathcal{D}^-_\alpha - \left( \mathcal{D}^-_\alpha\lambda_{sc}^{++} \right) \mathcal{D}^{--}
\\  &-\left(\mathcal{D}^+_\alpha \lambda_{sc}^{-\beta} \right) \mathcal{D}^-_\beta
        -
        \frac{1}{2}
        \left( \mathcal{D}^-_\beta  \lambda_{sc}^{-\beta} \right) \mathcal{D}^+_\alpha.
    \end{split}
\end{equation}
Besides the homogeneous terms $\sim \left(\mathcal{D}^+_\alpha \lambda^{-\beta}_{sc} \right) $ and $\sim (\mathcal{D}^{--}\lambda^{++}_{sc})$,
this transformation law contains  two inhomogeneous contributions.

Using the last line of \eqref{eq: rel cov derivatives}  and the properties \eqref{eq: sc der prop}, we can find the superconformal variation of the vector derivative:
\begin{equation}\label{eq: part sc}
    \begin{split}
        \delta_{sc} \partial_{\alpha\dot{\alpha}} = &
        \frac{1}{4i} \left\{ \delta_{sc} \mathcal{D}^-_\alpha, \bar{\mathcal{D}}^+_{\dot{\alpha}} \right\}
        +
         \frac{1}{4i} \left\{  \mathcal{D}^-_\alpha, \delta_{sc} \bar{\mathcal{D}}^+_{\dot{\alpha}} \right\}
        \\
        =&
        -
        (\mathcal{D}^+_\alpha \lambda_{sc}^{-\beta}) \partial_{\beta\dot{\alpha}}
        -
        (\bar{\mathcal{D}}^+_{\dot{\alpha}} \bar{\lambda}_{sc}^{-\dot{\beta}}) \partial_{\alpha\dot{\beta}}
        -
        (\mathcal{D}^{--}\lambda^{++}_{sc}) \partial_{\alpha\dot{\alpha}}
        \\
        &
        -
        \frac{i}{4} \left( \bar{\mathcal{D}}^-_{\dot{\alpha}} \lambda_{sc}^{++} \right)
        \mathcal{D}^-_\alpha
        +
        \frac{i}{4} \left( \mathcal{D}^-_\alpha \lambda_{sc}^{++} \right) \bar{\mathcal{D}}^-_{\dot{\alpha}}
        \\
        &
        - \frac{1}{2} \left( \partial_{\beta\dot{\alpha}} \lambda_{sc}^{-\beta} \right)  \mathcal{D}^+_\alpha
        -
        \frac{1}{2}
       \left( \partial_{\alpha\dot{\beta}} \bar{\lambda}_{sc}^{-\dot{\beta}} \right) \bar{\mathcal{D}}^+_{\dot{\alpha}}.
    \end{split}
\end{equation}
This transformation law is more complicated  as compared to that of spinor derivatives. Here the first line are the homogeneous terms,
while the second and third lines collect the inhomogeneous contributions proportional to the spinor derivatives.

\smallskip

It is important to note that, as a consequence of the algebra of covariant derivatives, all transformation laws are expressed exclusively in terms of the parameters $\lambda^{++}_{sc}$ and $\lambda^{-\hat{\alpha}}_{sc}$.

\medskip

We define $\mathcal{N}=2$ \textit{chiral superfield} $\Phi_{ch}$ by the set of covariant conditions:
\begin{equation}\label{eq: ch cond}
    \bar{\mathcal{D}}^\pm_{\dot\alpha} \Phi_{ch} = 0,
    \qquad
 \mathcal{D}^{\pm\pm} \Phi_{ch} = 0,
    \qquad
    \mathcal{D}^0  \Phi_{ch} = 0.
\end{equation}
Since under the superconformal group the spinor derivative $\bar{\mathcal{D}}^-_{\dot{\alpha}}$ transforms through both $\bar{\mathcal{D}}^+_{\dot{\alpha}}$ and $\mathcal{D}^{--}$,
the standard Grassmann chirality conditions in \eqref{eq: ch cond} are superconformally covariant only with taking account of the extra harmonic constraints.

\subsection{Superconformal transformations of $\mathcal{N}=2$ superspaces measures}

To construct the superconformally invariant   actions, we need to know the superconformal transformation laws of the integration measures.
In compliance with the existence of 3 types of invariant $\mathcal{N}=2$ superspaces,  there are three types of  superfield actions: those in the full harmonic superspace,  in the analytic harmonic superspace
and in chiral superspaces. In this subsection we present the explicit $\mathcal{N}=2$ superconformal transformation laws for the relevant integration measures.

\medskip
\noindent$\bullet$ \textit{\underline{Full superspace}} integration measure:
\begin{equation}
    \begin{split}
    \delta_{sc}(d^4 x d^8\theta du) =& \Big[\partial_{\alpha\dot{\alpha}} \lambda^{\alpha\dot{\alpha}}
    -
    \partial^-_{\hat{\alpha}} \lambda^{+\hat{\alpha}}
    -
    \partial^+_{\hat{\alpha}} \lambda^{-\hat{\alpha}}
    +
    \underbrace{\frac{\partial (\lambda^{++}u^-_i)}{\partial u^{+i}}}_{\partial^{--} \lambda^{++}} \Big] (d^4 x d^8\theta du )
   \\ :=&\,
    \Omega_{full}^{(sc)}
    (d^4 x d^8\theta du) \,.
        \end{split}
\end{equation}
After substituting the superconformal parameters \eqref{eq:superconformal symmetry}, the corresponding weight factor becomes:
\begin{equation}\label{eq: fss Omega}
    \begin{split}
        \Omega^{(sc)}_{full} &= 2\lambda^{+-}
        + 4i (\theta^{-\alpha}\bar{\theta}^{+\dot{\alpha}} + \theta^{+\alpha}\bar{\theta}^{-\dot{\alpha}}) k_{\alpha\dot{\alpha}}
        + 4i (\theta^+\eta^- + \theta^-\eta^+)
        +
        4i (\bar{\theta}^+ \bar{\eta}^- + \bar{\theta}^- \bar{\eta}^+)
        \\
        &= \mathcal{D}^{--} \lambda^{++}_{sc}.
    \end{split}
\end{equation}
Here we introduced the short-cut  notations $\lambda^{+-} = \lambda^{(ij)}u^+_i u^-_j$, $\eta^\pm_\alpha = \eta^i_\alpha u^\pm_i$, $\theta^+ \eta^- = \theta^{+\alpha} \eta^-_\alpha$, $\bar{\theta}^+ \bar{\eta}^{-}  = \bar{\theta}^+_{\dot{\alpha}} \bar{\eta}^{-\dot{\alpha}} $, etc.
For the integral of the harmonic-independent Lagrangian density $\mathcal{L}(x, \theta^\pm, \bar{\theta}^\pm,  u^\pm_i)$,
\begin{equation}
    \mathcal{D}^{++} \mathcal{L} = 0,
    \qquad
    \mathcal{D}^{--} \mathcal{L} = 0\,,
\end{equation}
we have
 \begin{equation}
    \int \delta_{sc}  \left( d^4 x d^8\theta du  \right)  \mathcal{L} =
        \int d^4 x d^8\theta du \; \mathcal{D}^{--} \lambda^{++}_{sc}  \mathcal{L}
        =
        -
            \int d^4 x d^8\theta du \;  \lambda^{++}_{sc}  \mathcal{D}^{--} \mathcal{L} = 0,
 \end{equation}
which amounts to the well-known property that the supervolume of superspace $\mathbb{R}^{4|8}$ is $\mathfrak{su}(2,2|2)$ invariant \cite{Ferber:1977qx}.

 \medskip

\noindent$\bullet$ \textit{\underline{Analytic superspace}} measure:
\begin{equation}
    \begin{split}
    \delta_{sc} (d^4 x d^4\theta^+ du)  =& \Big[\partial_{\alpha\dot{\alpha}} \lambda^{\alpha\dot{\alpha}}
    -
    \partial^-_{\hat{\alpha}} \lambda^{+\hat{\alpha}}
    +
    \partial^{--} \lambda^{++} \Big] (d^4 x d^4\theta^+ du)
    \\ :=&\, \Omega^{(sc)}_A \,(d^4 x d^4\theta^+ du)\,.
    \end{split}
\end{equation}
The corresponding weight factor is:
\begin{equation}
    \Omega^{(sc)}_A = 2{\rm d} - 2 \lambda^{+-} + 2 (k\cdot x) - 8i \left(  \theta^+ \eta^- + \bar{\theta}^+ \bar{\eta}^- \right).
\end{equation}
This weight factor satisfies a notable equality:
\begin{equation}\label{eq: analytic measure rel}
    \mathcal{D}^{++}   \Omega^{(sc)}_A
    =
    - 2 \lambda^{++}_{sc}.
\end{equation}

\noindent$\bullet$ \textit{\underline{Chiral superspace}} measure:

\begin{equation}
    \delta_{sc} (d^4x_L d^4\theta)  =
    \left[ \partial^L_{\alpha\dot{\alpha}}\lambda_L^{\alpha\dot{\alpha}} - \partial_\alpha^i \lambda^\alpha_i  \right] (d^4x_L d^4\theta)
    := \Omega^{(sc)}_{Ch}(d^4x_L d^4\theta)\,.
\end{equation}
Here we have used the notation of Appendix \ref{App: sc conf} for the coordinates of the chiral basis.
The chiral weight factor is:
\begin{equation}
    \Omega^{(sc)}_{Ch}
    =
    2 ({\rm d}-i{\rm r}) + 2 (k\cdot x_L) + 8i (\theta^\alpha_i \eta^i_\alpha).
\end{equation}
In the analytic basis, this factor can be rewritten as:
\begin{equation}\label{eq: Omega chiral}
    \begin{split}
    \Omega^{(sc)}_{Ch}
    =&\,
    2({\rm d} - i{\rm r})
    +
    2(x^{\beta\dot{\beta}}
    +
    4i \theta^{-\beta}\bar{\theta}^{+\dot{\beta}})  k_{\beta\dot{\beta}}
    +
    4i (\theta^- \eta^+)
    -
    4i (\theta^+ \eta^-)
    \\
    = &\,   2\left( \bar{\mathcal{D}}^+_{\dot{\alpha}} \bar{\lambda}_{sc}^{-\dot{\alpha}}
    + \mathcal{D}^{--} \lambda_{sc}^{++} \right).
    \end{split}
\end{equation}
As a consequence of the properties \eqref{eq: sc equations} and \eqref{eq: sc e}, the weight factor $\Omega^{(sc)}_{Ch} $ satisfies
$\mathcal{N}=2$ chirality constraints  \eqref{eq: ch cond},
\begin{equation}\label{chircondOmega}
    \bar{\mathcal{D}}^\pm_{\dot\alpha} \Omega^{(sc)}_{Ch} = 0,
    \qquad
 \mathcal{D}^{\pm\pm} \Omega^{(sc)}_{Ch} = 0,
    \qquad
    \mathcal{D}^0  \Omega^{(sc)}_{Ch} = 0.
\end{equation}

\smallskip

It is worth to emphasize that $\mathcal{N}=2$ chiral superspace does not allow the superconformally-covariant addition of harmonic coordinates.
This amounts to non-invariance of the chirality conditions (see eq. \eqref{eq: D- sc}) for the superfields having a non-trivial dependence on harmonics.
The consistency of $\mathcal{N}=2$ chirality with the superconformal invariance  can be achieved
only for the superfields satisfying the full set of conditions \eqref{eq: ch cond}. In the Appendix  \ref{App: sc conf}
we list all possible $\mathcal{N}=2$ superspaces which admit the superconformal symmetry.

\section{$\mathcal{N}=2$ Maxwell action: superconformal invariance }\label{eq: sec 3}

Before passing to $\mathcal{N}=2$ conformal supergravity, we will dwell on  $\mathcal{N}=2$ Maxwell theory as an instructive example of
$\mathcal{N}=2$ superconformally invariant theory. A few features of this theory admit a direct generalization to the conformal supergravity.

\medskip

In harmonic superspace, the free $\mathcal{N}=2$ Maxwell supermultiplet is described by the action
\begin{equation}\label{eq: Maxwell action}
    S_{\text{Max}} = \int d^4xd^8
    \theta du \; V^{++} V^{--} ,
\end{equation}
where $V^{++}$ is the analytic gauge prepotential which collects all degrees of freedom of the theory. The non-analytic gauge superfield $V^{--}$
is defined as a solution of the harmonic zero-curvature condition:
\begin{equation}\label{eq: spin 1 zc}
    \mathcal{D}^{++} V^{--} = \mathcal{D}^{--} V^{++} .
\end{equation}
This equation uniquely specifies $V^{--}$ in terms of $V^{++}$.
The action and zero-curvature condition are invariant under the abelian gauge transformations:
\begin{equation}
    \delta_\lambda V^{++} = \mathcal{D}^{++} \lambda,
    \qquad
    \delta_\lambda V^{--} =  \mathcal{D}^{--} \lambda\,,
\end{equation}
with an arbitrary analytic parameter $\lambda(\zeta, u)$.

Under $\mathcal{N}=2$ \textit{superconformal transformations} the analytic prepotential transforms as
\begin{equation}\label{eq: v++ sc}
    \delta_{sc} V^{++} (\zeta, u)  \simeq V^{\prime ++}(\zeta^\prime, u^\prime)  - V^{++}(\zeta, u) =0.
\end{equation}
The transformation of $V^{--}$ can be deduced from the requirement of superconformal covariance of the zero-curvature equation \eqref{eq: spin 1 zc},
\begin{equation}\label{eq: v-- sc}
    \delta_{sc} V^{--} = - (\mathcal{D}^{--}\lambda_{sc}^{++}) V^{--}.
\end{equation}

The superconformal invariance of $\mathcal{N}=2$ Maxwell action is checked by performing the transformations \eqref{eq: fss Omega},
\eqref{eq: v++ sc} and \eqref{eq: v-- sc} in \eqref{eq: Maxwell action}:
\begin{equation}
    \delta_{sc} S_{\text{Max}} = \int  \underbrace{\delta_{sc} \left\{ d^4xd^8
        \theta du \right\}}_{\left\{ d^4xd^8
        \theta du \right\} (\mathcal{D}^{--}\lambda^{++}_{sc}) }  \; V^{++} V^{--}
    +
    \int d^4xd^8
    \theta du \; \underbrace{\delta_{sc} \left\{ V^{++} V^{--} \right\}}_{- (\mathcal{D}^{--}\lambda_{sc}^{++}) V^{++}V^{--}}
    =
    0.
\end{equation}

\smallskip

\noindent \textbf{Chiral representation}

\smallskip

The action \eqref{eq: Maxwell action} is an integral over the total $\mathcal{N}=2$ harmonic superspace.
For what follows, it is instructive to rewrite this action as an integral over $\mathcal{N}=2$ chiral superspace
and to study its superconformal properties in such an equivalent formulation.

As the first step, we pass to the chiral integration measure using
$d^4 \bar{\theta} = \frac{1}{16} (\bar{\mathcal{D}}^-)^2  (\bar{\mathcal{D}}^+)^2$ and analyticity of $V^{++}$:
\begin{equation}\label{eq: Max - ch}
    S_{\text{Max}} = \int d^4xd^8
    \theta du
    \; V^{++} V^{--}
    =
    \frac{1}{16}\int d^4x d^4 \theta du
    \; (\bar{\mathcal{D}}^-)^2 \left\{ V^{++}  (\bar{\mathcal{D}}^+)^2 V^{--} \right\}.
\end{equation}
Then we introduce $\mathcal{N}=2$ covariant gauge-invariant superstrengths
\begin{equation}
    \mathcal{W} := (\bar{\mathcal{D}}^+)^2 V^{--},
    \qquad
    \overline{\mathcal{W}} :=  (\mathcal{D}^+)^2 V^{--}\,,\label{DefW}
\end{equation}
\begin{equation}
    \delta_\lambda \mathcal{W} = 0,
    \qquad
    \delta_\lambda \overline{\mathcal{W}} = 0.
\end{equation}

The superstrengths satisfy the covariant  \textit{harmonic-independence} conditions:
\begin{equation}\label{eq: W hi}
    \mathcal{D}^{\pm\pm} \mathcal{W} = 0,
    \qquad
    \mathcal{D}^{\pm\pm} \overline{\mathcal{W}} = 0,
\end{equation}
and, as a consequence of \eqref{eq: W hi} and the definition \eqref{DefW}, the \textit{chirality} conditions
\begin{equation}\label{eq: W chir}
    \mathcal{D}^\pm_{\alpha} \overline{\mathcal{W}} = 0,
    \qquad
    \bar{\mathcal{D}}^\pm_{\dot{\alpha}} \mathcal{W} = 0
\end{equation}
and \textit{Bianchi identities}:
\begin{equation}
    \left(\mathcal{D}^+\right)^2 \mathcal{W} = \left(\bar{\mathcal{D}}^+ \right)^2 \overline{\mathcal{W}},
    \qquad
    \left(\mathcal{D}^+ \mathcal{D}^-\right) \mathcal{W}
    =
     \left(\bar{\mathcal{D}}^+ \bar{\mathcal{D}}^- \right) \overline{\mathcal{W}},
    \qquad
    \left(\mathcal{D}^-\right)^2 \mathcal{W} = \left(\bar{\mathcal{D}}^- \right)^2 \overline{\mathcal{W}}.
\end{equation}

One can rewrite the action \eqref{eq: Maxwell action} in terms of $\mathcal{W}$ and $\overline{\mathcal{W}}$.
To this end, we need to pass to the central basis in harmonic superspace.
Then, using integrations by parts for harmonic derivatives, the  algebra of covariant derivatives \eqref{eq: rel cov derivatives}, eqs. \eqref{eq: W hi}, \eqref{eq: W chir}
and the zero-curvature condition \eqref{eq: spin 1 zc}, we obtain:
\begin{equation}\label{eq: Max - ch 2}
    \begin{split}
        S_{Max} =&\; \frac{1}{16}
        \int d^4x d^4 \theta du
        \; (\bar{\mathcal{D}}^-)^2  \,V^{++} \mathcal{W}
        =
        \frac{1}{16}
        \int d^4x d^4 \theta du
        \; \mathcal{D}^{--}(\bar{\mathcal{D}}^+\bar{\mathcal{D}}^-)  \, V^{++} \mathcal{W}
        \\
        =&\;
        \frac{1}{16}
        \int d^4x d^4 \theta du
        \; (\bar{\mathcal{D}}^+)^2  \,V^{--} \mathcal{W}
        =
        \frac{1}{16}
        \int d^4x d^4 \theta du
        \;  \mathcal{W}^2
          =
        \frac{1}{16}
        \int d^4x_L d^4 \theta
        \;  \mathcal{W}^2.
    \end{split}
\end{equation}
Thereby, the $\mathcal{N}=2$  Maxwell action \eqref{eq: Maxwell action} can be equally written as an integral over $\mathcal{N}=2$ chiral superspace.

 Alternatively, a chiral form of the Maxwell action can be achieved using the representation
\begin{equation}\label{eq: alt repr}
    \mathcal{W} = (\bar{\mathcal{D}}^+)^2 V^{--}
    =
    \int du_1 (\bar{\mathcal{D}}^+)^2 \frac{V^{++}(u_1)}{(u^+u^+_1)^2}
    =
    \int du (\bar{\mathcal{D}}^-)^2 V^{++},
\end{equation}
which  follows from the relations in the central basis:
\begin{subequations}
\begin{equation}
    \bar{\mathcal{D}}^+_{\dot{\alpha}} (u):= \mathcal{D}^i_{\dot{\alpha}} u^+_i =
    \bar{\mathcal{D}}^-_{\dot{\alpha}}(u_1) (u^+_1 u^+) - \bar{\mathcal{D}}^+_{\dot{\alpha}}(u_1) (u^-_1u^+) ,
\end{equation}
\begin{equation}
    \left(\bar{\mathcal{D}}^+(u)\right)^2  V^{++}(u_1)=
    \left(\bar{\mathcal{D}}^- (u_1)\right)^2  V^{++}(u_1)  (u^+_1 u^+)^2.
\end{equation}
\end{subequations}
The harmonic distribution appearing in \eqref{eq: alt repr} is defined by the relations \cite{18, Galperin:1985bj}
\begin{equation}\label{eq: harmonic distrib}
    \partial^{++}_1  \frac{1}{(u^+_1u^+_2)}  = \delta^{(1,-1)}(u_1, u_2),
    \qquad
        \partial^{++}_1  \frac{1}{(u^+_1u^+_2)^2}  = \partial^{--}_1 \delta^{(2,-2)}(u_1, u_2),
\end{equation}
where $\delta^{(n,-n)}(u_1, u_2)$ are the appropriate harmonic delta-functions.

\smallskip
\noindent\textbf{$\mathcal{N}=2$ superconformal transformations in chiral representation}
\smallskip

\noindent We now turn to $\mathcal{N}=2$ superconformal invariance in the chiral representation.
Using the transformation laws \eqref{eq: D+ sc} and \eqref{eq: v-- sc}, we obtain:
\begin{equation}
    \begin{split}
        \delta_{sc} \mathcal{W} =&
        -
        \left( \bar{\mathcal{D}}^+_{\dot{\alpha}} \bar{\lambda}_{sc}^{-\dot{\alpha}}
        + \mathcal{D}^{--} \lambda_{sc}^{++} \right) \mathcal{W}
        \\&
        -
        \left[ (\bar{\mathcal{D}}^+)^2 \bar{\lambda}_{sc}^{-\dot{\alpha}}
        +
        2 \bar{\mathcal{D}}^{-\dot{\alpha}} \lambda_{sc}^{++} \right]  \bar{\mathcal{D}}^+_{\dot{\alpha}} V^{--}.
    \end{split}
\end{equation}
In the second line we highlighted the inhomogeneous terms.
Using the relation \eqref{eq: sc e} for the rigid superconformal
transformation,
\begin{equation}
    (\bar{\mathcal{D}}^+)^2 \bar{\lambda}_{sc}^{-\dot{\alpha}}
    =
    - 2 \bar{\mathcal{D}}^{-\dot{\alpha}} \lambda_{sc}^{++}
    =
    2\bar{\partial}^{-\dot{\alpha}} \lambda_{sc}^{++}
    =
    8i \theta^+_\alpha k^{\alpha\dot{\alpha}} - 8i \bar{\eta}^{\dot{\alpha}+}\,,
\end{equation}
we observe that inhomogeneous terms are canceled  and
we are finally left with the homogeneous transformation law for $\mathcal{W}$:
\begin{equation}\label{eq: sc W Maxwell th}
    \delta_{sc} \mathcal{W} =
    -
    \left( \bar{\mathcal{D}}^+_{\dot{\alpha}} \bar{\lambda}_{sc}^{-\dot{\alpha}}
    +
     \mathcal{D}^{--} \lambda_{sc}^{++} \right) \mathcal{W}.
\end{equation}
Thus the transformation of the Lagrangian density $\mathcal{L}_{ch} = \mathcal{W}^2$ cancels
the chiral weight factor \eqref{eq: Omega chiral} coming from the transformation of the chiral
 measure\footnote{Note that under general gauge transformations of conformal $\mathcal{N}=2$ supergravity
$\mathcal{W}$ possesses an inhomogeneous transformation law.}.

It is instructive to derive the superconformal transformation law \eqref{eq: sc W Maxwell th} in a different way, starting from the representation \eqref{eq: alt repr}.

\begin{equation}\label{eq: alt var}
    \delta_{sc} \mathcal{W}
    =
    \int  \delta_{sc} \left\{du  \right\} (\bar{\mathcal{D}}^-)^2 V^{++}
    +
    \int du \, \delta_{sc} \left\{(\bar{\mathcal{D}}^-)^2 V^{++} \right\}.
\end{equation}
Harmonic measure transforms as
\begin{equation}
     \delta_{sc} \left\{du  \right\}
     =
     \left\{du  \right\} \left( \mathcal{D}^{--} \lambda^{++} \right)
\end{equation}
Superconformal transformation of $(\bar{\mathcal{D}}^-)^2$ is given by
\begin{equation}\label{eq: D^-2 var}
    \begin{split}
    \delta_{sc} (\bar{\mathcal{D}}^-)^2
    =&
    -
    \left[ 2 (\mathcal{D}^{--}\lambda^{++})
    +
    (\bar{\mathcal{D}}^-_{\dot{\beta}} \bar{\lambda}_{sc}^{-\dot{\beta}}) \right]
    (\bar{\mathcal{D}}^-)^2
    \\
    &
    -
    (\bar{\mathcal{D}}^-_{\dot{\beta}} \bar{\lambda}_{sc}^{-\dot{\beta}} ) (\bar{\mathcal{D}}^- \bar{\mathcal{D}}^+ )
    -
    \frac{1}{2} \left( (\bar{\mathcal{D}}^-)^2 \bar{\lambda}_{sc}^{-\dot{\alpha}} \right) \bar{\mathcal{D}}^{+}_{\dot{\alpha}}
        \\
    &
    -
    \left(  (\bar{\mathcal{D}}^-)^2
    \lambda_{sc}^{++}  \right) \mathcal{D}^{--}
    +
    2 \left( \mathcal{D}^{--} \bar{\mathcal{D}}^{-\dot{\alpha}} \lambda^{++}_{sc}  \right) \bar{\mathcal{D}}^-_{\dot{\alpha}}
    +
    2 \left(  \bar{\mathcal{D}}^{-\dot{\alpha}} \lambda^{++}_{sc}  \right)
    \mathcal{D}^{--} \bar{\mathcal{D}}^-_{\dot{\alpha}}.
    \end{split}
\end{equation}
Taking into account the analyticity of the prepotential and its trivial superconformal transformation law \eqref{eq: v++ sc}, we obtain
\begin{equation}
    \begin{split}
    \delta_{sc} \left\{(\bar{\mathcal{D}}^-)^2 V^{++} \right\}
    =&
        -
    \left[ 2 (\mathcal{D}^{--}\lambda^{++})
    +
    (\bar{\mathcal{D}}^-_{\dot{\beta}} \bar{\lambda}^{-\dot{\beta}}) \right]
    \left\{(\bar{\mathcal{D}}^-)^2 V^{++} \right\}
    \\
    &
    - \mathcal{D}^{++} \Big\{   \left(  (\bar{\mathcal{D}}^-)^2
    \lambda^{++}  \right)  V^{--}  \Big\}
    +
    2 \mathcal{D}^{--} \Big\{
    \left(  \bar{\mathcal{D}}^{-\dot{\alpha}} \lambda^{++}  \right)
     \bar{\mathcal{D}}^-_{\dot{\alpha}}  V^{++} \Big\}.
    \end{split}
\end{equation}
Terms that are total harmonic derivatives do not contribute to the variation \eqref{eq: alt var}, and we recover the superconformal transformation law \eqref{eq: sc W Maxwell th}.

\section{$\mathcal{N}=2$ conformal supergravity}\label{eq: sec 4}

$\mathcal{N}=2$ off-shell conformal supergravity multiplet has a concise geometric superfield formulation in harmonic superspace \cite{Galperin:1987em, 18, Ivanov:2022vwc}.
It naturally originates from the analytic localization of rigid superconformal transformations
$\lambda^M_{sc}\to \lambda^M(\zeta,u)$, $\lambda^{++}_{sc} \to \lambda^{++}(\zeta,u)$, $\lambda^{-\hat{\alpha}}_{sc} \to \lambda^{-\hat{\alpha}}(\zeta,\theta^-,u)$.
The covariance of harmonic derivative $\mathcal{D}^{++}$ under the localized superconformal transformations is achieved by introducing the
supervielbeins:
\begin{equation}\label{eq: mathfrak D}
    \mathcal{D}^{++}
    \quad
    \to
    \quad
    \mathfrak{D}^{++} = \partial^{++} + \mathcal{H}^{++\alpha\dot{\alpha}} \partial_{\alpha\dot{\alpha}}
    +
    \mathcal{H}^{++\hat{\alpha}+} \partial^-_{\hat{\alpha}}
    +
    \mathcal{H}^{++\hat{\alpha}-} \partial^+_{\hat{\alpha}}
    +
    \mathcal{H}^{(+4)} \partial^{--}.
\end{equation}
All $\mathcal{H}^{++}$-prepotentials, apart from $\mathcal{H}^{++\hat{\alpha}-}$, are analytic and otherwise unconstrained.
Compared to the minimal version of Einstein's supergravity \cite{Galperin:1987ek}, the harmonic derivative contains an extra $\mathcal{H}^{(+4)}$ vielbein.

We require that under localized superconformal transformations $\mathfrak{D}^{++}$ transforms as under the rigid ${\cal N}=2$ superconformal group,
\begin{equation}\label{eq: spin 2 ct}
    \delta_\lambda \mathfrak{D}^{++} = - \lambda^{++} \mathcal{D}^0.
\end{equation}
This transformation law implies the following transformation rules for prepotentials:
\begin{equation}\label{eq: H transf}
    \begin{split}
        \delta_\lambda \mathcal{H}^{++\alpha\dot{\alpha}} &= \mathfrak{D}^{++}  \lambda^{\alpha\dot{\alpha}},
        \\
        \delta_\lambda \mathcal{H}^{++\hat{\alpha}+} &= \mathfrak{D}^{++} \lambda^{+\hat{\alpha}}
        -
        \lambda^{++} \theta^{+\hat{\alpha}},
        \\
        \delta_\lambda \mathcal{H}^{++\hat{\alpha}-} &= \mathfrak{D}^{++} \lambda^{-\hat{\alpha}} + \lambda^{++} \theta^{-\hat{\alpha}},
        \\
        \delta_\lambda
        \mathcal{H}^{(+4)} &= \mathfrak{D}^{++} \lambda^{++}.
    \end{split}
\end{equation}
All parameters are analytic, with the exception of $\lambda^{-\hat{\alpha}}$.
By making use of this non-analytic parameter, one can fix the \textit{analytic gauge} $\mathcal{H}^{++\hat{\alpha}-} = \theta^{+\hat{\alpha}}$.
This means that the superfield $\mathcal{H}^{++\hat{\alpha}-}$ does not contain the  physical degrees of freedom which are carried solely by the remaining analytic supervielbeins.
 In the analytic gauge,  the harmonic derivative $\mathfrak{D}^{++}$ satisfies the analyticity condition
 $[\mathcal{D}^+_{\hat{\alpha}}, \mathfrak{D}_{a.g.}^{++}] = 0$.  Moreover, in this gauge, the non-analytic parameter is given in terms of the analytic ones as a solution of the equation:
 \begin{equation}
    \mathfrak{D}^{++} \lambda^{-\hat{\alpha}} = \lambda^{+\hat{\alpha}} -\lambda^{++} \theta^{-\hat{\alpha}}.
 \end{equation}

In the flat $\mathcal{N}=2$ super-background $\mathfrak{D}^{++} \rightarrow \mathfrak{D}^{++}_{flat}=\mathcal{D}^{++}$, and the supervielbeins are reduced to:
\begin{equation}
    \mathcal{H}^{++\alpha\dot{\alpha}}_{flat} = - 4i \theta^{+\alpha} \bar{\theta}^{+\dot{\alpha}},
    \qquad
    \mathcal{H}_{flat}^{++\hat{\alpha}+} = 0,
    \qquad
    \mathcal{H}_{flat}^{++\hat{\alpha}-} = \theta^{+\hat{\alpha}},
    \qquad
    \mathcal{H}_{flat}^{(+4)} = 0.
\end{equation}
Denoting the deviations of the prepotentials  from their flat background values as
\begin{equation}
    h^{++\alpha\dot{\alpha}},
    \qquad
    h^{++\hat{\alpha}+},
    \qquad
    h^{++\hat{\alpha}-},
    \qquad
    h^{(+4)},
\end{equation}
 one can determine from eq. \eqref{eq: H transf}  the linearized gauge transformation laws for these superfields, as well as
their rigid $\mathcal{N}=2$ superconformal transformations.  We will present them later in the form most convenient for our purposes.

Using the linearized gauge transformations, it is possible to impose the Wess-Zumino (WZ) type gauge:
\begin{equation}
    \begin{split}
        h_{WZ}^{++\alpha\dot{\alpha}}
        &=
        -4i \theta^{+\beta}\bar{\theta}^{+\dot{\beta}} \Phi^{\alpha\dot{\alpha}}_{\beta\dot{\beta}}
        +  (\bar{\theta}^+)^2 \theta^{+\beta} \psi^{\alpha\dot{\alpha}i}_\beta u^-_i + (\theta^+)^2 \bar{\theta}^{+\dot{\rho}} \bar{\psi}^{\alpha\dot{\alpha}i}_{\dot{\rho}}u_i^-
     +  (\theta^+)^4 V^{\alpha\dot{\alpha}(ij)}u^-_iu^-_j\,,
        \\
        h_{WZ}^{++\alpha+} &=(\theta^+)^2 \bar{\theta}^+_{\dot{\alpha}} P^{\alpha\dot{\alpha}}
        + \left(\bar{\theta}^+\right)^2 \theta^+_\beta T^{(\alpha\beta)}
        +  (\theta^+)^4 \chi^{\alpha i}u^-_i, \;\;\;
        \\
        h_{WZ}^{++\dot{\alpha}+} &= \widetilde{h_{WZ}^{++\alpha+}}\,,
        \\
        h_{WZ}^{(+4)} \;\;\,&= (\theta^+)^4 D\,.
    \end{split}
\end{equation}
This gauge is defined up to a residual gauge freedom which is properly realized on the component fields. For instance,
$\delta \Phi_{\beta\dot{\beta}}^{\alpha\dot{\alpha}} = \partial_{\beta\dot{\beta}}a^{\alpha\dot{\alpha}} + l_{(\alpha}^{\;\;\beta)}\delta_{\dot{\alpha}}^{\dot{\beta}}
+ \delta_\alpha^{\beta} \bar{l}_{(\dot{\alpha}}^{\;\;\dot{\beta})} + d \delta_{\alpha}^{\beta} \delta_{\dot{\alpha}}^{\dot{\beta}} $, where the gauge parameters
correspond to local translations ($a^{\alpha\dot{\alpha}}$), local Lorentz transformations ($l_{(\alpha\beta)}, \bar{l}_{(\dot{\alpha}\dot{\beta})}$) and local scale transformations ($d$).
 The set of fields in WZ gauge, together with their residual gauge freedom, reproduce the content of $\mathcal{N}=2$ Weyl multiplet.
 For further details see, e.g., \cite{Buchbinder:2024pjm, Buchbinder:2025ceg}. It is worth emphasizing that the form of WZ-type gauge remains the same
 also in the complete nonlinear case.

\medskip
The basic aim of the rest of this section is to construct \textit{the linearized $\mathcal{N}=2$  conformal supergravity action}
in terms of  $\mathcal{N}=2$ Weyl supermultiplet analytic prepotentials and to study its transformation properties under rigid superconformal symmetry.

\subsection{Covariant superfields}

To deal with the linearized conformal supergravity,  it will be useful to pass to the  superfields exhibiting manifest invariance under  rigid background supersymmetry.
The covariant harmonic derivative~\eqref{eq: mathfrak D} is invariant under rigid $\mathcal{N}=2$ supersymmetry.
Then, following refs. \cite{Zupnik:1998td, Kuzenko:1999pi}, we  expand the harmonic derivative $  \mathfrak{D}^{++}$ over the basis of supersymmetry-invariant covariant derivatives:
\begin{equation}
    \mathfrak{D}^{++} = \mathcal{D}^{++} + G^{++\alpha\dot{\alpha}}
    \partial_{\alpha\dot{\alpha}}
    +
    G^{++\hat{\alpha}-} \mathcal{D}^+_{\hat{\alpha}}
    +
    G^{++\hat{\alpha}+} \mathcal{D}^-_{\hat{\alpha}}
    +
    G^{(+4)} \mathcal{D}^{--}.
\end{equation}
Here $G^{++}$-vielbeins are expressed through the analytic $h^{++}$-vielbeins by the relations:
\begin{equation}\label{Gsuperpot}
    \begin{split}
        &
        G^{++\alpha\dot{\alpha}} = h^{++\alpha\dot{\alpha}}
        +
        4i h^{++\alpha+} \bar{\theta}^{-\dot{\alpha}}
        +
        4i \theta^{-\alpha} h^{++\dot{\alpha}+}
        -
        4i h^{(+4)} \theta^{-\alpha} \bar{\theta}^{-\dot{\alpha}},
        \\
        &
        G^{++\hat{\alpha}-}
        =
        h^{++\hat{\alpha}-},
        \\
        &
        G^{++\hat{\alpha}+} = - h^{++\hat{\alpha}+} + h^{(+4)} \theta^{-\hat{\alpha}},
        \\
        &
        G^{(+4)} \;\,\,= h^{(+4)}.
    \end{split}
\end{equation}
The basic feature of $G^{++}$-superfields is their manifest invariance under supersymmetry.
These superfields (except for $G^{(+4)}$) are non-analytic, but, as a consequence of the analyticity of the $h^{++}$-vielbeins,
satisfy the constraints:
\begin{equation}\label{eq: G++ constr}
    \begin{split}
        \bar{\mathcal{D}}^+_{\dot{\beta}} G^{++\alpha\dot{\alpha}} = 4i G^{++\alpha+} \delta^{\dot{\alpha}}_{\dot{\beta}}, \qquad
        &\mathcal{D}^{+}_{\beta} G^{++\alpha\dot{\alpha}} = -4i G^{++\dot{\alpha}+} \delta^\alpha_\beta,
        \\
        \bar{\mathcal{D}}^+_{\dot{\beta}} G^{++\alpha+} = 0,
        \qquad
        &
        \mathcal{D}^+_{\beta} G^{++\alpha+} = G^{(+4)} \delta^\alpha_\beta,
        \\
        \bar{\mathcal{D}}^+_{\dot{\beta}} G^{++\dot{\alpha}+} = G^{(+4)}\delta_{\dot{\beta}}^{\dot{\alpha}},
        \qquad
        &
        \mathcal{D}^+_{\beta} G^{++\dot{\alpha}+}= 0,
        \\
        \bar{\mathcal{D}}^+_{\dot{\beta}} G^{(+4)}  = 0,
        \qquad
        &
        \mathcal{D}^+_{\beta} G^{(+4)} = 0.
    \end{split}
\end{equation}
This set of constraints can be treated as an alternative definition of the  $\mathcal{N}=2$ Weyl multiplet prepotentials. Their gauge transformations read:
\begin{equation}\label{eq: G trans full}
        \begin{split}
            \delta_\lambda G^{++\alpha\dot{\alpha}}
            =&\;
            \mathfrak{D}^{++} \Lambda^{\alpha\dot{\alpha}}
            -
            4i  \left( G^{++\alpha+} \Lambda^{-\dot{\alpha}} +  \Lambda^{-\alpha} G^{++\dot{\alpha}+} \right)
            \\
            &\qquad\qquad\;+
            4i  \left( G^{++\alpha-} \Lambda^{+\dot{\alpha}} +  \Lambda^{+\alpha} G^{++\dot{\alpha}-}\right),
\\
        \delta_\lambda G^{++\hat{\alpha}+}
        =
        &\;
        \mathfrak{D}^{++} \Lambda^{+\hat{\alpha}}
        +
        G^{(+4)} \Lambda^{-\hat{\alpha}}
        -
        G^{++\hat{\alpha}-}\Lambda^{++} ,
\\
        \delta_\lambda G^{++\hat{\alpha}-}
        = &\;
        \mathfrak{D}^{++} \Lambda^{-\hat{\alpha}}
        +
        \Lambda^{+\hat{\alpha}},
\\
\delta_\lambda G^{(+4)} = &\; \mathfrak{D}^{++} \Lambda^{++}.
        \end{split}
    \end{equation}
    Here the gauge $\Lambda$-parameters are defined through the expansion of the first-rank superdifferential operator $\hat{\Lambda}$ (see eq. \eqref{eq: hat Lambda})
    over covariant derivatives:
    \begin{subequations}\label{eq: Lambda and lambda}
    \begin{equation}
     \hat{\Lambda}
    =
    \Lambda^{\alpha\dot{\alpha}} \partial_{\alpha\dot{\alpha}}
    +
    \Lambda^{+\hat{\alpha}} \mathcal{D}^-_{\hat{\alpha}}
    +
    \Lambda^{-\hat{\alpha}} \mathcal{D}^+_{\hat{\alpha}}
    +
    \Lambda^{++} \mathcal{D}^{--}.
  \end{equation}
  Their explicit connection with the analytic $\lambda$-parameters is given by the relations:
    \begin{equation}
        \begin{split}
    & \Lambda^{\alpha\dot{\alpha}} \,= \lambda^{\alpha\dot{\alpha}}
            +
            4i \lambda^{\alpha+} \bar{\theta}^{-\dot{\alpha}}
            +
            4i \theta^{-\alpha} \bar{\lambda}^{+\dot{\alpha}}
            -
            4i \lambda^{++} \theta^{-\alpha} \bar{\theta}^{-\dot{\alpha}},
                        \\
            &
            \Lambda^{+\hat{\alpha}} = - \lambda^{+\hat{\alpha}} + \lambda^{++} \theta^{-\hat{\alpha}},
            \\
            & \Lambda^{-\hat{\alpha}}
            =
            \lambda^{-\hat{\alpha}},
            \\
            &
            \Lambda^{++} = \lambda^{++}.
        \end{split}
    \end{equation}
    \end{subequations}
    The corresponding analyticity conditions mirror eqs. \eqref{eq: G++ constr}:
    \begin{equation}\label{eq: Lambda constr}
        \begin{split}
            \bar{\mathcal{D}}^+_{\dot{\beta}} \Lambda^{\alpha\dot{\alpha}} = 4i \Lambda^{\alpha+} \delta^{\dot{\alpha}}_{\dot{\beta}}, \qquad
            &\mathcal{D}^{+}_{\beta} \Lambda^{\alpha\dot{\alpha}} = -4i \bar{\Lambda}^{\dot{\alpha}+} \delta^\alpha_\beta,
            \\
            \bar{\mathcal{D}}^+_{\dot{\beta}} \Lambda^{\alpha+} = 0,
            \qquad
            &
            \mathcal{D}^+_{\beta} \Lambda^{\alpha+} = \Lambda^{++} \delta^\alpha_\beta,
            \\
            \bar{\mathcal{D}}^+_{\dot{\beta}} \bar{\Lambda}^{\dot{\alpha}+} = \Lambda^{++}\delta_{\dot{\beta}}^{\dot{\alpha}},
            \qquad
            &
            \mathcal{D}^+_{\beta} \bar{\Lambda}^{\dot{\alpha}+}= 0,
            \\
            \bar{\mathcal{D}}^+_{\dot{\beta}} \Lambda^{++}  = 0,
            \qquad
            &
            \mathcal{D}^+_{\beta} \Lambda^{++} = 0.
        \end{split}
    \end{equation}
Note that, using the definitions \eqref{Gsuperpot}, these gauge transformations may be equivalently rewritten
    in terms of the  analytic $h^{++}$- prepotentials. In what follows, it will be more convenient for us
    to deal with  $G^{++}$-superfields.

From eqs. \eqref{eq: G trans full} we deduce the linearized gauge transformations of these superfields
\begin{equation}\label{eq: G++ GF}
    \begin{split}
        &\delta_\lambda G^{++\alpha\dot{\alpha}}
        =
        \mathcal{D}^{++} \Lambda^{\alpha\dot{\alpha}},
        \\
        &\delta_\lambda G^{++\hat{\alpha}+} = \mathcal{D}^{++} \Lambda^{+\hat{\alpha}},
           \\
        &\delta_\lambda G^{++\hat{\alpha}-} = \mathcal{D}^{++}\Lambda^{-\hat{\alpha}} + \Lambda^{+\hat{\alpha}},
        \\
        &\delta_\lambda G^{(+4)} = \mathcal{D}^{++} \Lambda^{++}.
    \end{split}
\end{equation}
Like in $\mathcal{N}=2$ Maxwell theory, introducing the appropriate zero-curvature conditions, we can construct gauge superfields with negative harmonic charges:
\begin{equation}\label{ZeroG}
    \begin{split}
        \mathcal{D}^{++} G^{--\alpha\dot{\alpha}}
        &=
        \mathcal{D}^{--} G^{++\alpha\dot{\alpha}},
        \\
        \mathcal{D}^{++} G^{--\hat{\alpha}+}
        &=
        \mathcal{D}^{--} G^{++\hat{\alpha}+},
        \\
        \mathcal{D}^{++} G^{--\hat{\alpha}-} &= - G^{--\hat{\alpha}+}.
    \end{split}
\end{equation}
The negatively-charged gauge potentials are uniquely specified by \eqref{ZeroG}
in terms of the $G^{++}$ superfields \eqref{Gsuperpot}. This can be directly checked, e.g.,  in WZ gauge.

From these equations, various useful relations follow, in particular,
\begin{subequations}\label{eq: cons of zc}
\begin{equation}
    \mathcal{D}^{++} \left( 8i G^{--\alpha+} - \bar{\mathcal{D}}^+_{\dot{\beta}} G^{--\alpha\dot{\beta}} \right)
    =
    \bar{\mathcal{D}}^-_{\dot{\beta}} G^{++\alpha\dot{\beta}},
\end{equation}
\begin{equation}
    \mathcal{D}^{++} \mathcal{D}^+_\beta G^{--\alpha+}
    =
    -
    \mathcal{D}^-_\beta G^{++\alpha+}
    +
    \delta^\alpha_\beta
    \mathcal{D}^{--} G^{(+4)},
\end{equation}
\begin{equation}
    \partial_{\alpha}^{\dot{\beta}} G^{++}_{\beta \dot{\beta}}
    =
    \mathcal{D}^{++} \left\{ \frac{1}{4i} \mathcal{D}^+_\alpha \bar{\mathcal{D}}^{+\beta} G^{--}_{\beta\dot{\beta}}
    +
    2 \mathcal{D}^+_{(\alpha} G^{--+}_{\beta)} \right\},
\end{equation}
\begin{equation}
    \mathcal{D}^-_{(\alpha} G^{+++}_{\beta)}
    =
    - \mathcal{D}^{++} \mathcal{D}^+_{(\alpha} G^{--+}_{\beta)},
\end{equation}
\begin{equation}
    \mathcal{D}^+_{(\alpha} \mathcal{D}^{--} G^{---}_{\beta)} = 0\,.
\end{equation}
\end{subequations}

\subsection{Gauge-invariant action and $\mathcal{N}=2$ Weyl supertensor}

Following ref. \cite{Ivanov:2024gjo}, we consider the complex dimensionless superfield:
\begin{equation}\label{eq: H++ alpha beta}
    \mathcal{H}^{++}_{(\alpha\beta)} = \partial_{(\alpha}^{\dot{\beta}} G^{++}_{\beta)\dot{\beta}}
    +
    \mathcal{D}^{-}_{(\alpha} G^{+++}_{\beta)},
\end{equation}
which satisfies the ``half-analyticity'' condition\footnote{Such a condition was introduced in the curved harmonic superspace in ref. \cite{Galperin:1987em}.
There, the authors explicitly show that it is possible to define a ``hybrid basis'' in HSS, such that the half-analyticity gets manifest. We will return to this issue in Sec. \ref{sec 5}. }:
\begin{equation}\label{eq: HA cond}
    \bar{\mathcal{D}}^+_{\dot{\rho}}    \mathcal{H}^{++}_{(\alpha\beta)} = 0.
\end{equation}
This condition fixes the relative coefficient in \eqref{eq: H++ alpha beta}.
The superfield  $\mathcal{H}^{++}_{(\alpha\beta)}$ has a simple gauge transformation law:
\begin{equation}\label{eq: H alpha beta gauge}
    \delta_\lambda \mathcal{H}^{++}_{(\alpha\beta)} = \mathcal{D}^{++} \Lambda_{(\alpha\beta)}
    +
    \underbrace{\mathcal{D}^+_{(\alpha} \Lambda^+_{\beta)}}_{0},
    \qquad
    \Lambda_{(\alpha\beta)} := \partial_{(\alpha}^{\dot{\beta}} \Lambda_{\beta)\dot{\beta}}
    +
    \mathcal{D}^-_{(\alpha} \Lambda^+_{\beta)}.
\end{equation}
The composite gauge parameter $   \Lambda_{(\alpha\beta)}$ is also subject to the half-analyticity condition,  $\bar{\mathcal{D}}^+_{\dot{\rho}} \Lambda_{(\alpha\beta)} = 0$.

\smallskip

By virtue of the relations \eqref{eq: cons of zc}, the superfield can be represented as a total harmonic derivative
\begin{equation}
    \mathcal{H}^{++}_{(\alpha\beta)}
    =
    \mathcal{D}^{++}  \mathcal{D}^+_{(\alpha} \left\{
    \frac{1}{4i} \bar{\mathcal{D}}^{+\dot{\beta}} G^{--}_{\beta)\dot{\beta}}  + G^{--+}_{\beta)}   \right\}.
\end{equation}
It is worth to note that the expression on which the harmonic derivative acts is not half-analytic.

\medskip
Using the zero-curvature equation
\begin{equation}\label{eq: zero-curv eq}
    \mathcal{D}^{++} \mathcal{H}^{--}_{(\alpha\beta)} =     \mathcal{D}^{--} \mathcal{H}^{++}_{(\alpha\beta)}\,,
\end{equation}
one can define a negatively charged superfield,
\begin{equation}
    \mathcal{H}^{--}_{(\alpha\beta)} = \partial_{(\alpha}^{\dot{\beta}} G^{--}_{\beta)\dot{\beta}}
    +
    \mathcal{D}^-_{(\alpha} G^{--+}_{\beta)}
    +
     \mathcal{D}^+_{(\alpha} G^{---}_{\beta)},
\end{equation}
with the gauge transformation law $\delta_\lambda \mathcal{H}^{--}_{(\alpha\beta)} = \mathcal{D}^{--} \Lambda_{(\alpha\beta)}$.

Now one can construct the linearized action:
\begin{equation}\label{eq: Weyl action 0}
    S_{Weyl} = \int d^4xd^8\theta du \; \mathcal{H}^{++(\alpha\beta)} \mathcal{H}^{--}_{(\alpha\beta)} + {\rm c.c.}\;,
\end{equation}
the gauge invariance of which can be easily checked:
\begin{equation}
    \begin{split}
    \delta_\lambda  S_{Weyl} =& \int d^4xd^8\theta du \; \left\{ \mathcal{D}^{++} \Lambda^{(\alpha\beta)} \mathcal{H}^{--}_{(\alpha\beta)}
    +
    \mathcal{H}^{++(\alpha\beta)} \mathcal{D}^{--} \Lambda_{(\alpha\beta)}
        \right\}
        \\
        =&\;
        2 \int d^4xd^8\theta du \;
        \mathcal{H}^{++(\alpha\beta)} \mathcal{D}^{--} \Lambda_{(\alpha\beta)}
        \\
        = &\;
         \frac{1}{8} \int d^4xd^4\theta du \;
       \left( \bar{\mathcal{D}}^- \right)^2 \Big\{ \mathcal{H}^{++(\alpha\beta)}   \underbrace{ \left(\bar{\mathcal{D}}^+ \right)^2 \left[ \mathcal{D}^{--} \Lambda_{(\alpha\beta)} \right] }_{0}
       \Big\} = 0
           \end{split}
\end{equation}
(the same check can be performed for the complex-conjugated part of the action \eqref{eq: Weyl action 0}).
Thereby, the invariance of the action is ensured by the half-analyticity condition. This is analogous to what happens in $\mathcal{N}=2$ Maxwell theory, where
the gauge invariance is secured due to the full harmonic analyticity.

\smallskip

\smallskip

\noindent \textbf{Chiral representation}

\smallskip

Analogously to the super Maxwell theory action, the action \eqref{eq: Weyl action 0} admits a chiral representation.
Using $\mathcal{H}^{--}_{(\alpha\beta)}$ one can construct the gauge-invariant supertensor
\begin{equation}\label{eq: linearized Weyl}
    \mathcal{W}_{(\alpha\beta)} = (\bar{\mathcal{D}}^+)^2 \mathcal{H}^{--}_{(\alpha\beta)},
\end{equation}
which is gauge invariant, thanks to the half-analyticity condition. Being written in terms of components, this supertensor contains the linearized Weyl tensor
\begin{equation}
    \mathcal{W}_{(\alpha\beta)} \sim  \theta^{+\gamma} \theta^{-\delta} C_{(\alpha\beta\gamma\delta)} + \dots,
    \qquad
    C_{(\alpha\beta\gamma\delta)}
    =
    \partial_{(\alpha}^{\dot{\alpha}}
    \partial_{\beta}^{\dot{\beta}} \Phi_{\gamma\delta) (\dot{\alpha}\dot{\beta})}.
\end{equation}
So we can identify $\mathcal{W}_{(\alpha\beta)} $  with the holomorphic part of the linearized $\mathcal{N}=2$ super Weyl tensor.
The relation between this supertensor and the super Weyl tensor defined in the standard superspace $\mathbb{R}^{4+2|8}$ will be established
in Appendix \ref{Appendix}, after imposing a harmonic-independent gauge.

As another consequence of the half-analyticity,    $\mathcal{W}_{(\alpha\beta)}$ is $\mathcal{N}=2$ chiral superfield.
Also, $\mathcal{N}=2$ super Weyl  tensor satisfies the Bianchi identity:
\begin{equation}
        (\mathcal{D}^{+\alpha} \mathcal{D}^{-\beta}) \mathcal{W}_{(\alpha\beta)}
        = ( \bar{\mathcal{D}}^{+\dot{\alpha}} \bar{\mathcal{D}}^{-\dot{\beta}} ) \overline{\mathcal{W}}_{(\dot{\alpha}\dot{\beta})}
\end{equation}
proved in Appendix \ref{app: A2}.

By performing a chain of transformations completely analogous  to those used in eqs. \eqref{eq: Max - ch} and \eqref{eq: Max - ch 2},
and using the half-analyticity  property (or, alternatively, the identity similar to \eqref{eq: alt repr}), it is easy to find the  chiral
representation for the action \eqref{eq: Weyl action 0}:
\begin{equation}\label{eq: Weyl action}
    S_{Weyl} = \int d^4xd^8\theta du \; \mathcal{H}^{++(\alpha\beta)} \mathcal{H}^{--}_{(\alpha\beta)} +{\rm c.c.}
    =
    \frac{1}{16}\int d^4x_L d^4\theta \, \mathcal{W}^{(\alpha\beta)} \mathcal{W}_{(\alpha\beta)}+{\rm c.c.}\,.
\end{equation}
After passing to components, this action involves the square of the linearized Weyl tensor, so it indeed describes the linearized $\mathcal{N}=2$ conformal supergravity.
The action \eqref{eq: Weyl action} has been constructed on a flat background. It should be emphasized that an analogous construction
can be carried out as well on an arbitrary conformally flat background, as demonstrated in \cite{Kuzenko:2021pqm} by making use of the $\mathcal{N}=2$ conformal superspace techniques \cite{Butter:2011sr}.
A generalization of this construction to harmonic superspace still remains to be done.

\medskip

To close this subsection, we highlight the remarkable analogy between the spin $\mathbf{2}$ gauge superfields discussed here and the spin $\mathbf{1}$ (Maxwell) gauge superfields:
\begin{equation}\label{eq: analogy}
    V^{\pm\pm} \;\; \leftrightarrow \;\; \mathcal{H}^{\pm\pm}_{(\alpha\beta)},
    \qquad
    \mathcal{D}^+_{\hat{\rho}}V^{\pm\pm}=0
     \;\;
      \leftrightarrow
       \;\; \bar{\mathcal{D}}^+_{\dot{\rho}}\mathcal{H}^{\pm\pm}_{(\alpha\beta)}=0,
    \qquad
    \mathcal{W}
     \;\;
     \leftrightarrow
     \;\; \mathcal{W}_{(\alpha\beta)}.
\end{equation}
This analogy may serve as a hint in constructing the complete non-linear $\mathcal{N}=2$ Weyl supergravity action in harmonic superspace.
Some further reasonings on this subject are contained in the final Section.

\subsection{Superconformal invariance of $\mathcal{N}=2$ conformal supergravity action}

Substituting the superconformal parameters into the transformation law
 \eqref{eq: G trans full}, we deduce  the superconformal transformation laws of $G^{++}$-superfields (see eq.  \eqref{eq: G++ trans in Lambda})\footnote{In this section we deal only with rigid $\mathcal{N}=2$ superconformal transformations.
Therefore, for sake of brevity, in most calculations we omit the subscript ``$sc$''  on $\lambda$-parameters.}:
\begin{equation}\label{eq: sc G++M}
    \begin{split}
        &\delta_{sc} G^{++\alpha\dot{\alpha}} =
        G^{++\alpha\dot{\beta}} \left( \bar{\mathcal{D}}^+_{\dot{\beta}} \bar{\lambda}_{}^{-\dot{\alpha}} \right)
        +
        G^{++\beta\dot{\alpha}} \left( \mathcal{D}^+_\beta \lambda_{}^{-\alpha} \right)
        +
       \left( \mathcal{D}^{--} \lambda^{++} \right) G^{++\alpha\dot{\alpha}} ,
        \\
        &
        \delta_{sc} G^{++\alpha+}
        =
        \frac{i}{4} (\bar{\mathcal{D}}^-_{\dot{\beta}} \lambda^{++})  G^{++\alpha\dot{\beta}}
        +
        G^{++\beta+} (\mathcal{D}^+_\beta \lambda^{-\alpha})
        +
         \left( \mathcal{D}^{--} \lambda^{++} \right) G^{++\alpha+},
        \\
        &
        \delta_{sc} G^{++\alpha-} =
        G^{++\beta\dot{\beta}} \left(\partial_{\beta\dot{\beta}} \lambda^{-\alpha} \right)
        +
       G^{++\beta-}   (\mathcal{D}^+_\beta \lambda^{-\alpha})
        +
        G^{++\beta+} (\mathcal{D}^-_\beta\lambda^{-\alpha}),
        \\
        &
        \delta_{sc} G^{(+4)} =
               G^{++\hat{\beta}+} (\mathcal{D}^-_{\hat{\beta}} \lambda^{++} )
              +  G^{(+4)}  (\mathcal{D}^{--}\lambda^{++}).
    \end{split}
\end{equation}
Note that these transformations can be written in various equivalent forms, as discussed in the Appendix \ref{eq: App C}. Here we choose the form that is most convenient
for further analysis. Using the relations \eqref{Gsuperpot},   one can derive from eqs.~\eqref{eq: sc G++M} also the superconformal transformation laws of analytic $h^{++}$-superfields.

The transformation of $G^{++\hat{\alpha}-}$ is inhomogeneous, which implies that for preserving the analytic
gauge $G^{++\hat{\alpha}-}_{a.g.} = 0$ one needs to  compensate the superconformal transformations by the appropriate
gauge transformations with a field-dependent parameter $\lambda^{-\hat{\alpha}}(G)$. So,
in the analytic gauge, the superconformal transformations of $G^{++}$-superfields become non-linear.
However, since $G^{++\hat{\alpha}-}$ is not present  in the dynamical action \eqref{eq: Weyl action} at all,
in what follows we are allowed  not to fix the analytic gauge and to deal solely with the analytic supervielbeins.
This is supported by the fact that the superfield $G^{++\alpha-}$ does not enter the superconformal transformation
laws of the remaining vielbeins.

\smallskip

Using the transformations given above, we can study the superconformal properties of the theory defined by the action \eqref{eq: Weyl action}.
It is convenient to consider the superconformal transformation laws of the basic object $\mathcal{H}^{++}_{(\alpha\beta)}$ in various $\lambda$-sectors. Using relations \eqref{eq: sc der prop} we derive:

\noindent\underline{$\lambda^{-\beta }$ sector:}

\begin{equation}
    \delta_{sc}^{(1)} \mathcal{H}^{++}_{(\alpha\beta)}
    =
    - \left( \mathcal{D}^+_{(\alpha} \lambda^{-\rho)} \right) \mathcal{H}^{++}_{\rho\beta}
        -
        \left( \mathcal{D}^+_{(\beta} \lambda^{-\rho)} \right) \mathcal{H}^{++}_{\alpha\rho}.
\end{equation}

\noindent\underline{$\bar{\lambda}^{-\dot{\beta}}$ sector:}

\begin{equation}
    \delta_{sc}^{(2)} \mathcal{H}^{++}_{(\alpha\beta)}
    =
    -
    \left(\partial_{(\alpha}^{\dot{\rho}} \bar{\mathcal{D}}^{+\dot{\beta}} \bar{\lambda}^-_{\dot{\rho}} \right) G^{++}_{\beta)\dot{\beta}}
    +
    4i
    \left(\partial_{(\alpha\dot{\beta}} \bar{\lambda}^{-\dot{\beta}} \right) G^{+++}_{\beta)}.
\end{equation}
Using equation \eqref{eq: sc f} we express this variation in terms of parameter $\lambda^{++}$
\begin{equation}
    \delta_{sc}^{(2)} \mathcal{H}^{++}_{(\alpha\beta)}
    =
   -
   2 \left( \mathcal{D}^-_{(\alpha} \mathcal{D}^{--} \lambda^{++} \right)
   G^{+++}_{\beta)}
   +
   \frac{i}{2 }
   \left( \mathcal{D}^-_{(\alpha} \bar{\mathcal{D}}^{-\dot{\beta}} \lambda^{++} \right) G^{++}_{\beta)\dot{\beta}}.
 \end{equation}

\smallskip

\noindent\underline{$\lambda^{++}$ sector:}
\begin{equation}\label{eq: lambda ++}
    \begin{split}
        \delta^{(3)}_{sc} \mathcal{H}^{++}_{(\alpha\beta)}
        = \;&
   +
\frac{i}{4} \left( \mathcal{D}^-_{(\alpha}  \lambda^{++} \right) \bar{\mathcal{D}}^{-\dot{\beta}} G^{++}_{\beta)\dot{\beta}}
-
\frac{i}{4} \left(\mathcal{D}^-_{(\alpha} \bar{\mathcal{D}}^{-\dot{\beta}} \lambda^{++} \right) G^{++}_{\beta)\dot{\beta}}
        \\&
           -
        \left(\mathcal{D}^-_{(\alpha} \lambda^{++} \right) \mathcal{D}^{--} G^{+++}_{\beta)}
        +
         \left(\mathcal{D}^-_{(\alpha} \mathcal{D}^{--}  \lambda^{++} \right) G^{+++}_{\beta)}
    .
    \end{split}
\end{equation}

The variation $\delta_{sc}^{(1)}$ yields a homogeneous transformation law. The sum of the two other contributions takes the form:
\begin{equation}\label{eq: inhom terms}
    \begin{split}
    \left( \delta^{(2)}_{sc} + \delta^{(3)}_{sc}  \right) \mathcal{H}^{++}_{(\alpha\beta)}
    =&
    -
    \mathcal{D}^{--} \left\{   \left(\mathcal{D}^-_{(\alpha} \lambda^{++} \right)  G^{+++}_{\beta)}  \right\}
    -
    \frac{i}{4}
    \bar{\mathcal{D}}^{-\dot{\beta}}
    \left\{
     \left( \mathcal{D}^-_{(\alpha}  \lambda^{++} \right)  G^{++}_{\beta)\dot{\beta}}  \right\}
     \\
     =&
     + \mathcal{D}^{--} \left\{   \left(\mathcal{D}^-_{(\alpha} \lambda^{++} \right)  G^{+++}_{\beta)}  \right\}
     +
     \frac{i}{4} \bar{\mathcal{D}}^{+\dot{\beta}} \mathcal{D}^{--}
        \left\{
     \left( \mathcal{D}^-_{(\alpha}  \lambda^{++} \right)  G^{++}_{\beta)\dot{\beta}}  \right\}.
     \end{split}
\end{equation}
For further use, let us denote the expressions standing under the harmonic derivative $\mathcal{D}^{--}$ in the last line  as
\begin{equation}
    X^{(+4)}_{(\alpha\beta)} = \left(\mathcal{D}^-_{(\alpha} \lambda^{++} \right)  G^{+++}_{\beta)},
    \qquad
    X^{(+3)}_{(\alpha\beta)\dot{\beta}}
    =
     \left( \mathcal{D}^-_{(\alpha}  \lambda^{++} \right)  G^{++}_{\beta)\dot{\beta}}.
\end{equation}
Introducing analogs of the harmonic zero-curvature equations, we can relate $X^{(+4)}_{(\alpha\beta)}$ and  $X^{(+3)}_{(\alpha\beta)\dot{\beta}}$ to the newly defined superfields $Y_{(\alpha\beta)}$ and
$Y^{-}_{(\alpha\beta)\dot{\beta}}$ as
\begin{equation} \lb{zeroCurvNew}
        \mathcal{D}^{--} X^{(+4)}_{(\alpha\beta)}
        =
        \mathcal{D}^{++} Y_{(\alpha\beta)}
        ,
        \qquad
            \mathcal{D}^{--}    X^{(+3)}_{(\alpha\beta)\dot{\beta}}
            =
                \mathcal{D}^{++} Y^{-}_{(\alpha\beta)\dot{\beta}}\,.
\end{equation}
The solution to the first equation is not unique: it is defined up to adding a term $Z_{(\alpha\beta)}$
annihilated by $\mathcal{D}^{++}$,  $\mathcal{D}^{++}Z_{(\alpha\beta)} =0$.

As a result, the variation \eqref{eq: inhom terms} takes the form:
\begin{equation}
    \left( \delta^{(2)}_{sc} + \delta^{(3)}_{sc}  \right) \mathcal{H}^{++}_{(\alpha\beta)}
    =
    \mathcal{D}^{++} \Lambda_{(\alpha\beta)}^{(sc)} ,
\end{equation}
where $\Lambda_{(\alpha\beta)}^{(sc)}$ is a field-dependent gauge-like parameter,
\begin{equation} \lb{ExpanLambda}
\Lambda_{(\alpha\beta)}^{(sc)} : = Y_{(\alpha\beta)}
+
Z_{(\alpha\beta)} + \frac{i}{4} \bar{\mathcal{D}}^{+\dot{\beta}} Y^{-}_{(\alpha\beta)\dot{\beta}}.
\end{equation}

\smallskip
\textbf{\begin{flushleft}
        \underline{The net result}
\end{flushleft}}
\smallskip
As the result of manipulations described above, restoring the subscript ``$sc$'', and  using  the relations \eqref{eq: Lor factors},
we  obtain
\begin{equation}\label{eq: result variantion}
\boxed{ \delta_{sc} \mathcal{H}^{++}_{(\alpha\beta)}
    =
    - \left(\mathcal{D}^{+}_{(\alpha}\lambda_{sc}^{-\rho)}\right) \mathcal{H}^{++}_{(\rho\beta)}
    - \left(\mathcal{D}^{+}_{(\beta}\lambda_{sc}^{-\rho)}\right) \mathcal{H}^{++}_{(\alpha\rho)}
    +
    \mathcal{D}^{++} \Lambda_{(\alpha\beta)}^{(sc)}.}
\end{equation}
Now we are going to show that, using the freedom in the definition of $Y_{(\alpha\beta)}$, the field-dependent parameter $\Lambda_{(\alpha\beta)}^{(sc)}$ can be chosen to be half-analytic.

The defining property of the superfield $\mathcal{H}^{++}_{(\alpha\beta)}$ is its harmonic half-analyticity.
Since $\bar{\mathcal{D}}^+_{\dot{\rho}} \lambda^{-\alpha}_{sc} = 0$, the first two terms in \eqref{eq: result variantion} obey the half-analyticity condition.
However, this property is not obvious for the gauge-transformation term.  To show that it also respects this property, we consider:
\begin{equation}
    0 = \delta_{sc} \left( \bar{\mathcal{D}}^+_{\dot{\alpha}} \mathcal{H}^{++}_{(\alpha\beta)}  \right)
    =
    - (\bar{\mathcal{D}}^+_{\dot{\alpha}} \lambda_{sc}^{-\dot{\beta}}) \underbrace{\bar{\mathcal{D}}^+_{\dot{\beta}} \mathcal{H}^{++}_{(\alpha\beta)} }_{0}
    +
    \bar{\mathcal{D}}^+_{\dot{\alpha}} \left(\delta_{sc}  \mathcal{H}^{++}_{(\alpha\beta)} \right)
    \quad
    \Rightarrow
    \quad
        \bar{\mathcal{D}}^+_{\dot{\alpha}} \left(\delta_{sc}  \mathcal{H}^{++}_{(\alpha\beta)} \right) = 0  .
\end{equation}
Thus the superconformal variation of $\mathcal{H}^{++}_{(\alpha\beta)}$ is also half-analytic.
This property, together with eq. \eqref{eq: result variantion}, lead us to the conclusion that $\mathcal{D}^{++}\Lambda^{(sc)}_{(\alpha\beta)}$ satisfies  the half-analyticity condition:
\begin{equation}\label{eq: cong on compos}
    \bar{\mathcal{D}}^+_{\dot{\alpha}}  \mathcal{D}^{++} \Lambda^{(sc)}_{(\alpha\beta)} = 0.
\end{equation}

In order to analyze this condition, we pass to the chiral basis
\eqref{eq: N=2 bases}, where $\bar{\mathcal{D}}^+_{\dot{\alpha}} = u^+_i \bar{\mathcal{D}}^i_{\dot{\alpha}}$.
The zero-curvature equations \eqref{zeroCurvNew} specify the harmonic dependence of different pieces of the gauge parameter \eqref{ExpanLambda}.
We will present them in a schematic form, focusing on their first terms:
\begin{subequations}
\begin{equation}
    Y_{(\alpha\beta)} \sim u^+ u^+ u^- u^-
    +
    \dots,
\end{equation}
\begin{equation}
    Y^-_{(\alpha\beta)\dot{\beta}} \sim  u^+ u^- u^-
    +
    \dots,
    \qquad
    \bar{\mathcal{D}}^+ Y^-_{(\alpha\beta)\dot{\beta}} \sim  u^+ u^-
    +
    \dots.
\end{equation}
\end{subequations}
Then, schematically,
\begin{equation} \lb{ExpanY}
\bar{\mathcal{D}}^+_{\dot{\alpha}} \left(Y_{(\alpha\beta)}
+
 \frac{i}{4} \bar{\mathcal{D}}^{+\dot{\beta}} Y^{-}_{(\alpha\beta)\dot{\beta}}
  \right)
  \sim
  u^+ +u^+u^+u^- + \dots.
\end{equation}
The equation \eqref{eq: cong on compos} implies that all terms in \eqref{ExpanY}, except for the first one, vanish.
The problematic first term can be eliminated by a suitable choice of the harmonic-independent zero mode $Z_{(\alpha\beta)}$.

\smallskip

This means that the term $  \mathcal{D}^{++} \Lambda_{(\alpha\beta)}^{(sc)}$ in the transformation law \eqref{eq: result variantion} has the form of half-analytic gauge transformations,
$\bar{\mathcal{D}}^+_{\dot{\alpha}}\Lambda_{(\alpha\beta)}^{(sc)} = 0$.
Since we have already proved that the Weyl action \eqref{eq: Weyl action} is gauge-invariant,
it is legitimate to ignore such a term.

\smallskip

The superconformal variation of the zero-curvature equation \eqref{eq: zero-curv eq} leads to the following equation for the superconformal variation $\delta_{sc} \mathcal{H}^{--}_{(\alpha\beta)}$:
\begin{equation}
    \mathcal{D}^{++} \delta_{sc} \mathcal{H}^{--}_{(\alpha\beta)}
    +
    2 \lambda^{++}_{sc}
    \mathcal{H}^{--}_{(\alpha\beta)}
    =
    -
    \left(\mathcal{D}^{--} \lambda^{++}_{sc}\right) \mathcal{D}^{--}    \mathcal{H}^{++}_{(\alpha\beta)}
    +
    \mathcal{D}^{--} \delta_{sc} \mathcal{H}^{++}_{(\alpha\beta)}.
\end{equation}
As its unique solution we obtain:
\begin{equation}\label{eq: H-- sc}
    \boxed{\delta_{sc}  \mathcal{H}^{--}_{(\alpha\beta)}
    =
        - \left(\mathcal{D}^{+}_{(\alpha}\lambda_{sc}^{-\rho)}\right) \mathcal{H}^{--}_{(\rho\beta)}
    - \left(\mathcal{D}^{+}_{(\beta}\lambda_{sc}^{-\rho)}\right) \mathcal{H}^{--}_{(\alpha\rho)}
        -
    \left(\mathcal{D}^{--} \lambda_{sc}^{++}\right)  \mathcal{H}^{--}_{(\alpha\beta)}. }
\end{equation}
The resulting formula for $\delta_{sc}    \mathcal{H}^{--}_{(\alpha\beta)}$  reminds the structure of transformations
$\delta_{sc}V^{--}$ in $\mathcal{N}=2$ Maxwell's theory (see eq. \eqref{eq: v-- sc}), with additional Lorentz-rotation type terms for each spinor index.
The last term involves a weight factor intended to compensate the non-invariance of the harmonic superspace measure \eqref{eq: fss Omega}.
We expect that the same mechanism works for the complete non-linear $\mathcal{N}=2$ conformal supergravity action in harmonic superspace,
with the rigid superconformal transformations being promoted to the local ones (see Sect 5 for some further details).

\medskip

The  Lagrange  density $\mathcal{H}^{++(\alpha\beta)} \mathcal{H}^{--}_{(\alpha\beta)}$ is a Lorentz scalar and it is transformed under the rigid ${\cal N}=2$ superconformal group just like the
Lagrange density $V^{++}V^{--}$ in $\mathcal{N}=2$ Maxwell theory. This immediately implies the superconformal invariance
of the linearized $\mathcal{N}=2$ Weyl theory action~\eqref{eq: Weyl action}.

\medskip

Analogously to the superconformal transformation of the superstrength $\mathcal{W}$ in $\mathcal{N}=2$ Maxwell theory, for the superconformal variation
of $\mathcal{N}=2$ super-Weyl tensor \eqref{eq: linearized Weyl} we obtain:
\begin{equation}\label{eq: Weyl sc}
    \begin{split}
    \delta_{sc} \mathcal{W}_{(\alpha\beta)} =   &
    -\left(\mathcal{D}^{+}_{(\alpha}\lambda_{sc}^{-\rho)}\right) \mathcal{W}_{(\rho\beta)}
    - \left(\mathcal{D}^{+}_{(\beta}\lambda_{sc}^{-\rho)}\right) \mathcal{W}_{(\alpha\rho)}
    \\&-
    \left( \bar{\mathcal{D}}^+_{\dot{\beta}} \bar{\lambda}_{sc}^{-\dot{\beta}} \right) \mathcal{W}_{(\alpha\beta)}
    -
    \left(\mathcal{D}^{--} \lambda_{sc}^{++}\right)  \mathcal{W}_{(\alpha\beta)}.
    \end{split}
\end{equation}
The second line is fully analogous to eq. \eqref{eq: sc W Maxwell th} for the Maxwell case.
The first line appeared due to non-trivial Lorentz properties of $\mathcal{W}_{(\alpha\beta)}$.
These terms are canceled in the variation of the Lagrangian density $\mathcal{W}^{(\alpha\beta)} \mathcal{W}_{(\alpha\beta)}$. The remaining terms
cancel the chiral weight factor \eqref{eq: Omega chiral}, thus providing the superconformal invariance of the chiral form of $\mathcal{N}=2$ Weyl action in \eqref{eq: Weyl action}.

\medskip
 Alternatively, the transformation law \eqref{eq: Weyl sc} can be derived by representing the Weyl tensor  as
 \begin{equation}
    \mathcal{W}_{(\alpha\beta)} = \int du \, \left( \bar{\mathcal{D}}^- \right)^2 \mathcal{H}^{++}_{(\alpha\beta)}.
 \end{equation}
 Considering the superconformal variation of  $\mathcal{W}_{(\alpha\beta)}$ in this form, the inhomogeneous terms \eqref{eq: inhom terms}  lead to a total $\mathcal{D}^{--}$ harmonic derivative:
 \begin{equation}
 \left( \bar{\mathcal{D}}^- \right)^2   \left( \delta^{(2)}_{sc} + \delta^{(3)}_{sc}  \right) \mathcal{H}^{++}_{(\alpha\beta)}
 =
    -
 \mathcal{D}^{--} \left\{    \left( \bar{\mathcal{D}}^- \right)^2 \left(\mathcal{D}^-_{(\alpha} \lambda^{++}_{sc} \right)  G^{+++}_{\beta)}  \right\}.
 \end{equation}
 Similarly to the case of the Maxwell superfield strength, only the homogeneous term survives from the variation \eqref{eq: D^-2 var} of $\left( \bar{\mathcal{D}}^- \right)^2$ , just because
 of  the half-analyticity of $\mathcal{H}^{++}_{(\alpha\beta)}$:
 \begin{equation}
    \begin{split}
    \delta_{sc} \left( \bar{\mathcal{D}}^- \right)^2 \mathcal{H}^{++}_{(\alpha\beta)}
        =&
    -
    \left[ 2 (\mathcal{D}^{--}\lambda^{++})
    +
    (\bar{\mathcal{D}}^-_{\dot{\beta}} \bar{\lambda}_{sc}^{-\dot{\beta}}) \right]
    (\bar{\mathcal{D}}^-)^2 \mathcal{H}^{++}_{(\alpha\beta)}
    \\
    &
    -
  \mathcal{D}^{++} \Big\{   \left(  (\bar{\mathcal{D}}^-)^2
    \lambda_{sc}^{++}  \right)  \mathcal{H}^{--}_{(\alpha\beta)} \Big\}
 +
    2   \mathcal{D}^{--}
    \Big\{
    \left(  \bar{\mathcal{D}}^{-\dot{\alpha}} \lambda^{++}_{sc}  \right)
 \bar{\mathcal{D}}^-_{\dot{\alpha}} \mathcal{H}^{++}_{(\alpha\beta)}
  \Big\}.
    \end{split}
 \end{equation}
 Then, taking into account the variation of the harmonic measure, we reproduce the transformation law~\eqref{eq: Weyl sc}.

\medskip

To conclude this Section, it is instructive to explicitly present the component structure of the factor appearing in the superconformal transformations \eqref{eq: result variantion} and \eqref{eq: H-- sc}:
\begin{equation}
\mathcal{D}^{+}_{(\alpha}\lambda_{sc}^{-\beta)}
=
l_{(\alpha}^{\;\beta)}
+
k_{(\alpha\dot{\beta}} x^{\beta)\dot{\beta}}
+
4i
k_{(\alpha\dot{\beta}} \theta^{-\beta)} \bar{\theta}^{+\dot{\beta}}
-
4i \eta^i_{(\alpha} \left( \theta^{-\beta)}u^+_i - \theta^{+\beta)} u^-_i \right).
\end{equation}
We note also one more interesting relation for the parameter $\Lambda_{(\alpha\beta)}$ (defined in eq. \eqref{eq: H alpha beta gauge})which is valid in the rigid superconformal limit:
\begin{equation}
     \Lambda_{(\alpha}^{\;\;\,\beta)}\Big|_{sc}
    =
    - \partial_{(\alpha\dot{\beta}} \Lambda_{sc}^{\dot{\beta}\beta)}
    +
    \mathcal{D}^-_{(\alpha} \Lambda^{+\beta)}_{sc}
    \overset{
        \eqref{eq: partial alpha dot alpha}
    }{=}
        -2
    \mathcal{D}^+_{(\alpha} \Lambda^{-\beta)}_{sc}
        +
    \mathcal{D}^-_{(\alpha} \Lambda^{+\beta)}_{sc}
    \overset{\eqref{eq: Lor factors}}{=}
        -
    \mathcal{D}^+_{(\alpha} \lambda^{-\beta)}_{sc}.
\end{equation}

\section{Towards non-linear Weyl action}\label{sec 5}

The analogy \eqref{eq: analogy} between $\mathcal{N}=2$ Maxwell theory and the linearized $\mathcal{N}=2$ Weyl theory
suggests a guess  on the structure of nonlinear $\mathcal{N}=2$ Weyl action in harmonic superspace.
In this section, we provide a discussion of the structure of such a conjectural nonlinear action of $\mathcal{N}=2$ conformal supergravity in the harmonic superspace formulation.

\subsection{Remarks on  $\mathcal{N}=2$ Yang-Mills theory}

Our starting point is the action of $\mathcal{N}=2$ supersymmetric Yang-Mills theory \cite{Zupnik:1987vm}:
\begin{equation}\label{eq: SYM}
    S_{\text{SYM}}^{\mathcal{N}=2} =
    \frac{1}{2} \sum_{n=2}^\infty \frac{(-i)^n}{n} \text{Tr} \int d^4x d^8\theta du_1 \dots du_n\; \frac{V^{++}(x, \theta, u_1) \dots V^{++}(x, \theta, u_n) }{(u^+_1u^+_2)\dots (u^+_nu^+_1)}.
\end{equation}
This action is invariant with respect to non-Abelian gauge transformations,
\begin{equation}
    \delta_\lambda V^{++} = \mathcal{D}^{++} \lambda + i[V^{++}, \lambda], \lb{V++Gauge}
\end{equation}
where the analytic superfields $V^{++}$ and $\lambda$ are in the adjoint representation of the gauge algebra. The proof of invariance uses the simple
general formula for the variation of $S_{\text{SYM}}^{\mathcal{N}=2}$
\be
\delta S_{\text{SYM}}^{\mathcal{N}=2} = \frac12\text{Tr} \int d^4x d^8\theta du \delta V^{++} V^{--}\,, \lb{genVar}
\ee
where
\begin{equation} \label{eq: V--}
V^{--}(u)
=
\sum_{n=1}^\infty (-i)^{n+1} \int du_1 \dots du_n \; \frac{V^{++}(x, \theta, u_1) \dots V^{++}(x, \theta, u_n) }{(u^+u_1^+)(u^+_1u^+_2)\dots (u^+_nu^+_1)(u^+_1 u^+)}
\end{equation}
 is the solution of the zero-curvature equation:
\begin{equation}\label{eq: YM zc}
    [\mathcal{D}^{++} + iV^{++}, \mathcal{D}^{--} + i V^{--}] = \mathcal{D}^0
        \quad
    \Leftrightarrow
    \quad
    \mathcal{D}^{++} V^{--} -   \mathcal{D}^{--} V^{++}
    +
    i [V^{++}, V^{--}] = 0.
\end{equation}

The invariance is ensured by the fact that, using  \eqref{V++Gauge}, \eqref{genVar}, \eqref{eq: YM zc} and integrating by parts with respect to harmonic derivatives,
the gauge variation of the action can be represented as
\begin{equation}\label{eq: YM var}
    \delta_\lambda  S_{\text{SYM}}^{\mathcal{N}=2}
    =
    \frac{1}{2}
    \text{Tr} \int d^4x d^8\theta du\; \delta_\lambda V^{++} V^{--}
    =
    -   \frac{1}{2} \text{Tr} \int d^4x d^8\theta du\;  V^{++} \mathcal{D}^{--} \lambda.
\end{equation}
Then, using the Grassmann harmonic analyticity, the variation \eqref{eq: YM var} can be shown to vanish:
\begin{equation}
    \left( \mathcal{D}^{+} \right)^4  \left\{ V^{++} \mathcal{D}^{--} \lambda\right\} = 0.
\end{equation}

Let us highlight most significant features of the above consideration:

\medskip

\noindent \textbf{1.} The variation \eqref{genVar} of the action  has the very simple form, involving the solution of the zero-curvature equation for $V^{--}$, which is the gauge connection
for the  harmonic derivative $\mathcal{D}^{--}$.  The action itself is constructed solely from the analytic connection $V^{++}$. Moreover, the action itself can be restored by postulating
the variation formula \eqref{genVar}. Namely, making the substitution
\be
V^{++} \,\rightarrow\,  tV^{++}\,, \quad \delta V^{++} = \delta{t}\,V^{++}
\ee
both in \eqref{genVar} and in \eqref{eq: YM zc}, we find the very simple equation for the action as a function of $t$,
\bea
&&\frac{\partial}{\partial t} S_{\text{SYM}}^{\mathcal{N}=2}(t) = \frac12\text{Tr} \int d^4x d^8\theta du\, V^{++}V^{--}(tV^{++})\,, \label{Eq-forS} \\
&& S_{\text{SYM}}^{\mathcal{N}=2}(t=0) = 0\,, \quad
S_{\text{SYM}}^{\mathcal{N}=2}(t=1) = S_{\text{SYM}}^{\mathcal{N}=2}\,. \nonumber
\eea
Solving it, we find the expression for $S_{\text{SYM}}^{\mathcal{N}=2}$
\bea
S_{\text{SYM}}^{\mathcal{N}=2} = \frac12\text{Tr} \int d^4x d^8\theta du\, V^{++}\int_0^1 dt V^{--}(tV^{++})\,.\lb{Act-int-t}
\eea
Expanding the series \eqref{eq: V--} for $V^{--}(tV^{++})$ over powers of $t$ and doing $t$ integrals in each term,
we come just to the action \eqref{eq: SYM}. In this way of deriving the ${\cal N}=2$ SYM action we need to know beforehand only the solution \eqref{eq: V--} for $V^{--}$.


\medskip
\noindent \textbf{2.} The gauge invariance of the action is a consequence of the analyticity of the prepotential and the supergauge parameter.

\medskip

\noindent \textbf{3.}  The action \eqref{eq: SYM} is written in the central basis $\{x_C, \theta^i_\alpha, \bar{\theta}^i_{\dot{\alpha}}, u^\pm\}$
of harmonic superspace, where the harmonic derivative has the simple ``short'' form ($\mathcal{D}^{++}_{CB} = \partial^{++}$) and the zero-curvature equations can be solved
using the harmonic distributions \cite{Galperin:1985bj} (defined by eqs. \eqref{eq: harmonic distrib}).

\medskip
\noindent It is natural to expect that all these special features are transferred to the full nonlinear $\mathcal{N}=2$ supergravity Weyl action.

\subsection{Hybrid basis in curved HSS and covariant half-analyticity condition}

Let us first recall that the key feature in the construction of the linearized superconformal theory was
the homogeneous transformation  low \eqref{eq: D+ sc} for one of the spinor derivatives:
\begin{equation}
    \delta_{sc} \bar{\mathcal{D}}^+_{\dot{\alpha}}
    =
    - \left(\bar{\mathcal{D}}^+_{\dot{\alpha}} \bar{\lambda}^{-\dot{\beta}}_{sc} \right) \bar{\mathcal{D}}^+_{\dot{\beta}},
\end{equation}
which is a consequence of the fact that $\bar{\mathcal{D}}^+_{\dot{\alpha}} \lambda^{-\beta}_{sc} = 0$ for the rigid $\mathcal{N}=2$ superconformal transformations.
This property ensures the covariance of the notion of half-analytic superfields with respect to rigid superconformal symmetry. However, for general analyticity-preserving coordinate
transformations \eqref{eq: an diff} this property fails to hold:
 \begin{equation}
    \delta_{\lambda} \bar{\partial} ^+_{\dot{\alpha}}
    =
    -
    \left( \bar{\partial}^+_{\dot{\alpha}} \lambda^{-\beta}\right)
    \partial^+_{\beta}
    -
     \left(\bar{\partial}^+_{\dot{\alpha}} \bar{\lambda}^{-\dot{\beta}} \right) \bar{\partial}^+_{\dot{\beta}}.
 \end{equation}
It was noticed in ref.  \cite{Galperin:1987em}  that there exists a \textit{hybrid basis} in the curved harmonic superspace,
\begin{equation} \label{Hybrid}
    Z_H = \{ x_H^{\alpha\dot{\alpha}}, \theta_H^{\pm\hat{\alpha}}, u^\pm_i \}
    =
        \{ x_A^{\alpha\dot{\alpha}}, \theta_A^{+\hat{\alpha}}, \bar{\theta}_A^{-\dot{\alpha}},  \theta_H^{-\alpha} := \theta_A^{-\alpha} - \psi^{-\alpha}, u^\pm_i \},
 \end{equation}
such that  the property
 \begin{equation}
        \delta_{\lambda} \bar{\Delta} ^+_{\dot{\alpha}}
    =
    -
    \left(\bar{\Delta}^+_{\dot{\alpha}} \bar{\lambda}^{-\dot{\beta}} \right) \bar{\Delta}^+_{\dot{\beta}}\,,
    \qquad
   \bar{\Delta}^+_{\dot{\alpha}}
    \lambda^{-\beta} = 0\,,
 \end{equation}
  is satisfied for the shortened derivative\footnote{For the explicit construction of the bridge $\psi^{-\alpha}$ in \eqref{Hybrid} see the original article \cite{Galperin:1987em}.}:
\begin{equation}
  \bar{\Delta}^+_{\dot{\alpha}} :=
 \bar{ \partial}^+_{H\dot{\alpha}}
  =
   \frac{\partial}{\partial \bar{\theta}_H^{-\dot{\alpha}}}.
 \end{equation}
This property makes it possible to introduce (and resolve) a \textit{covariantization  of the half-analyticity condition} \eqref{eq: HA cond}:
\begin{equation}
      \bar{\Delta}^+_{\dot{\alpha}} \phi(Z) = 0
      \qquad
      \Rightarrow
      \qquad
      \phi(Z) =
      \phi (x_H, \theta^{+\hat{\alpha}}_H, \theta^{-\alpha}_H, u).
\end{equation}

 \subsection{Superfields $\mathcal{H}^{\pm\pm}_{(\alpha\beta)}$, zero curvature conditions and the variation of action}

Building on our observations in the linearized case and the properties of the hybrid basis just mentioned,
we assume the existence of a curved generalization of the half-analytic superfield \eqref{eq: H++ alpha beta},
such that it is transformed according to
\begin{equation}\label{eq: NL H++}
    \delta_\lambda \mathcal{H}^{++}_{(\alpha\beta)}
    =
    \mathfrak{D}^{++} \Lambda_{(\alpha\beta)}
    +
    \Lambda_{(\alpha}^{\;\;\rho)}  \mathcal{H}^{++}_{(\rho\beta)}
    +
    \Lambda_{(\beta}^{\;\;\rho)}  \mathcal{H}^{++}_{(\alpha\rho)}
\end{equation}
and satisfies the covariant half-analyticity condition:
\begin{equation}\label{eq: cov half an}
    \bar{\Delta}^+_{\dot{\beta}} \mathcal{H}^{++}_{(\alpha\beta)} = 0,
    \qquad
        \bar{\Delta}^+_{\dot{\beta}} \Lambda_{(\alpha\beta)} = 0.
\end{equation}
Since all off-shell degrees of freedom of $\mathcal{N}=2$ conformal supergravity are carried out by analytic prepotentials,
we also assume that the superfield $\mathcal{H}^{++}_{(\alpha\beta)}$ can be properly constructed in terms of them. Similarly,
the transformation parameter $\Lambda_{(\alpha\beta)}$ should be ultimately expressible in terms of the original analytic parameters.
In the linearized limit, all these superfield objects have to be  reduced to \eqref{eq: H++ alpha beta} and \eqref{eq: H alpha beta gauge}.
We will return to this point in some detail in the next subsection.

To construct the dynamical action and the Weyl tensor in the nonlinear case, we also need to define the negatively-charged superfield $\mathcal{H}^{--}_{(\alpha\beta)}$.

In $\mathcal{N}=2$ conformal supergravity the covariant zero-curvature equation is known to be:
\begin{equation}\label{eq: SG full harm}
    [\mathfrak{D}^{++} - \mathcal{H}^{(+4)}\mathfrak{D}^{--} ,  \mathfrak{D}^{--}]
    =
    \mathcal{D}^0.
\end{equation}
Unlike the analogous harmonic equations in the Yang-Mills case, this condition amounts to a set of nonlinear equations,
making it difficult to immediately find for them the explicit solutions of the type \eqref{eq: V--}.
Nevertheless, the zero-curvature condition \eqref{eq: SG full harm} is covariant  under the transformations
\begin{equation}\label{eq: mathfrak D trans}
    \begin{split}
    &\delta_\lambda \mathfrak{D}^{++}
    =
    -\lambda^{++} \mathcal{D}^0,
    \\
        &\delta_\lambda \mathfrak{D}^{--}
        =
        -\left( \mathfrak{D}^{--} \lambda^{++}\right) \mathfrak{D}^{--},
    \\
    &   \delta_\lambda \mathcal{H}^{(+4)}
        =
        \mathfrak{D}^{++} \lambda^{++}.
        \end{split}
\end{equation}
Using  the covariant derivatives $\mathfrak{D}^{++}$ we can  define the negatively-charged superfield $\mathcal{H}^{--}_{(\alpha\beta)}$
as the solution of the modified zero-curvature equation:
\begin{equation}\label{eq: NL ZC equation H}
    \mathfrak{D}^{++}   \mathcal{H}^{--}_{(\alpha\beta)}
    -
    \mathfrak{D}^{--}   \mathcal{H}^{++}_{(\alpha\beta)}
    +
    \left(\mathfrak{D}^{--}   \mathcal{H}^{(+4)}\right) \mathcal{H}^{--}_{(\alpha\beta)}
    +
    2\mathcal{H}^{++\rho}_{(\alpha} \mathcal{H}^{--}_{\rho\beta)}
    =
    0,
\end{equation}
which resembles the SYM zero-curvature condition \eqref{eq: YM zc}\footnote{In contrast to \eqref{eq: YM zc}, we have no idea so far how to solve
eq. \eqref{eq: NL ZC equation H} for $\mathcal{H}^{--}_{\rho\beta)}$ (even perturbatively) because of its intrinsic non-linearity related to the presence of conformal supergravity vielbeins in  $\mathfrak{D}^{\pm\pm}$.}.
This equation is covariant under the transformations \eqref{eq: NL H++}, \eqref{eq: mathfrak D trans} accompanied  with
\begin{equation}\label{eq: H-- transf}
    \delta_\lambda \mathcal{H}^{--}_{(\alpha\beta)}
    =
    \mathfrak{D}^{--} \Lambda_{(\alpha\beta)}
    +
    \Lambda_{(\alpha}^{\;\;\rho)}  \mathcal{H}^{--}_{(\rho\beta)}
    +
    \Lambda_{(\beta}^{\;\;\rho)}  \mathcal{H}^{--}_{(\alpha\rho)}
    -(\mathfrak{D}^{--} \lambda^{++})
    \mathcal{H}^{--}_{(\alpha\beta)}.
\end{equation}

By analogy with \eqref{eq: YM var}, we will suppose that variation of the hypothetical action (which we are not yet able to explicitly exhibit) is given by:
\begin{equation}\label{eq: Weyl variation}
    \begin{split}
    \delta_\lambda S_{\text{Weyl}}
    =&
    \int d^4x d^8 \theta du E \; \delta_\Lambda \mathcal{H}^{++}_{(\alpha\beta)} \mathcal{H}^{--(\alpha\beta)}
    \\
    \overset{\eqref{eq: NL H++}}{=}&
        \int d^4x d^8 \theta du E \; \left(\mathfrak{D}^{++} \Lambda_{(\alpha\beta)} \mathcal{H}^{--(\alpha\beta)}
        +
        2
    \Lambda_{(\alpha}^{\;\;\rho}    \mathcal{H}^{++}_{\rho\beta)}
        \mathcal{H}^{--(\alpha\beta)}
        \right).
    \end{split}
\end{equation}
For the first term we obtain:
\begin{equation}
    \begin{split}
    &\int d^4x d^8 \theta du E \;  \mathfrak{D}^{++} \Lambda_{(\alpha\beta)} \mathcal{H}^{--(\alpha\beta)}
    \\&\qquad=
        \int d^4x d^8 \theta du E \;
        \left[          \mathfrak{D}^{++} \left(    \Lambda_{(\alpha\beta)}
 \mathcal{H}^{--(\alpha\beta)} \right)
        -
        \Lambda_{(\alpha\beta)}
    \mathfrak{D}^{++} \mathcal{H}^{--(\alpha\beta)} \right]
    \\&\quad\;\,\overset{\eqref{eq: NL ZC equation H}}{=}
    -   \int d^4x d^8 \theta du E \;    \Lambda^{(\alpha\beta)}
    \left\{
    \mathfrak{D}^{--} \mathcal{H}^{++}_{(\alpha\beta)}
    -
    2\mathcal{H}^{++\rho}_{(\alpha} \mathcal{H}^{--}_{\rho\beta)}
    \right\}
        \\&\qquad\quad\;\;-    \int d^4x d^8 \theta du E\;
        \Lambda^{(\alpha\beta)}
        \left[ \mathfrak{D}^{++} \ln E+ (-1)^M\partial_M \mathcal{H}^{++M}  -\left(\mathfrak{D}^{--}   \mathcal{H}^{(+4)}\right)
        \right] \mathcal{H}^{--}_{(\alpha\beta)}.
\end{split}
\end{equation}
The last term in the third line  cancels the second term in the variation \eqref{eq: Weyl variation},
while the expression in the square brackets (fourth line) is zero in virtue of reasoning in \cite{18},
\begin{equation}\label{Gamma}
        \left[ \mathfrak{D}^{++} \ln E+ \Gamma^{++} -\left(\mathfrak{D}^{--}   \mathcal{H}^{(+4)}\right)
    \right] = 0,
    \qquad
    \Gamma^{++} := (-1)^M\partial_M \mathcal{H}^{++M}\,.
\end{equation}
So we are left with the variation:
\begin{equation}
    \delta_\Lambda S_{Weyl}
    =
        -   \int d^4x d^8 \theta du E \;    \Lambda^{(\alpha\beta)}
    \mathfrak{D}^{--} \mathcal{H}^{++}_{(\alpha\beta)} .
\end{equation}
To show that it vanishes, we will employ the arguments from \cite{Galperin:1987em}\footnote{See eq. (5.28)  there.}.
In the hybrid basis the following relations hold
\begin{equation}\label{eq: hebt prop}
    \begin{split}
        & (\bar{\Delta}^+)^2  \left( E  J \right) =  \left( E  J \right) \left[(\bar{\Delta}^+)^2 \left( \ln E  J \right) + 2 \bar{\Delta}_{\dot{\alpha}}   \left( \ln E  J \right)  \bar{\Delta}^{\dot{\alpha}}  \left( \ln E  J \right)  \right]  = 0,
        \\
&(\bar{\Delta}^+)^2 \mathcal{H}^{--\alpha\dot{\alpha}, \hat{\alpha}} - 2\bar{\Delta}^{+\dot{\alpha}} \left( \ln E  J \right)  \bar{\Delta}_{\dot{\alpha}}^{+} \mathcal{H}^{--\alpha\dot{\alpha}, \hat{\alpha}} = 0,
\end{split}
\end{equation}
where $J := \text{Ber} \frac{\partial Z_A}{\partial Z_H}$ is the Berezinian of the transformation to the hybrid basis in HSS.
These relations imply
\begin{equation}
    \begin{split}
(\bar{\Delta}^+)^2 \left[ E J  \mathcal{H}^{--\alpha\dot{\alpha}, \hat{\alpha}}  \right]
=
EJ
\Big[ & \left\{ (\bar{\Delta}^+)^2 \left( \ln E  J \right)
+
2 \bar{\Delta}_{\dot{\alpha}}   \left( \ln E  J \right)  \bar{\Delta}^{\dot{\alpha}}  \left( \ln E  J \right)
 \right\}
  \mathcal{H}^{--\alpha\dot{\alpha}, \hat{\alpha}}
\\& +
(\bar{\Delta}^+)^2 \mathcal{H}^{--\alpha\dot{\alpha}, \hat{\alpha}} - 2\bar{\Delta}^{+\dot{\alpha}} \left( \ln E  J \right)  \bar{\Delta}_{\dot{\alpha}}^{+} \mathcal{H}^{--\alpha\dot{\alpha}, \hat{\alpha}}  \Big] = 0.
\end{split}
\end{equation}
Thus we find,
\begin{equation}
    \delta_\Lambda S_{Weyl}
    =
        -   \int d^4x_H d^4 \theta^+_H d^2 \theta^-_H du \;  (\bar{\Delta}^+)^2 \left[ E J \,    \Lambda^{(\alpha\beta)}
    \mathfrak{D}^{--} \mathcal{H}^{++}_{(\alpha\beta)}   \right]
    =
    0
\end{equation}
as a consequence of the  covariant half-analyticity condition \eqref{eq: cov half an}.

\subsection*{Conjecture on the complete non-linear action}

Like in the case of $\mathcal{N}=2$ Yang-Mills theory, by integrating the variation of the action one can hope to derive the full action.
The main difficulty in our case is that in conformal $\mathcal{N}=2$  supergravity for the time being we do not know how to write a closed solution
to the zero-curvature equations. Nevertheless, we will present here a number of guesses  regarding the possible structure of such an action.

For a fixed background of $\mathcal{N}=2$ conformal supergravity, eq. \eqref{eq: NL ZC equation H} can be formally solved as a perturbation series:
\begin{equation} \lb{eq: NL ZC equation HSol}
    \mathcal{H}^{--}_{(\alpha\beta)} (u) = \sum_{n=1}^\infty \int du_1\dots du_n \; \frac{ (\mathcal{H}^{++}(u_1))_{(\alpha}^{\;\;\beta_1)}
        (\mathcal{H}^{++}(u_2))_{(\beta_1}^{\;\;\beta_2)}
        \dots
        (\mathcal{H}^{++}(u_n))_{\beta_{n-1}\beta)} }{[u^+u_1^+][u_1^+u_2^+]\dots [u_n^+u^+]}.
\end{equation}
The symbols $\frac{1}{[u^+u_1^+]}$ stand for  a proper generalization of harmonic distributions $\frac{1}{(u^+u_1^+)}$ to curved case.
Though we are not still aware of any explicit expressions for them, crucial for our purposes will be the following property.
Due to the presence of the homogeneity term in the superfield transformation \eqref{eq: H-- transf},
each term in the gauge transformation of the  series \eqref{eq: NL ZC equation HSol} has to contain a homogeneous piece with the weight factor $-(\mathfrak{D}^{--} \lambda^{++}) $.
This term in the transformation law plays an important role.

Under the conformal supergravity transformations, the harmonic superspace measure transforms non-trivially.
So in order to construct an  invariant action one needs to covariantize the superspace measure as \cite{18, Galperin:1987ek, Galperin:1987em}:
\begin{equation}\label{eq: measure transf}
    d^4x d^8\theta du \Rightarrow  d^4x d^8\theta du \, E\,, \quad   \delta_\lambda \left(d^4x d^8\theta du \, E\right)
    =
    \left(\mathfrak{D}^{--} \lambda^{++} \right) \left(d^4x d^8\theta du \, E\right).
\end{equation}
For the invariance of the full action, the  variation of the Weyl Lagrangian superfield density must cancel this weight factor.

Now we consider the ``trial'' action functional
\begin{equation}\label{eq: N=2 Weyl full}
    S_{Weyl} = \int  d^4x d^8\theta du \, E \sum_{n=2}^\infty \int du_1\dots du_n \; \frac{1}{n}  \frac{ (\mathcal{H}^{++}(u))_{(\alpha}^{\;\;\beta_1)}
        (\mathcal{H}^{++}(u_1))_{(\beta_1}^{\;\;\beta_2)}
        \dots
        (\mathcal{H}^{++}(u_n))^{\quad\alpha)}_{(\beta_{n-1}} }{[u^+u_1^+][u_1^+u_2^+]\dots [u_n^+u^+]}.
\end{equation}
The weight factor $ \left(\mathfrak{D}^{--} \lambda^{++} \right)$ from the transformation of the measure \eqref{eq: measure transf} will be precisely canceled by the factor coming from the Lagrangian density, and the variation of the action
will take almost the same form as in the $\mathcal{N}=2$ Yang-Mills case \eqref{eq: Weyl variation}. We hope to work out this opportunity elsewhere.

\subsection{$\mathcal{N}=2$ Weyl supertensor}

In this subsection, we will provide the evidence  for the existence of the superfields $\mathcal{H}^{\pm\pm}_{(\alpha\beta)}$  at the full nonlinear level, although
explicitly expressing them in terms of the analytic prepotentials \eqref{eq: H transf} still remains a challenge.



Let is consider a non-linear generalization of the Weyl tensor \eqref{eq: linearized Weyl}:
\begin{equation}\label{eq: nl Weyl}
    \mathcal{W}_{(\alpha\beta)} = (\bar{\Delta}^+)^2  \left(  EJ\, \mathcal{H}^{--}_{(\alpha\beta)} \right).
\end{equation}
Here, besides the superfield $\mathcal{H}^{--}_{(\alpha\beta)}$, we have inserted the harmonic superspace measure and the Berezinian of the transformation
into hybrid basis, since just these quantities provide the properties required, as we will demonstrate now.

\smallskip

We start by computing the harmonic derivative of \eqref{eq: nl Weyl}:
\begin{equation}
    \begin{split}
    \mathfrak{D}^{++}   \mathcal{W}_{(\alpha\beta)}
    =&\;
    (\bar{\Delta}^+)^2
     EJ \Bigg( \underbrace{\left[  \mathfrak{D}^{++} \ln (EJ)  -
     \left(\mathfrak{D}^{--}   \mathcal{H}^{(+4)}\right) \right]}_{-\Gamma^{++}} \mathcal{H}^{--}_{(\alpha\beta)}
     +
        \mathfrak{D}^{--}   \mathcal{H}^{++}_{(\alpha\beta)}
     -
     2\mathcal{H}^{++\rho}_{(\alpha} \mathcal{H}^{--}_{\rho\beta)}
      \Bigg)
      \\
      =&- \Gamma^{++} \mathcal{W}_{(\alpha\beta)}  - 2\, \mathcal{H}^{++\rho}_{(\alpha} \mathcal{W}_{\rho\beta)}.
      \end{split}
\end{equation}
Here  we made use of \eqref{eq: hebt prop} and the half-analyticity property \eqref{eq: cov half an}.
As a result, we observe that the Weyl tensor satisfies a generalized condition of harmonic independence:
\begin{equation}\label{eq: gen HI cond}
        \mathfrak{D}^{++}   \mathcal{W}_{(\alpha\beta)}
        +
         \Gamma^{++} \mathcal{W}_{(\alpha\beta)}
         +
          2\, \mathcal{H}^{++\rho}_{(\alpha} \mathcal{W}_{\rho\beta)} = 0.
\end{equation}
The second term mimics the analogous term in the equations of motion of the hypermultiplet in the background of $\mathcal{N}=2$ conformal supergravity (see, e.g. \cite{18}):
\begin{equation}
    \left( \mathfrak{D}^{++} + \frac{1}{2} \Gamma^{++} \right) q^+_a = 0,
\end{equation}
where $\Gamma^{++}$ is the connection term defined in \eqref{Gamma} and needed to compensate the conformal weight. In a similar way,
the superfield $\mathcal{H}^{++}_{(\alpha\beta)}$ acquires the interpretation of the connection intended to compensate Lorentz-like
gauge transformations.

Thus, the generalized harmonic-independence condition \eqref{eq: gen HI cond} and the chirality condition $\bar{\Delta}^+_{\dot{\alpha}}  \mathcal{W}_{(\alpha\beta)} = 0 $ demonstrate,
that the supertensor \eqref{eq: nl Weyl} satisfies  the same  constraints in the nonlinear case as $\mathcal{N}=2$ Weyl supertensor in the linearized case.
As a consequence, the existence of the Weyl tensor guarantees the presence of the superfield $\mathcal{H}^{--}_{(\alpha\beta)}$,
and, in virtue of the zero-curvature equations and harmonic independence, implies the existence of the half-analytic superfield $\mathcal{H}^{++}_{(\alpha\beta)}$.

One can also compute the transformation law of the Weyl supertensor \eqref{eq: nl Weyl}.
To this end, we employ the relations:
\begin{subequations}
\begin{equation}
    \delta_\lambda (\bar{\Delta}^+)^2
    =
    - \left(\bar{\Delta}^+_{\dot{\alpha}} \bar{\lambda}^{-\dot{\alpha}} \right) (\bar{\Delta}^+)^2
    -
    (\bar{\Delta}^+)^2  \bar{\lambda}^{-\dot{\beta}} \bar{\Delta}^+_{\dot{\beta}},
\end{equation}
\begin{equation}
    (\bar{\Delta}^+)^2 \left(\bar{\Delta}^+_{\dot{\alpha}} \bar{\lambda}^{-\dot{\alpha}} \right)
    =
    (\bar{\Delta}^+)^2 \bar{\lambda}^{-\dot{\alpha}} \bar{\Delta}^+_{\dot{\alpha}}
    +
    \left(\bar{\Delta}^+_{\dot{\alpha}} \bar{\lambda}^{-\dot{\alpha}} \right)(\bar{\Delta}^+)^2,
\end{equation}
\begin{equation}
    \begin{split}
    \delta_{\lambda} \left(EJ\right)
    &=
    - \left(\partial_{\alpha\dot{\alpha}} \lambda^{\alpha\dot{\alpha}}
    -
     \partial^{-}_{\hat{\alpha}} \lambda^{+\hat{\alpha}}
    -
         \partial^{+}_{\alpha} \lambda^{-\alpha}
         -
             \bar{\Delta}^{+}_{\dot{\alpha}} \bar{\lambda}^{-\dot{\alpha}}
             +
             \partial^{--} \lambda^{++}
             -
             \mathfrak{D}^{--} \lambda^{++}
      \right)  \left(EJ\right)
      \\
      &=-
     \left(  \Omega_H  -
     \bar{\Delta}^{+}_{\dot{\alpha}} \bar{\lambda}^{-\dot{\alpha}}
     -
     \mathfrak{D}^{--} \lambda^{++} \right)  \left(EJ\right).
\end{split}
\end{equation}
\end{subequations}
Employing these explicit variations, we derive the homogeneous transformation law for ${\cal W}_{(\alpha\beta)}$,
\begin{equation}
    \delta_\lambda \mathcal{W}_{(\alpha\beta)}
    =
    \Lambda_{(\alpha}^{\;\;\rho)}
    \mathcal{W}_{(\rho\beta)}
    +
        \Lambda_{(\beta}^{\;\;\rho)}
    \mathcal{W}_{(\alpha\rho)}
    -
    \Omega_H \mathcal{W}_{(\alpha\beta)},
\end{equation}
where
\begin{equation}
\Omega_H :=  \partial_{\alpha\dot{\alpha}} \lambda^{\alpha\dot{\alpha}}
-
\partial^{-}_{\hat{\alpha}} \lambda^{+\hat{\alpha}}
-
\partial^{+}_{\alpha} \lambda^{-\alpha}
+
\partial^{--} \lambda^{++}.
\end{equation}


\section{Conclusions} \label{sec 6}

In this work we have constructed a harmonic gauge-invariant linearized action for $\mathcal{N}=2$ Weyl multiplet, proceeding from
the description of the latter through analytic vielbeins of the harmonic derivative $\mathfrak{D}^{++}$.
The action has a simple structure basically repeating the structure of $\mathcal{N}=2$ Maxwell theory action,
the role of the analyticity condition being replaced by the ``half-analyticity'' condition. We also derived a chiral representation for this action.
Using the realization of rigid superconformal transformations in harmonic superspace, we have shown that the  actions constructed
are invariant under rigid $\mathcal{N}=2$ superconformal transformations. Based on the observation that the notion of half-analyticity admits
a generalization to the curved harmonic superspace, we proposed the nonlinear extension of the linearized $\mathcal{N}=2$ Weyl action and proved a number of noticeable
features of the latter, including its invariance under local $\mathcal{N}=2$ superconformal gauge transformations. To complete the theory, it remained to solve the difficult problem
of expressing the basic half-analytic superfield $\mathcal{H}^{++}_{(\alpha\beta)}$ through the underlying  analytic harmonic vielbeins of conformal $\mathcal{N}=2$  supergravity.

\smallskip

Beside this issue, our results offer a few further  possible generalizations.

\medskip

 $\bullet$ \textit{$\mathcal{N}=2$ higher-spin superconformal theory in harmonic superspace}

\smallskip
$\mathcal{N}=2$ Weyl multiplet is the simplest superconformal higher-spin multiplet \cite{Buchbinder:2024pjm}.
The natural task is to generalize the spin 2 $\mathcal{N}=2$ Weyl theory to an arbitrary integer $\mathcal{N}=2$ higher spin.
The main difficulty of such a generalization is that the relevant multiplets are described by a larger set of superfields,
which are necessary for the linear realization of rigid superconformal transformations.

As was shown in \cite{Buchbinder:2024pjm}, there exists a gauge in which, for arbitrary superconformal spins $\mathbf{s}$,  only few analytic prepotentials survive, {\it viz:}
\begin{equation}
    h^{++\alpha(s-1)\dot{\alpha}(s-1)},
    \quad
    h^{++\alpha(s-1)\dot{\alpha}(s-2)},
    \quad
    h^{++\alpha(s-2)\dot{\alpha}(s-1)},
    \quad
    h^{++\alpha(s-2)\dot{\alpha}(s-2)}.
\end{equation}
In this gauge one can easily construct (as in Ref. \cite{Zaigraev:2024ryg}) the higher-spin counterpart $\mathcal{H}^{++}_{\alpha(2s-2)}$ of $\mathcal{H}^{++}_{(\alpha\beta)}$,
as well as the higher-spin Weyl supertensor $\mathcal{W}_{\alpha(2s-2)}$ and, using these objects, set up a gauge-invariant action. These results will be presented elsewhere.


\medskip

$\bullet$ \textit{ $\mathcal{N}=2$ cubic vertices in harmonic superspace}

\smallskip

Based on the conjectured action \eqref{eq: N=2 Weyl full}, it would be of interest to explore the explicit structure of the cubic vertex in terms of analytic prepotentials.
This may provide a useful guide in constructing non-Abelian
cubic vertices \cite{Berends:1984wp, Fradkin:1987ks, Zinoviev:2008ck} in $\mathcal{N}=2$ higher-spin theory.

\medskip

$\bullet$ \textit{$\mathcal{N}=2$ AdS supergravity}

\smallskip

The transformation laws of $\mathcal{N}=2$  conformal supergravity superfields \eqref{eq: sc G++M}
with respect to the rigid superconformal transformations fully specify their transformation laws with respect
to $\mathcal{N}=2$ AdS supersymmetry $\mathfrak{osp}(2|4) \subset \mathfrak{su}(2,2|2)$.
This property, combined with the ideas  of a recent work \cite{Ivanov:2025jdp} (see also \cite{Gargett:2025xcg}), can be utilized to construct
$\mathcal{N}=2$ AdS supergravity in harmonic superspace\footnote{The superfield formulation for $\mathcal{N} = 2$
	AdS supergravity in the conventional superspace was first given in \cite{Butter:2010jm}.
	The linearised superfield formulation for $\mathcal{N} = 2$ AdS supergravity was
	derived in \cite{Butter:2011ym}.}.
	We expect that such a formulation admits a straightforward generalization to $\mathcal{N}=2$ higher-spin theories, as in flat $4D$ space \cite{Buchbinder:2021ite}.



\acknowledgments

This work was partially supported by
the Foundation for the Advancement of Theoretical Physics and
Mathematics ``BASIS'', grant \verb|#| 25-1-1-10-4 .

\appendix

\section{Superconformal transformations in different bases}\label{App: sc conf}

In this article we used the analytic basis of ${\cal N}=2$ HSS as the preferred one. In particular, the superconformal transformations
were given in this basis, see eq. \eqref{eq:superconformal symmetry}. The goal of this Appendix is to establish
the precise relations between coordinates in diverse bases and find the appropriate coordinate realizations of $\mathcal{N}=2$ superconformal group.

We use the following  notations for coordinates in different bases:
\begin{equation}\label{eq: N=2 bases}
    \begin{split}
        \text{Analytic basis:}&\qquad
        \left\{x^{\alpha\dot{\alpha}}, \theta^{\pm}_\alpha, \bar{\theta}^\pm_{\dot{\alpha}}, u_i^\pm\right\},
        \\
        \text{Central basis:}&\qquad
    \left\{x_{C}^{\alpha\dot{\alpha}}, \theta^{i}_\alpha, \bar{\theta}^i_{\dot{\alpha}}, u^\pm_i \right\},
        \\
        \text{Chiral basis:}&
        \qquad
        \left\{ x_L^{\alpha\dot{\alpha}  }, \theta^{i}_\alpha, \bar{\theta}^i_{\dot{\alpha}},  u^\pm_i  \right\},
            \\
        \text{Anti-Chiral basis:}&
            \qquad
        \left\{ x_R^{\alpha\dot{\alpha}  }, \theta^{i}_\alpha, \bar{\theta}^i_{\dot{\alpha}} ,  u^\pm_i  \right\}.
    \end{split}
\end{equation}
The analytic coordinates are defined in terms of the central basis coordinates  as
\begin{equation}
    \begin{split}
    &x^{\alpha\dot{\alpha}} = x_C^{\alpha\dot{\alpha}} - 4i \theta^{\alpha(i} \bar{\theta}^{\dot{\alpha}j)} u^+_i u^-_j,
    \\
    &
    \theta_{\alpha}^\pm = \theta^i_\alpha u^\pm_i,
    \qquad
    \bar{\theta}^\pm_{\dot\alpha}
    =
    \bar{\theta}^i_{\dot\alpha} u^\pm_i.
    \end{split}
\end{equation}
The inverse expressions for the central basis coordinates in terms of the analytic ones read:
\begin{equation}
    \begin{split}
    &x_C^{\alpha\dot{\alpha}} = x^{\alpha\dot{\alpha}}
    +
    2i \theta^{+\alpha} \bar{\theta}^{-\dot{\alpha}}
    +
    2i \theta^{-\alpha} \bar{\theta}^{+\dot{\alpha}},
    \\
    &
    \theta_{\alpha i} =
    \theta^{-}_{\alpha} u^+_i
    -
    \theta^{+}_{\alpha} u^-_i,
    \qquad
    \bar{\theta}_{\dot\alpha i} =
    \bar{\theta}^{-}_{\dot\alpha} u^+_i
    -
    \bar{\theta}^{+}_{\dot\alpha} u^-_i.
    \end{split}
\end{equation}

The left and right chiral bases are defined in terms of the central basis coordinates as
\begin{equation}\label{eq: C --> LR}
    x^{\alpha\dot{\alpha}}_L = x^{\alpha\dot{\alpha}}_C + 2i \theta^{\alpha}_i \bar{\theta}^{\dot{\alpha} i},
    \qquad
    x^{\alpha\dot{\alpha}}_R = x^{\alpha\dot{\alpha}}_C  - 2i \theta^{\alpha}_i \bar{\theta}^{\dot{\alpha} i}
\end{equation}
or, in terms of analytic coordinates, as
\begin{equation}
    \begin{split}
    &x_L^{\alpha\dot{\alpha}}
    =
    x^{\alpha\dot{\alpha}}
    +
    4i \theta^{-\alpha} \bar{\theta}^{+\dot{\alpha}}
    ,
    \qquad
        x_R^{\alpha\dot{\alpha}}
        =
        x^{\alpha\dot{\alpha}}
        +
        4i \theta^{+\alpha} \bar{\theta}^{-\dot{\alpha}}.
    \end{split}
\end{equation}

Being aware of the realization of $\mathcal{N}=2$ superconformal group in the analytic superspace \eqref{eq:superconformal symmetry}
and of the transition formulas given above, one can find how this supergroup acts on the coordinates of the central basis\footnote{For completeness,
we present also the $\mathcal{N}=2$ supersymmetry transformations:
$$
\delta_{\epsilon} x_C^{\alpha\dot{\alpha}} = 2i
\left( \epsilon^{\alpha i} \bar{\theta}_i^{\dot{\alpha}}
-
\theta^{\alpha i} \bar{\epsilon}^{\dot{\alpha}}_i
\right),
\quad
\delta_\epsilon x^{\alpha\dot{\alpha}}_L = -4i \theta^{\alpha i} \bar{\epsilon}^{\dot{\alpha}}_i,
\quad
\delta_\epsilon x_R^{\alpha\dot{\alpha}}
=
4i \epsilon^{\alpha i} \bar{\theta}^{\dot{\alpha}}_i,
\quad
\delta_\epsilon \theta^i_\alpha=
\epsilon^i_\alpha,
\quad
\delta_\epsilon \bar{\theta}^i_{\dot{\alpha}} =
\bar{\epsilon}^i_{\dot{\alpha}}.
$$}:
\begin{equation}
    \begin{split}
        & \delta_{sc} x_C^{\alpha\dot{\alpha}}
        =\, {\rm d} x_C^{\alpha\dot{\alpha}}
        +
         \frac{1}{2}
         \left(   x_L^{\dot{\alpha}\beta} k_{\beta\dot{\beta}} x_L^{\dot{\beta}\alpha}
         +
         x_R^{\dot{\alpha}\beta} k_{\beta\dot{\beta}} x_R^{\dot{\beta}\alpha}
         \right)
            +
         2i x_L^{\beta\dot{\alpha}} \theta^\alpha_i \eta^i_\beta
         +
         2i x_R^{\alpha\dot{\beta}} \bar{\eta}^i_{\dot{\beta}} \bar{\theta}^{-\dot{\alpha}},
        \\
    &\delta_{sc} \theta^\alpha_i =
    \frac{1}{2} ({\rm d}+i{\rm r})  \theta^\alpha_i
    +
    x_L^{\alpha\dot{\beta}} k_{\beta\dot{\beta}} \theta^\beta_i
    -
    \lambda_i^{\;j} \theta^\alpha_{j}
    -
    4i (\theta^\beta_i \eta^j_\beta) \theta^\alpha_j
    +
    x_L^{\alpha\dot{\beta}} \bar{\eta}_{\dot{\beta}i} ,
    \\
    & \delta_{sc} \bar{\theta}_i^{\dot{\alpha}}
    =
        \frac{1}{2} ({\rm d}-i{\rm r})  \bar{\theta}^{\dot\alpha}_i
    +
    x_R^{\dot{\alpha}\beta} k_{\beta\dot{\beta}} \bar{\theta}^{\dot\beta}_i
    -
    \lambda_{i}^{\;j}\bar{\theta}^{\dot{\alpha}}_j
    -
    4i (\bar{\eta}^j_{\dot{\beta}} \bar{\theta}^{\dot{\beta}}_i) \bar{\theta}^{\dot{\alpha}}_j
    +
    x_R^{\dot{\alpha}\beta} \eta_{\beta i}.
    \end{split}
\end{equation}
Analogously, we find for chiral and anti-chiral bases:
\begin{subequations}
\begin{equation}
    \delta_{sc} x_{L}^{\alpha\dot{\alpha}}
    =
    {\rm d} x_{L}^{\alpha\dot{\alpha}}
    +
       x_L^{\dot{\alpha}\beta} k_{\beta\dot{\beta}} x_L^{\dot{\beta}\alpha}
      +
      4i x_L^{\beta\dot{\alpha}} \theta^\alpha_i \eta^i_\beta,
\end{equation}
\begin{equation}
    \delta_{sc} x_{R}^{\alpha\dot{\alpha}}
    =
    {\rm d}
    x_{R}^{\alpha\dot{\alpha}}
    +
       x_R^{\dot{\alpha}\beta} k_{\beta\dot{\beta}} x_R^{\dot{\beta}\alpha}
    +
    4i x_R^{\alpha\dot{\beta}} \bar{\eta}^i_{\dot{\beta}} \bar{\theta}^{-\dot{\alpha}}.
\end{equation}
\end{subequations}
These realizations agree with those known in the literature (see, e.g., \cite{Park:1999pd}).
From them it immediately follows that the chiral superspace $\{x_L^{\alpha\dot{\alpha}}, \theta^i_\alpha \}$ and
the anti-chiral one $\{x_R^{\alpha\dot{\alpha}}, \bar{\theta}^i_{\dot{\alpha}} \}$ are closed  under $\mathcal{N}=2$ superconformal symmetry.

Harmonic variables in the central and chiral bases have the same transformation rules:
\begin{eqnarray}\label{eq: harm trans central}
\delta_{sc} u^{+}_i
    =
    \left\{\lambda^{kl}
    +
    4i \theta^{\alpha k} \bar{\theta}^{\dot{\alpha}l} k_{\alpha\dot{\alpha}}
    +
    4i
    (
    \theta^{\beta k} \eta^{l}_\beta
    +
    \bar{\eta}^k_{\dot{\beta}} \bar{\theta}^{\dot{\beta}l}
    )
    \right\}
    u^+_k u^+_l u^-_i, \quad \delta_{sc} u^-_i = 0.
\end{eqnarray}
It is evident that the harmonic coordinates do not respect chirality; therefore, in order to preserve chirality,
one must consider harmonic-independent superfields,
i.e. impose the harmonic-independence conditions:
\begin{equation}\label{eq: HI conditions}
    \mathcal{D}^{\pm\pm} \Phi = 0,
    \qquad
    \mathcal{D}^0 \Phi = 0.
\end{equation}
In the central and chiral bases the covariant harmonic derivatives are simplified to the partial ones:
\begin{equation}
    \mathcal{D}^{\pm\pm}_{C} = \partial^{\pm\pm},
    \qquad
    \mathcal{D}^0_{C} = \partial_0
\end{equation}
and the conditions \eqref{eq: HI conditions} are solved by the harmonic-independent
superfields $\Phi(x^{\alpha\dot{\alpha}}_{C}, \theta^i_\alpha, \bar{\theta}^i_{\dot{\alpha}})$. Analogously, the chirality conditions \eqref{eq: ch cond}
are solved by $\Phi_L \left( x_L^{\alpha\dot{\alpha}}, \theta^i_\alpha \right)$ and the anti-chirality ones by $\Phi_R \left( x_R^{\alpha\dot{\alpha}}, \bar{\theta}^i_{\dot\alpha} \right)$.

\medskip

In conclusion of this Appendix, we present a list of conventional $\mathcal{N}=2$ superspaces admitting the realization of superconformal group:
\begin{itemize}
    \item $\mathbb{HR}^{4+2|8} = \left\{x^{\alpha\dot{\alpha}}, \theta^{\pm}_\alpha, \bar{\theta}^\pm_{\dot{\alpha}}, u^{\pm}_i \right\}$ --- $\mathcal{N}=2$ harmonic superspace;
    \item $\mathbb{HA}^{4+2|4} = \left\{x^{\alpha\dot{\alpha}}, \theta^{+}_\alpha, \bar{\theta}^+_{\dot{\alpha}}, u^{\pm}_i \right\}$ --- $\mathcal{N}=2$ harmonic  analytic superspace;
    \item $\mathbb{R}^{4|8} = \left\{x_C^{\alpha\dot{\alpha}}, \theta^{i}_\alpha, \bar{\theta}^i_{\dot{\alpha}} \right\}$ --- $\mathcal{N}=2$ real superspace;
    \item $\mathbb{C}^{4|4} = \left\{x_L^{\alpha\dot{\alpha}}, \theta^{i}_\alpha \right\}$ --- $\mathcal{N}=2$ chiral superspace;
    \item $\bar{\mathbb{C}}^{4|4} = \left\{x_R^{\alpha\dot{\alpha}},  \bar{\theta}^i_{\dot{\alpha}} \right\}$ --- $\mathcal{N}=2$ anti-chiral superspace.
\end{itemize}
In contrast to chiral superspaces, the real superspace $\mathbb{R}^{4|8}$ can be extended by harmonic coordinates endowed with the transformation law \eqref{eq: harm trans central}.

Somewhat unexpectedly, there exist additional essentially complex superspaces which contain 3/4 Grassmann coordinates together with harmonic variables
and are closed under the superconformal group:
\begin{itemize}
    \item $\mathbb{HHA}^{4+2|6}
    = \left\{x^{\alpha\dot{\alpha}}, \theta^{\pm}_\alpha, \bar{\theta}^+_{\dot{\alpha}}, u_i^{\pm} \right\} = \left\{(\zeta^M, u_i^\pm), \theta^-_\alpha\right\} $ --- $\mathcal{N}=2$
    anti-half-analytic harmonic superspace;
    \item $\overline{\mathbb{HHA}}^{\,4+2|6}
    = \left\{x^{\alpha\dot{\alpha}}, \theta^{+}_\alpha, \bar{\theta}^\pm_{\dot{\alpha}}, u^{\pm}_i \right\}= \left\{(\zeta^M, u_i^\pm), \bar\theta^-_{\dot\alpha}\right\}$ ---
    $\mathcal{N}=2$ half-analytic harmonic superspace.
\end{itemize}
Superfields living on these superspaces solve the half-analyticity condition ($\bar{\mathcal{D}}^+_{\dot{\alpha}} \Phi = 0$ see eq.$\eqref{eq: HA cond}$) or
the anti-half-analyticity condition ($\mathcal{D}^+_{\alpha} \Phi = 0$), respectively. These are  precisely those conditions which are satisfied by
the superfields $\mathcal{H}^{++}_{(\alpha\beta)}$ and its conjugate appearing in the expressions for the linearized $\mathcal{N}=2$ Weyl supertensor.
Looking at the coordinate realization of $\mathcal{N}=2$ superconformal group in the analytic basis \eqref{eq: sc equations}, we observe that $\mathbb{HHA}^{4+2|6}$ and
$\overline{\mathbb{HHA}}^{\,4+2|6}$ are closed under this realization. In particular, each of the extra coordinates $\theta^-_\alpha$ and $\bar\theta^-_{\dot\alpha}$ is transformed through itself
and the coordinates of the analytic subspace $\mathbb{HA}^{4+2|4}$. These half-analytic $\mathcal{N}=2$ harmonic superspaces admit the full-fledged harmonic dependence
and could play an important (though for the time being unclear) role in the geometry of $\mathcal{N}=2$ conformal supergravity and its higher-spin analogs.
It is straightforward to find the superconformal transformations of the integration measures in these new complex invariant subspaces and the relevant superconformal factors.

\medskip

Finally, it is worth to point out once more that all $\mathcal{N}=2$ superspaces listed above are proper subspaces of $\mathcal{N}=2$ harmonic superspace $\mathbb{HR}^{4+2|8}$. We also emphasize that the notion of half-analytic superspace remains relevant in the curved case as well, since it is possible to choose a hybrid basis in curved HSS.

\section{Linearized $\mathcal{N}=2$ Weyl supergravity in harmonic-independent gauge}\label{Appendix}

In this appendix, for completeness, we present a discussion of Mezincescu-type prepotential and harmonic-independent gauge.
In deriving the harmonic-independent prepotentials we closely follow refs. \cite{ Zupnik:1998td,Kuzenko:1999pi, Butter:2010sc} (see also \cite{Buchbinder:2025yef} for the analogous $5D$ construction). This procedure was first carried out for the $\mathcal{N}=2$ Weyl multiplet in the work \cite{Kuzenko:1999pi}.
From a conceptual point of view, the idea of expressing constrained superfield gauge parameters in terms of unconstrained ones can be traced back to the paper by Siegel and Gates \cite{Siegel:1978mj}.

\subsection{Mezincescu-type prepotential and harmonic-independent gauge}

\medskip

One can represent the covariant harmonic derivative \eqref{eq: mathfrak D} as
\begin{equation}
    \mathfrak{D}^{++} = \mathcal{D}^{++} + \hat{\mathcal{H}}^{++},
\end{equation}
where the superfield differential operator $\hat{\mathcal{H}}^{++}$ is given by:
\begin{equation}
    \begin{split}
\hat{\mathcal{H}}^{++} :
=&\;
h^{++\alpha\dot{\alpha}} \partial_{\alpha\dot{\alpha}}
+
h^{++\hat{\alpha}-} \partial^+_{\hat{\alpha}}
+
h^{++\hat{\alpha}+} \partial^-_{\hat{\alpha}}
+
h^{(+4)} \partial^{--}
\\
=& \; G^{++\alpha\dot{\alpha}} \partial_{\alpha\dot{\alpha}}
+
G^{++\hat{\alpha}-} \mathcal{D}^+_{\hat{\alpha}}
+
G^{++\hat{\alpha}+} \mathcal{D}^-_{\hat{\alpha}}
+
G^{(+4)} \mathcal{D}^{--}.
\end{split}
\end{equation}
The superfields appearing in this expansion are treated as defining the linearized $\mathcal{N}=2$ conformal supergravity in a flat super-background.

Let us consider the action of differential operator $   \hat{\mathcal{H}}^{++}$ on an arbitrary analytic superfield $\Phi(\zeta, u)$.
The result is evidently analytic and can be represented as
\begin{equation}\label{eq: on pre-pre}
    \hat{\mathcal{H}}^{++} \Phi(\zeta, u) = \left(\mathcal{D}^+ \right)^4  \Big\{ \mathcal{H} \, \mathcal{D}^{--}   \Phi(\zeta, u) \Big\},
\end{equation}
where $\mathcal{D}^{--}$ is the flat harmonic derivative.
In \eqref{eq: on pre-pre} we have introduced the Mezincescu-type gauge prepotential $\mathcal{H}$, which is a general harmonic superfield. The $G^{++}$
superfields can be expressed in terms of
$\mathcal{H}$ as follows:
\begin{equation}
    \begin{split}
        G^{++\alpha\dot{\alpha}} &= i \mathcal{D}^{+\alpha} \bar{\mathcal{D}}^{+\dot{\alpha}} \mathcal{H},
        \\
        G^{++\alpha+} &= -\frac{1}{8} \mathcal{D}^{+\alpha} \left(\bar{\mathcal{D}}^+ \right)^2 \mathcal{H},
        \\
        \bar{G}^{++\dot{\alpha}+} &= \frac{1}{8} \bar{\mathcal{D}}^{+\dot{\alpha}} (\mathcal{D}^+)^2  \mathcal{H},
        \\
        G^{(+4)} &= (\mathcal{D}^+)^4  \mathcal{H}.
    \end{split}
\end{equation}
The prepotential $\mathcal{H}$ is defined up to the pre-gauge freedom
\begin{equation}
    \delta_{pg} \mathcal{H} = (\mathcal{D}^+)^2 \Omega^{--}  +   (\bar{\mathcal{D}}^+)^2 \widetilde{\Omega}^{--}\,,
\end{equation}
where $\Omega^{--}$ is an arbitrary harmonic superfield and $\widetilde{\Omega}^{--}$ is its ``tilde''-conjugated.
The original linearized gauge freedom \eqref{eq: G++ GF} is realized as
\begin{equation}\label{eq: gf cb}
    \delta \mathcal{H} = \mathcal{D}^{++} l^{--}\,,
\end{equation}
where the parameter $l^{--}$  is an arbitrary harmonic superfield.  The relation between this parameter and the analytic gauge parameters $\lambda$ is
 determined by the equality
\begin{equation}
        \hat{\Lambda} \Phi(\zeta, u) = \left(\mathcal{D}^+ \right)^4  \Big\{ l^{--}\, \mathcal{D}^{--}   \Phi(\zeta, u) \Big\},
\end{equation}
that is analogous to \eqref{eq: on pre-pre}. From this relation we obtain:
\begin{equation}
    \begin{split}
        \Lambda^{\alpha\dot{\alpha}} &= i \mathcal{D}^{+\alpha} \bar{\mathcal{D}}^{+\dot{\alpha}} l^{--},
        \\
        \Lambda^{+\alpha} &= -\frac{1}{8} \mathcal{D}^{+\alpha} \left(\bar{\mathcal{D}}^+ \right)^2 l^{--},
        \\
        \bar{\Lambda}^{+\dot{\alpha}} &= \frac{1}{8} \bar{\mathcal{D}}^{+\dot{\alpha}} (\mathcal{D}^+)^2  l^{--},
        \\
        \Lambda^{++} &= (\mathcal{D}^+)^4  l^{--}.
    \end{split}
\end{equation}
Using the relations \eqref{eq: Lambda and lambda}, one can find the corresponding expressions for the analytic gauge parameters
$\left\{\lambda^{\alpha\dot{\alpha}}, \lambda^{+\alpha}, \bar{\lambda}^{+\dot{\alpha}}, \lambda^{++} \right\}$.

\smallskip

Taking into account the gauge freedom \eqref{eq: gf cb} in the central basis, we may impose the gauge:
\begin{equation}
    \mathcal{H}(x^{\alpha\dot{\alpha}}, \theta^\pm_{\hat{\alpha}},u^\pm_i) = H(x^{\alpha\dot{\alpha}}_C, \theta^i_{\hat{\alpha}} ) \qquad \text{(harmonic-independent gauge)}.
\end{equation}
In this gauge we have the residual gauge freedom
\begin{equation}\label{eq: gauge transformations Mez}
    \delta H \sim \mathcal{D}_{ij} \Omega^{ij} +  \bar{\mathcal{D}}_{ij} \bar{\Omega}^{ij} .
\end{equation}
The Mesincescu-type prepotential defined here can be considered as the spin 2 generalization of  the $V_{ij}$ prepotential for $\mathcal{N}=2$
vector multiplet \cite{Mezincescu:1979af} (see also discussion in \cite{18}). Such a prepotential for the description of
linearized $\mathcal{N}=2$ conformal supergravity multiplet was firstly introduced in \cite{Howe:1981qj, Rivelles:1981qz}.

It is important to note that, while such prepotentials are sufficient for describing linearized theories (albeit at cost of losing the geometric interpretation of gauge transformations), they lead to considerable difficulties
when extended to the nonlinear level, in both  $\mathcal{N}=2$ Yang-Mills theory and $\mathcal{N}=2$ supergravity. In particular, it is
far from obvious how to generalize the linearized transformation \eqref{eq: gauge transformations Mez} to the
non-linear case.  In the harmonic approach, by contrast, such a generalization is straightforward, see eqs. \eqref{eq: H transf}

\subsection{$\mathcal{N}=2$ Weyl supertensor in harmonic-independent gauge}\label{app: A2}

Throughout this section we deal solely with the central basis, so we drop the subscript ``$C$'' on $x$ -coordinates.
In the central basis the covariant spinor derivatives are written as:
\begin{equation}
    \mathcal{D}^i_\alpha = \frac{\partial}{\partial \theta^\alpha_i} + 2i \bar{\theta}^{\dot{\alpha}i} \partial_{\alpha\dot{\alpha}},
    \qquad
    \bar{\mathcal{D}}_{\dot{\alpha} i} = - \frac{\partial}{\partial \bar{\theta}^{\dot{\alpha}i}} - 2i \theta^\alpha_i \partial_{\alpha\dot{\alpha}}\,
\end{equation}
and satisfy the algebra
\begin{equation}
    \left\{\mathcal{D}^i_\alpha, \bar{\mathcal{D}}_{\dot{\alpha}j} \right\}
    =
    -
        \left\{\mathcal{D}_{\alpha j}, \bar{\mathcal{D}}^i_{\dot{\alpha}} \right\}
    =
     -4i \delta^i_j \partial_{\alpha\dot{\alpha}}\,.
\end{equation}
The spinor derivatives in the analytic basis are then defined as projections $\mathcal{D}^\pm_\alpha = \mathcal{D}^i_\alpha u^\pm_i$, etc.

In terms of the harmonic-independent Mezincescu-type prepotential, the half-analytic superfield $\mathcal{H}^{++}_{(\alpha\beta)}$ can be written as:
\begin{equation}
    \mathcal{H}^{++}_{(\alpha\beta)} = i \partial_{(\alpha}^{\dot{\beta}} \mathcal{D}^+_{\beta)} \bar{\mathcal{D}}^+_{\dot{\beta}} H+ \frac{1}{8} \mathcal{D}^-_{(\alpha} \mathcal{D}^+_{\beta)} (\bar{\mathcal{D}}^+)^2 H
     =
     \frac{1}{8}
     (\bar{\mathcal{D}}^+)^2 \mathcal{D}^-_{(\alpha} \mathcal{D}^+_{\beta)} H,
\end{equation}
where we have used the relation $\left[\mathcal{D}^-_\alpha, \left(\bar{\mathcal{D}}^+\right)^2 \right] = 8i \partial_\alpha^{\dot{\alpha}}\, \bar{\mathcal{D}}^+_{\dot{\alpha}}$.

\smallskip

In the central basis, the following identities for the derivatives are valid:
\begin{equation}
\mathcal{D}^-_{(\alpha} \mathcal{D}^+_{\beta)}
=
\mathcal{D}^i_{(\alpha} \mathcal{D}^j_{\beta)} u^-_i u^+_j
=
\frac{1}{2}\mathcal{D}^i_{(\alpha} \mathcal{D}^j_{\beta)} \left(  u^-_i u^+_j -  u^-_j u^+_i  \right)
 :=
  -\frac{1}{2} \mathcal{D}_{(\alpha\beta)}.
\end{equation}
Then we find:
\begin{equation}
    \mathcal{H}^{++}_{(\alpha\beta)} = - \frac{1}{16}  (\bar{\mathcal{D}}^+)^2 \mathcal{D}_{(\alpha\beta)} H
    \qquad
    \Rightarrow
    \qquad
        \mathcal{H}^{--}_{(\alpha\beta)} = - \frac{1}{16}  (\bar{\mathcal{D}}^-)^2 \mathcal{D}_{(\alpha\beta)} H,
\end{equation}
and, for $\mathcal{N}=2$ super Weyl tensor,
\begin{equation}
    \mathcal{W}_{(\alpha\beta)}
    =
     \left( \bar{\mathcal{D}}^+ \right)^2   \mathcal{H}^{--}_{(\alpha\beta)}
    =
     - \frac{1}{16}   \left( \bar{\mathcal{D}}^+ \right)^2  (\bar{\mathcal{D}}^-)^2 \mathcal{D}_{(\alpha\beta)} H
     =
     - (\bar{\mathcal{D}})^4 \mathcal{D}_{(\alpha\beta)} H.
\end{equation}
Thereby,  we reproduced the definition of the $\mathcal{N}=2$ Weyl supertensor in the conventional ${\cal N}=2$ superspace
(see, e.g., \cite{Howe:1981qj, Kuzenko:2021pqm,  Butter:2010sc}).

Finally, note that in the harmonic-independent gauge we have:
\begin{equation}
    \mathcal{D}^{(\alpha\beta)} \mathcal{W}_{(\alpha\beta)} =   \bar{\mathcal{D}}^{(\dot{\alpha}\dot{\beta})} \overline{\mathcal{W}}_{(\dot{\alpha}\dot{\beta})}.
\end{equation}
Since this equality is valid for the gauge-invariant objects, we conclude that
\begin{equation}
    (\mathcal{D}^{+\alpha} \mathcal{D}^{-\beta}) \mathcal{W}_{(\alpha\beta)} =
  (  \bar{\mathcal{D}}^{+\dot{\alpha}} \bar{\mathcal{D}}^{-\dot{\beta}} ) \overline{\mathcal{W}}_{(\dot{\alpha}\dot{\beta})},
\end{equation}
regardless of the gauge-fixing.

\section{Rigid superconformal transformations}\label{eq: App C}

In this Appendix, we present the transformation laws for various derivatives and deduce useful relations involving the $\Lambda_{sc}$ parameters. We also derive rigid $\mathcal{N}=2$ superconformal transformations of the
$G^{++}$ prepotentials, proceeding  from eq. \eqref{eq: G trans full}.

Let us consider the superconformal  operator in the covariant basis:
\begin{equation}
    \hat{\Lambda}_{sc}
    =
    \Lambda_{sc}^{\alpha\dot{\alpha}} \partial_{\alpha\dot{\alpha}}
    +
    \Lambda_{sc}^{+\hat{\alpha}} \mathcal{D}^-_{\hat{\alpha}}
    +
    \Lambda_{sc}^{-\hat{\alpha}} \mathcal{D}^+_{\hat{\alpha}}
    +
    \Lambda^{++}_{sc} \mathcal{D}^{--}.
\end{equation}
Superconformal $\Lambda$-superparameters are expressed in terms of analytic ones \eqref{eq:superconformal symmetry} according to eq. \eqref{eq: Lambda and lambda}:
\begin{equation}\label{eq: Lambda and lambda1}
    \begin{split}
        &
        \Lambda_{sc}^{\alpha\dot{\alpha}} \,= \lambda_{sc}^{\alpha\dot{\alpha}}
        +
        4i \lambda_{sc}^{+\alpha} \bar{\theta}^{-\dot{\alpha}}
        +
        4i \theta^{-\alpha} \bar{\lambda}_{sc}^{+\dot{\alpha}}
        -
        4i \lambda_{sc}^{++} \theta^{-\alpha} \bar{\theta}^{-\dot{\alpha}},
        \\
        &
        \Lambda_{sc}^{-\hat{\alpha}}
        =
        \lambda_{sc}^{-\hat{\alpha}},
        \\
        &
        \Lambda_{sc}^{+\hat{\alpha}} = - \lambda_{sc}^{+\hat{\alpha}} + \lambda_{sc}^{++} \theta^{-\hat{\alpha}},
        \\
        &
        \Lambda_{sc}^{++} = \lambda^{++}_{sc}
    \end{split}
\end{equation}
and therefore satisfy the constraints \eqref{eq: Lambda constr}.

Below, we derive some relations satisfied by the parameters  $\Lambda_{sc}$ and write down the superconformal transformation laws for the covariant derivatives
and $G^{++}$ prepotentials in terms of these parameters.

\subsubsection*{Harmonic derivatives}

The condition \eqref{eq: N=2 sc},
\begin{equation}\label{eq: N=2 sc 1}
    \delta_{sc}\mathcal{D}^{++} := [ \hat{\Lambda}_{sc}, \mathcal{D}^{++}] = -  \Lambda^{++}_{sc} \mathcal{D}^0\,,
\end{equation}
is equivalent to a system of equations for the $\Lambda_{sc}$ parameters:
\begin{equation}
    \begin{cases}
    \mathcal{D}^{++}    \Lambda_{sc}^{\alpha\dot{\alpha}} = 0,
    \\
    \mathcal{D}^{++}    \Lambda_{sc}^{+\hat{\alpha}} = 0,
    \\
        \mathcal{D}^{++}    \Lambda_{sc}^{-\hat{\alpha}}
        +
        \Lambda_{sc}^{+\hat{\alpha}}
        =
        0,
    \\
    \mathcal{D}^{++} \Lambda_{sc}^{++} = 0.
    \end{cases}
\end{equation}

Likewise, the transformation law for the negatively charged harmonic derivative,
\begin{equation}
      \delta_{sc} \mathcal{D}^{--} := [\hat{\Lambda}_{sc}, \mathcal{D}^{--}]  =  - (\mathcal{D}^{--}\Lambda_{sc}^{++}) \mathcal{D}^{--}\,,
\end{equation}
amounts to the set of equations:
\begin{equation}\label{eq: d-- rel}
    \begin{cases}
        \mathcal{D}^{--}    \Lambda_{sc}^{\alpha\dot{\alpha}} = 0,
        \\
        \mathcal{D}^{--}    \Lambda_{sc}^{+\hat{\alpha}}
        +
        \Lambda_{sc}^{-\hat{\alpha}}  = 0,
        \\
        \mathcal{D}^{--}    \Lambda_{sc}^{-\hat{\alpha}}
        =
        0.
    \end{cases}
\end{equation}

From these relations some important identities for $\Lambda_{sc}^{\alpha\dot{\alpha}}$ follow\footnote{It is also worth noting the relation $\partial_{(\alpha}^{\dot{\beta}}\Lambda^{sc}_{\beta) \dot{\beta}} = -2\left(\mathcal{D}^-_{(\alpha} \Lambda^{+ sc}_{\beta)}\right)$.}:
\begin{subequations}
\begin{equation}\label{eq: D-beta}
\mathcal{D}^-_{\beta} \Lambda_{sc}^{\alpha\dot{\alpha}} = -4i \delta_\beta^\alpha \mathcal{D}^{--} \bar{\Lambda}_{sc}^{+\dot{\alpha}}
=
+4i \delta_\beta^\alpha \bar{\Lambda}_{sc}^{-\dot{\alpha}},
\end{equation}
\begin{equation}\label{eq: bar D- dot beta}
    \bar{\mathcal{D}}^-_{\dot{\beta}} \Lambda_{sc}^{\alpha\dot{\alpha}}
    =
+   4i \delta^{\dot{\alpha}}_{\dot{\beta}} \mathcal{D}^{--} \Lambda_{sc}^{+\alpha}
=
-
4i \delta^{\dot{\alpha}}_{\dot{\beta}} \Lambda_{sc}^{-\alpha},
\end{equation}
\begin{equation}\label{eq: partial alpha dot alpha}
    \partial_{\beta\dot{\beta}} \Lambda_{sc}^{\alpha\dot{\alpha}}
    =
    \delta^\alpha_\beta \left( \bar{\mathcal{D}}^-_{\dot{\beta}}  \bar{\Lambda}_{sc}^{+\dot{\alpha}}\right)
    +
    \delta^{\dot{\alpha}}_{\dot{\beta}} \left( \mathcal{D}^-_\beta \Lambda_{sc}^{+\alpha} \right) - \delta^\alpha_\beta \delta^{\dot{\alpha}}_{\dot{\beta}} \left( \mathcal{D}^{--} \Lambda_{sc}^{++}\right).
\end{equation}
\end{subequations}
These identities entail some interesting corollaries:

\medskip

\noindent$\bullet$ Action of $\mathcal{D}^+_\rho$ on eq. \eqref{eq: D-beta} implies:
\begin{equation}
    \delta^\alpha_\rho \mathcal{D}^-_\beta \bar{\Lambda}^{+\dot{\alpha}}_{sc} = \delta_\beta^\alpha \mathcal{D}^+_\rho \bar{\Lambda}^{-\dot{\alpha}}_{sc}
    \quad
    \Leftrightarrow
    \quad
    \begin{cases}
    \mathcal{D}^-_\beta \bar{\Lambda}^{+\dot{\alpha}}_{sc} = 0,
\\
    \mathcal{D}^+_\rho \bar{\Lambda}^{-\dot{\alpha}}_{sc} = 0,
\end{cases}
        \quad
    \overset{\eqref{eq: d-- rel}}{\Rightarrow}
    \quad
    \begin{cases}
        \mathcal{D}^\pm_\beta \bar{\Lambda}^{+\dot{\alpha}}_{sc} = 0,
    \\
    \mathcal{D}^\pm_\rho \bar{\Lambda}^{-\dot{\alpha}}_{sc} = 0.
    \end{cases}
\end{equation}
$\bullet$ Analogously, action of $\bar{\mathcal{D}}^+_{\dot{\rho}}$ on eq. \eqref{eq: bar D- dot beta} implies:
\begin{equation}
    \delta^{\dot{\alpha}}_{\dot{\rho}} \bar{\mathcal{D}}^-_{\dot{\beta}} \Lambda^{+\alpha}_{sc} = \delta_{\dot{\beta}}^{\dot{\alpha}} \bar{\mathcal{D}}^+_{\dot{\rho}} \Lambda^{-\alpha}_{sc}
    \quad
    \Leftrightarrow
    \quad
    \begin{cases}
        \bar{\mathcal{D}}^-_{\dot{\beta}} \Lambda^{+\alpha}_{sc} = 0,
        \\
        \bar{\mathcal{D}}^+_{\dot{\rho}} \Lambda^{-\alpha}_{sc} = 0,
    \end{cases}
    \quad
    \overset{\eqref{eq: d-- rel}}{\Rightarrow}
    \quad
    \begin{cases}
    \bar{\mathcal{D}}^\pm_{\dot{\beta}} \Lambda^{+\alpha}_{sc} = 0,
    \\
    \bar{\mathcal{D}}^\pm_{\dot{\rho}} \Lambda^{-\alpha}_{sc} = 0.
\end{cases}
\end{equation}
$\bullet$ Action of $\mathcal{D}^+_\alpha$ and $\bar{\mathcal{D}}^+_{\dot{\alpha}}$ on eq. \eqref{eq: partial alpha dot alpha} leads to:
\begin{subequations}
\begin{equation}
    4i\partial_{\beta\dot{\beta}} \Lambda^{+\alpha}_{sc} =- \delta_\beta^\alpha \bar{\mathcal{D}}^-_{\dot{\beta}} \Lambda^{++}_{sc}
    \qquad
    \overset{\eqref{eq: d-- rel}}{\Rightarrow}
    \qquad
    4i\partial_{\beta\dot{\beta}} \Lambda^{-\alpha}_{sc} =+ \delta_\beta^\alpha \bar{\mathcal{D}}^-_{\dot{\beta}} \mathcal{D}^{--}\Lambda^{++}_{sc},
\end{equation}
\begin{equation}\label{eq: vect bar Lambda}
    4i \partial_{\beta\dot{\beta}} \bar{\Lambda}^{+\dot{\alpha}}_{sc} = + \delta^{\dot{\alpha}}_{\dot{\beta}} \mathcal{D}^-_{\beta} \Lambda^{++}_{sc}
        \qquad
    \overset{\eqref{eq: d-- rel}}{\Rightarrow}
    \qquad
    4i \partial_{\beta\dot{\beta}} \bar{\Lambda}^{-\dot{\alpha}}_{sc} = - \delta^{\dot{\alpha}}_{\dot{\beta}} \mathcal{D}^-_{\beta} \mathcal{D}^{--}\Lambda^{++}_{sc}
    .
\end{equation}
\end{subequations}
Additional useful relations follow from  \eqref{eq: Lambda constr} and \eqref{eq: d-- rel}:
\begin{subequations}\label{eq: Lor factors}
\begin{equation}
    \mathcal{D}^-_\beta \Lambda^{+\alpha}
    -
    \mathcal{D}^+_\beta \Lambda^{-\alpha}
    =
    \delta_\beta^\alpha \left(\mathcal{D}^{--} \Lambda^{++}\right),
\end{equation}
\begin{equation}
    \bar{\mathcal{D}}^-_{\dot{\beta}} \bar{\Lambda}^{+\dot{\alpha}}
    -
    \bar{\mathcal{D}}^+_{\dot{\beta}} \bar{\Lambda}^{-\dot{\alpha}}
    =
    \delta_{\dot{\beta}}^{\dot{\alpha}} \left(\mathcal{D}^{--} \Lambda^{++}\right).
\end{equation}
\end{subequations}

\subsubsection*{Superconformal transformation laws of $G^{++}$-superfields}

Using eq. \eqref{eq: G trans full} for the general transformation law of $G^{++}$ superfields, one can deduce their rigid superconformal transformations in the analytic gauge:
\begin{equation}
    \begin{split}
    &\delta_{sc} G^{++\alpha\dot{\alpha}} = \hat{G}^{++} \Lambda^{\alpha\dot{\alpha}}_{sc}
    -
    4i \left( G^{++\alpha+} \Lambda_{sc}^{-\dot{\alpha}}
    +
    \Lambda_{sc}^{-\alpha} G^{++\dot{\alpha}+} \right),
    \\
    &
    \delta_{sc} G^{++\hat{\alpha}+} = \hat{G}^{++} \Lambda_{sc}^{+\hat{\alpha}}
    +
    G^{(+4)} \Lambda_{sc}^{-\hat{\alpha}},
    \\
    & \delta_{sc}G^{(+4)} = \hat{G}^{++} \Lambda_{sc}^{++}.
    \end{split}
\end{equation}
Employing various properties of $\Lambda_{sc}$-parameters, we can derive the relations:
\begin{subequations}
\begin{equation}
    \hat{G}^{++} \Lambda^{\alpha\dot{\alpha}}_{sc}
    =
    G^{++\beta\dot{\beta}} \partial_{\beta\dot{\beta}} \Lambda^{\alpha\dot{\alpha}}_{sc}
    +
    G^{++\beta+}
    \underbrace{\mathcal{D}^-_{\beta} \Lambda^{\alpha\dot{\alpha}}_{sc}}_{+4i \delta_\beta^\alpha \Lambda_{sc}^{-\dot{\alpha}} }
    +
    G^{++\dot{\beta}+} \underbrace{\bar{\mathcal{D}}^-_{\dot{\beta}} \Lambda^{\alpha\dot{\alpha}}_{sc}
    }_{-
        4i \delta^{\dot{\alpha}}_{\dot{\beta}} \Lambda_{sc}^{-\alpha}}  +
    G^{(+4)} \underbrace{\mathcal{D}^{--} \Lambda^{\alpha\dot{\alpha}}_{sc}}_{0},
\end{equation}
\begin{equation}
    \hat{G}^{++} \Lambda^{+\alpha}_{sc}
    =
    G^{++\beta\dot{\beta}} \partial_{\beta\dot{\beta}} \Lambda^{+\alpha}_{sc}
    +
    G^{++\beta+}
    \mathcal{D}^-_{\beta} \Lambda^{+\alpha}_{sc}
    +
    G^{++\dot{\beta}+} \underbrace{\bar{\mathcal{D}}^-_{\dot{\beta}} \Lambda^{+\alpha}_{sc} }_{0}
    +
    G^{(+4)} \underbrace{\mathcal{D}^{--} \Lambda^{+\alpha}_{sc}}_{-\Lambda_{sc}^{-\alpha}},
\end{equation}
\begin{equation}
    \hat{G}^{++} \Lambda^{++}_{sc}
    =
        G^{++\beta\dot{\beta}} \underbrace{\partial_{\beta\dot{\beta}} \Lambda^{++}_{sc} }_{0}
    +
    G^{++\beta+}
    \mathcal{D}^-_{\beta} \Lambda^{++}_{sc}
    +
    G^{++\dot{\beta}+} \bar{\mathcal{D}}^-_{\dot{\beta}} \Lambda^{++}_{sc}
    +
    G^{(+4)} \mathcal{D}^{--} \Lambda^{++}_{sc}.
\end{equation}
\end{subequations}
Exploiting these relations, we can find $\mathcal{N}=2$ superconformal transformations of $G^{++}$ superfields in different (though equivalent) forms:
\begin{subequations}\label{eq: G++ trans in Lambda}
\begin{equation}
    \begin{split}
    \delta_{sc} G^{++\alpha\dot{\alpha}} =& \; G^{++\beta\dot{\beta}} \partial_{\beta\dot{\beta}} \Lambda^{\alpha\dot{\alpha}}_{sc}
\\ \overset{\eqref{eq: partial alpha dot alpha}}{=}\;&
G^{++\alpha\dot{\beta}} \left( \bar{\mathcal{D}}^-_{\dot{\beta}} \bar{\Lambda}_{sc}^{+\dot{\alpha}} \right)
+
G^{++\beta\dot{\alpha}} \left( \mathcal{D}^-_\beta \Lambda_{sc}^{+\alpha} \right)
-
\left( \mathcal{D}^{--} \Lambda_{sc}^{++}\right)  G^{++\alpha\dot{\alpha}}
\\ \quad \overset{\eqref{eq: Lor factors}}{=}&\;
G^{++\alpha\dot{\beta}} \left( \bar{\mathcal{D}}^+_{\dot{\beta}} \bar{\Lambda}_{sc}^{-\dot{\alpha}} \right)
+
G^{++\beta\dot{\alpha}} \left( \mathcal{D}^+_\beta \Lambda_{sc}^{-\alpha} \right)
+
\left( \mathcal{D}^{--} \Lambda_{sc}^{++}\right)  G^{++\alpha\dot{\alpha}},
    \end{split}
\end{equation}
\begin{equation}
    \begin{split}
    \delta_{sc} G^{++\alpha+} = &\; G^{++\beta\dot{\beta}} \left( \partial_{\beta\dot{\beta}} \Lambda^{+\alpha}_{sc} \right)
    +
    G^{++\beta+}
\left( \mathcal{D}^-_{\beta} \Lambda^{+\alpha}_{sc} \right)
    \\
    =& \;\frac{i}{4}  G^{++\alpha\dot{\beta}} \left( \bar{\mathcal{D}}^-_{\dot{\beta}} \Lambda^{++}_{sc} \right)
    +
    G^{++\beta+}
    \left( \mathcal{D}^-_{\beta} \Lambda^{+\alpha}_{sc}\right),
    \\
    =&\;
    \frac{i}{4}  G^{++\alpha\dot{\beta}} \left( \bar{\mathcal{D}}^-_{\dot{\beta}} \Lambda^{++}_{sc} \right)
    +
    G^{++\beta+}
    \left( \mathcal{D}^+_{\beta} \Lambda^{-\alpha}_{sc}\right)
    +
    (\mathcal{D}^{--}\Lambda^{++}_{sc}) G^{++\alpha+}
    ,
        \end{split}
\end{equation}
\begin{equation}
    \delta_{sc} G^{(+4)} = G^{++\hat{\beta}+}
\left(  \mathcal{D}^-_{\hat{\beta}} \Lambda^{++}_{sc} \right)
    +
    G^{(+4)}
    \left( \mathcal{D}^{--} \Lambda^{++}_{sc}\right).
\end{equation}
\end{subequations}




\begin{thebibliography}{99}

\bibitem{Galperin:1984av}
A.~Galperin, E.~Ivanov, S.~Kalitzin, V.~Ogievetsky and E.~Sokatchev,
{\it Unconstrained $\mathcal{N}=2$ Matter, Yang-Mills and Supergravity Theories in Harmonic Superspace},
\href{https://doi.org/10.1088/0264-9381/1/5/004}{Class. Quant. Grav. \textbf{1} (1984), 469-498}
[erratum: \href{https://doi.org/10.1088/0264-9381/2/1/512}{Class. Quant. Grav. \textbf{2} (1985), 127}].

\bibitem{18} A.~S.~Galperin, E.~A.~Ivanov, V.~I.~Ogievetsky, E.~S.~Sokatchev,
{\it Harmonic superspace}, Cambridge Monographs on Mathematical
Physics, Cambridge University Press, 2001, 306 p.

\bibitem{Galperin:1985zv}
A.~Galperin, E.~Ivanov, V.~Ogievetsky and E.~Sokatchev,
{\it Conformal invariance in harmonic superspace},
Dubna preprint JINR E2-85-363, (1985), published in "Quantum Field
theory and Quantum Statistics", eds. I. Batalin, C. J. Isham, G. Vilkovisky, vol.2,
Adam Hilger, Bristol, 1987, p. 233–248.

\bibitem{Eden:2000qp}
B.~U.~Eden, P.~S.~Howe, A.~Pickering, E.~Sokatchev and P.~C.~West,
{\it Four point functions in $\mathcal{N}=2$ superconformal field theories},
\href{https://doi.org/10.1016/S0550-3213(00)00218-2}{Nucl. Phys. B \textbf{581} (2000), 523-558}
[arXiv:hep-th/0001138].

\bibitem{Ivanov:2005qf}
E.~A.~Ivanov, A.~V.~Smilga and B.~M.~Zupnik,
{\it Renormalizable supersymmetric gauge theory in six dimensions},
\href{https://doi.org/10.1016/j.nuclphysb.2005.08.014}{Nucl. Phys. B \textbf{726} (2005), 131-148}
[arXiv:hep-th/0505082].

\bibitem{Buchbinder:2024pjm}
I.~Buchbinder, E.~Ivanov and N.~Zaigraev,
{\it $\mathcal{N}=2$ superconformal higher-spin multiplets and their hypermultiplet couplings},
\href{https://doi.org/10.1007/JHEP08(2024)120}{JHEP \textbf{08} (2024), 120}
[arXiv:2404.19016 [hep-th]].



\bibitem{Buchbinder:2021ite}
I.~Buchbinder, E.~Ivanov and N.~Zaigraev,
{\it Unconstrained off-shell superfield formulation of $4D,  \mathcal{N}=2$ supersymmetric higher spins},
\href{https://doi.org/10.1007/JHEP12(2021)016}{JHEP \textbf{12} (2021), 016}
[arXiv:2109.07639 [hep-th]].

\bibitem{Buchbinder:2022kzl}
I.~Buchbinder, E.~Ivanov and N.~Zaigraev,
{\it Off-shell cubic hypermultiplet couplings to $\mathcal{N}=2$ higher spin gauge superfields},
\href{https://doi.org/10.1007/JHEP05(2022)104}{JHEP \textbf{05} (2022), 104}
[arXiv:2202.08196 [hep-th]].


\bibitem{Buchbinder:2022vra}
I.~Buchbinder, E.~Ivanov and N.~Zaigraev,
{\it $\mathcal{N}=2$ higher spins: superfield equations of motion, the hypermultiplet supercurrents, and the component structure},
\href{https://doi.org/10.1007/JHEP03(2023)036}{JHEP \textbf{03} (2023), 036}
[arXiv:2212.14114 [hep-th]].

\bibitem{Buchbinder:2025ceg}
I.~Buchbinder, E.~Ivanov and N.~Zaigraev,
{\it Towards $\mathcal{N}=2$ higher-spin supergravity},
[arXiv:2503.02438 [hep-th]].


\bibitem{Galperin:1987ek}
A.~S.~Galperin, E.~A.~Ivanov, V.~I.~Ogievetsky and E.~Sokatchev,
{\it $\mathcal{N}=2$ Supergravity in Superspace: Different Versions and Matter Couplings},
\href{https://doi.org/10.1088/0264-9381/4/5/023}{Class. Quant. Grav. \textbf{4} (1987), 1255}.

\bibitem{Kuzenko:2021pqm}
S.~M.~Kuzenko and E.~S.~N.~Raptakis,
{\it Extended superconformal higher-spin gauge theories in four dimensions},
\href{https://doi.org/10.1007/JHEP12(2021)210}{JHEP \textbf{12} (2021), 210}
[arXiv:2104.10416 [hep-th]].

\bibitem{Hutchings:2023iza}
D.~Hutchings, S.~M.~Kuzenko and E.~S.~N.~Raptakis,
{\it The $\mathcal{N}=2$ superconformal gravitino multiplet},
\href{https://doi.org/10.1016/j.physletb.2023.138132}{Phys. Lett. B \textbf{845} (2023), 138132}
[arXiv:2305.16029 [hep-th]].

\bibitem{Ivanov:2024bsb}
E.~Ivanov and N.~Zaigraev,
{\it $\mathcal{N}=2$ superconformal gravitino in harmonic superspace},
\href{https://doi.org/10.1016/j.physletb.2025.139333}{Phys. Lett. B \textbf{862} (2025), 139333}
[arXiv:2412.14822 [hep-th]].

\bibitem{Galperin:1987em}
A.~S.~Galperin, N.~A.~Ky and E.~Sokatchev,
{\it $\mathcal{N}=2$ Supergravity in Superspace: Solution to the Constraints},
\href{https://doi.org/10.1088/0264-9381/4/5/022}{Class. Quant. Grav. \textbf{4} (1987), 1235}.

\bibitem{Sokatchev:1988aa}
E.~Sokatchev,
{\it Off-shell Six-dimensional Supergravity in Harmonic Superspace},
\href{https://doi.org/10.1088/0264-9381/5/11/009}{Class. Quant. Grav. \textbf{5} (1988), 1459-1471}.

\bibitem{Ivanov:2022vwc}
E.~Ivanov,
{\it $\mathcal{N}=2$ Supergravities in Harmonic Superspace}, in: Bambi, C., Modesto, L., Shapiro, I. (eds) \href{https://doi.org/10.1007/978-981-19-3079-9\_43-1}{Handbook of Quantum Gravity, Springer, 2024}
[arXiv:2212.07925 [hep-th]].


\bibitem{Kuzenko:2024vms}
S.~M.~Kuzenko and E.~S.~N.~Raptakis,
{\it Towards $ \mathcal{N}=2$ superconformal higher-spin theory},
\href{https://doi.org/10.1007/JHEP11(2024)013}{JHEP \textbf{11} (2024), 013}
[arXiv:2407.21573 [hep-th]].


\bibitem{Butter:2013lta}
D.~Butter, B.~de Wit, S.~M.~Kuzenko and I.~Lodato,
{\it New higher-derivative invariants in $\mathcal{N}=2$ supergravity and the Gauss-Bonnet term},
\href{https://doi.org/10.1007/JHEP12(2013)062}{JHEP \textbf{12} (2013), 062}
[arXiv:1307.6546 [hep-th]].

\bibitem{Butter:2015nza}
D.~Butter,
{\it On conformal supergravity and harmonic superspace},
\href{https://doi.org/10.1007/JHEP03(2016)107}{JHEP \textbf{03} (2016), 107}
[arXiv:1508.07718 [hep-th]].

\bibitem{Butter:2016qkx}
D.~Butter, S.~M.~Kuzenko, J.~Novak and S.~Theisen,
{\it Invariants for minimal conformal supergravity in six dimensions},
\href{https://doi.org/10.1007/JHEP12(2016)072}{JHEP \textbf{12} (2016), 072}
[arXiv:1606.02921 [hep-th]].



\bibitem{Ivanov:2024gjo}
E.~Ivanov and N.~Zaigraev,
{\it Off-shell invariants of linearized $4D,\mathcal{N}=2$ supergravity in the harmonic approach},
\href{https://doi.org/10.1103/PhysRevD.110.066020}{Phys. Rev. D \textbf{110} (2024) no.6, 066020}
[arXiv:2407.08524 [hep-th]].

\bibitem{Ivanov:2025mld}
E.~A.~Ivanov and N.~M.~Zaigraev,
{\it $\mathcal{N} = 2$ Supergravity and Harmonic Superspace: Linearized Supercurvatures},
\href{https://doi.org/10.1134/S106377962470148X}{Phys. Part. Nucl. \textbf{56} (2025) no.2, 247-252}.



\bibitem{Howe:1981qj}
P.~S.~Howe, K.~S.~Stelle and P.~K.~Townsend,
{\it Supercurrents},
\href{https://doi.org/10.1016/0550-3213(81)90429-6}{Nucl. Phys. B \textbf{192} (1981), 332-352}.

\bibitem{Gates:1981qq}
S.~J.~Gates, Jr. and W.~Siegel,
{\it Linearized $\mathcal{N}=2$ superfield supergravity,},
\href{https://doi.org/10.1016/0550-3213(82)90047-5}{Nucl. Phys. B \textbf{195} (1982), 39-60}.

\bibitem{Howe:1981gz}
P.~S.~Howe,
{\it Supergravity in Superspace},
\href{https://doi.org/10.1016/0550-3213(82)90349-2}{Nucl. Phys. B \textbf{199} (1982), 309-364}.




\bibitem{Fradkin:1985am}
E.~S.~Fradkin and A.~A.~Tseytlin,
{\it Conformal supergravity},
\href{https://doi.org/10.1016/0370-1573(85)90138-3}{Phys. Rept. \textbf{119} (1985), 233-362}.

\bibitem{Muller:1989uhj}
M.~M\"uller,
{\it Consistent Classical Supergravity Theories},
\href{https://doi.org/10.1007/3-540-51427-9}{
Lect.Notes Phys. 336 (1989)}.

\bibitem{Kuzenko:2008ep}
S.~M.~Kuzenko, U.~Lindstrom, M.~Rocek and G.~Tartaglino-Mazzucchelli,
{\it $4D$ $\mathcal{N} = 2$ Supergravity and Projective Superspace},
\href{https://doi.org/10.1088/1126-6708/2008/09/051}{JHEP \textbf{09} (2008), 051}
[arXiv:0805.4683 [hep-th]].





\bibitem{Butter:2011sr}
D.~Butter,
{\it $\mathcal{N}=2$ Conformal Superspace in Four Dimensions},
\href{https://doi.org/10.1007/JHEP10(2011)030}{JHEP \textbf{10} (2011), 030}
[arXiv:1103.5914 [hep-th]].




\bibitem{Ferber:1977qx}
A.~Ferber,
{\it Supertwistors and Conformal Supersymmetry},
\href{https://doi.org/10.1016/0550-3213(78)90257-2}{Nucl. Phys. B \textbf{132} (1978), 55-64}.



\bibitem{Zupnik:1998td}
B.~M.~Zupnik,
{\it Background harmonic superfields in $\mathcal{N}=2$ supergravity},
\href{https://doi.org/10.1007/BF02557138}{Theor. Math. Phys. \textbf{116} (1998), 964-977}
[arXiv:hep-th/9803202 [hep-th]].

\bibitem{Kuzenko:1999pi}
S.~M.~Kuzenko and S.~Theisen,
{\it Correlation functions of conserved currents in $\mathcal{N}=2$ superconformal theory},
\href{https://doi.org/10.1088/0264-9381/17/3/307}{Class. Quant. Grav. \textbf{17} (2000), 665-696}
[arXiv:hep-th/9907107].

\bibitem{Galperin:1985bj}
A.~Galperin, E.~A.~Ivanov, V.~Ogievetsky and E.~Sokatchev,
{\it Harmonic Supergraphs. Green Functions},
\href{https://doi.org/10.1088/0264-9381/2/5/004}{Class. Quant. Grav. \textbf{2} (1985), 601}.



\bibitem{Zaigraev:2024ryg}
N.~Zaigraev,
{\it $\mathcal{N}=2$ higher-spin supercurrents},
\href{https://doi.org/10.1016/j.physletb.2024.139056}{Phys. Lett. B \textbf{858} (2024), 139056}
[arXiv:2408.00668 [hep-th]].

\bibitem{Zupnik:1987vm}
B.~M.~Zupnik,
{\it The Action of the Supersymmetric $\mathcal{N}=2$ Gauge Theory in Harmonic Superspace},
\href{https://doi.org/10.1016/0370-2693(87)90433-3}{Phys. Lett. B \textbf{183} (1987), 175-176}.

\bibitem{Berends:1984wp}
F.~A.~Berends, G.~J.~H.~Burgers and H.~Van Dam,
{\it  On Spin Three Self Interactions},
\href{https://doi.org/10.1007/BF01410362}{Z. Phys. C \textbf{24} (1984), 247-254}.

\bibitem{Fradkin:1987ks}
E.~S.~Fradkin and M.~A.~Vasiliev,
{\it On the Gravitational Interaction of Massless Higher Spin Fields},
\href{https://doi.org/10.1016/0370-2693(87)91275-5}{Phys. Lett. B \textbf{189} (1987), 89-95}.

\bibitem{Zinoviev:2008ck}
Y.~M.~Zinoviev,
{\cal On spin 3 interacting with gravity},
\href{https://doi.org/10.1088/0264-9381/26/3/035022}{Class. Quant. Grav. \textbf{26} (2009), 035022}
[arXiv:0805.2226 [hep-th]].



\bibitem{Ivanov:2025jdp}
E.~Ivanov and N.~Zaigraev,
{\it $\mathcal{N}=2$ AdS hypermultiplets in harmonic superspace},
\href{https://doi.org/10.1016/j.physletb.2025.139964}{Phys. Lett. B \textbf{871} (2025), 139964}
[arXiv:2509.01406 [hep-th]].


\bibitem{Gargett:2025xcg}
T.~Gargett and I.~Samsonov,
{\it Analytic action principle in $\mathcal{N}=2$ $AdS_4$ harmonic superspace},
\href{https://doi.org/10.1103/jbh9-c8gg}{Phys. Rev. D \textbf{112} (2025) no.12, 125031}
[arXiv:2510.08905 [hep-th]].

\bibitem{Butter:2010jm}
D.~Butter and S.~M.~Kuzenko,
{\it New higher-derivative couplings in $4D \mathcal{N} = 2$ supergravity},
\href{https://doi.org/10.1007/JHEP03(2011)047}{JHEP \textbf{03} (2011), 047}
[arXiv:1012.5153 [hep-th]].

\bibitem{Butter:2011ym}
D.~Butter and S.~M.~Kuzenko,
{\it $\mathcal{N}=2$ AdS supergravity and supercurrents},
\href{https://doi.org/10.1007/JHEP07(2011)081}{JHEP \textbf{07} (2011), 081}
[arXiv:1104.2153 [hep-th]].

\bibitem{Park:1999pd}
J.~H.~Park,
{\it Superconformal symmetry and correlation functions},
\href{https://doi.org/10.1016/S0550-3213(99)00432-0}{Nucl. Phys. B \textbf{559} (1999), 455-501}
[arXiv:hep-th/9903230].

\bibitem{Butter:2010sc}
D.~Butter and S.~M.~Kuzenko,
{\it $\mathcal{N}=2$ supergravity and supercurrents},
\href{https://doi.org/10.1007/JHEP12(2010)080}{JHEP \textbf{12} (2010), 080}
[arXiv:1011.0339 [hep-th]].

\bibitem{Buchbinder:2025yef}
E.~I.~Buchbinder, S.~M.~Kuzenko and I.~B.~Samsonov,
{\it Massless higher-spin supermultiplets in 5D harmonic superspace},
\href{https://doi.org/10.1007/JHEP02(2026)122}{JHEP \textbf{02} (2026), 122}
[arXiv:2509.06604 [hep-th]].

\bibitem{Siegel:1978mj}
W.~Siegel and S.~J.~Gates, Jr.,
{\it Superfield Supergravity},
\href{https://doi.org/10.1016/0550-3213(79)90416-4}{Nucl. Phys. B \textbf{147} (1979), 77-104}.

\bibitem{Mezincescu:1979af}
L.~Mezincescu,
{\it On the superfield formulation of $O(2)$-supersymmetry
},
Dubna preprint JINR-P2-12572 (1979).


\bibitem{Rivelles:1981qz}
V.~O.~Rivelles and J.~G.~Taylor,
{\it Linearised $\mathcal{N} = 2$ superfield supergravity},
\href{https://doi.org/10.1088/0305-4470/15/1/025}{J. Phys. A \textbf{15} (1982), 163}.








\end{thebibliography}
\end{document}